\documentstyle[12pt]{article}

\large
\newcommand{\be}{\begin{eqnarray}}
\newcommand{\ee}{\end{eqnarray}}
\newcommand\del{\partial}

\newcommand\noi {\noindent}

\begin{document}
\setlength{\baselineskip}{21pt}
\pagestyle{empty}  
\vfill

\eject
\begin{flushright}                                                             
SUNY-NTG-95-5
\end{flushright}

\vskip 2.0cm 
\centerline{\large {\bf Hadrons and QCD Instantons :}}
\centerline{\bf A Bosonized View}
\vskip 2.0 cm

\centerline{M. Kacir, M. Prakash and I. Zahed} 
\vskip 1cm
\vskip .5cm
\centerline{Department of Physics}
\centerline{State University of New York at Stony Brook}
\centerline {Stony Brook, New York 11794-3800}
\vskip 2cm                                                                   
                                                                               
\centerline{\bf Abstract}

In a dilute system of instantons and antiinstantons, the
U$_{\rm A}$(1) and scale anomalies are shown to be directly related to the bulk
susceptibility and compressibility of the system. Using $1/N_c$ (where $N_c$ is
the number of  colors) as a book-keeping argument, mesonic, baryonic and
gluonic correlators are worked out in p-space and Fourier transformed to
x-space for a comparison with recently simulated correlators. The results are
in overall agreement with simulations and lattice calculations, for distances
up to 1.5 fm, despite the fact that some channels lack the  necessary physical
singularities.  We analyze various space-like form factors of the nucleon and 
show that they are amenable to constituent quark form factors to leading
order in $1/N_c$.  Issues related to the lack of confinement in the model and
its consequence on the various correlation functions and form factors are also
discussed.

\vfill
\noindent
\begin{flushleft}
SUNY-NTG-95-5\\
December 1995
\end{flushleft}
\eject
\pagestyle{plain}

\section{Introduction.}

An outstanding problem in QCD is the understanding of the hadronic spectrum 
from first principles. Decades of dedicated lattice simulations
 have shown that the problem is difficult when all QCD 
degrees of freedom are taken into account.
Through the years there have been numerous proposals, both theoretical and 
numerical, which suggest that only some relevant degrees of freedom may be 
important for the bulk aspects of the hadronic spectrum. The proposals range 
from lattice cooling procedures \cite{LATTICE} to semi-classical techniques
\cite{CLASSICAL}.

Recent lattice simulations based on cooling procedures have suggested that 
instanton and antiinstanton configurations may account for a large part of  the
hadronic correlations \cite{COOLING,NEGELE}, although the local character of
the cooling algorithms may not totally rule out persistent quantum effects at
large distances \cite{TEPER}. On a periodic lattice (without twists), 
instanton or antiinstanton configurations are necessarily singular 
\cite{POLONYI}. Their continuum analogs are the BPST instantons in singular 
gauge, a point of some recent concern \cite{PROBLEM}. Notwithstanding such
concerns,  an impressive amount of results, both from  cooled lattice 
simulations \cite{COOLING,NEGELE} and from random instanton simulations
\cite{SHU}, seem to indicate that the basic features of the hadronic  spectrum
may emerge from a dilute ensemble of singular instantons and  antiinstantons. 

Sometime ago,  't Hooft's suggested \cite{THOOFT} that instantons provide the
answer to  the axial U$_{\rm A}$(1) problem. In the presence of instantons or 
antiinstantons, light quarks acquire zero modes, which bunch into
flavour-singlet configurations ('t Hooft's vertices) thereby dynamically
breaking the axial U$_{\rm A}$(1) symmetry. At low energy,  't Hooft's 
interactions provide interesting correlations in various hadronic channels, as 
noted by Callan, Dashen and Gross \cite{CLASSICAL}, and analyzed using QCD sum
rules \cite{SUM}, resummation procedures \cite{RESUM,DYAKONOV},  instanton
simulations \cite{SHURYAK}, and bosonisation  techniques
\cite{DYAKONOV1,NVZ1,ALKOFER,BOSON}.

In this paper, we will assume that the QCD partition function simplifies into 
a grand canonical ensemble composed of 't Hooft's vertices, with an apriorily 
unspecified measure for the instanton-antiinstanton interactions. We will 
further assume that the ensemble is dilute with a screened topological charge,
as discussed in Refs. \cite{BOSON,NOWAK,NOWAK1}. The screening is
expected from the feedback of the light quarks on the instantons and
antiinstantons  in the 
vacuum \cite{ZAHED,VERSHU}. In this respect, problems related
to the original  choice of the instanton-antiinstanton ansatz
\cite{DYAKONOV2},  as well as the limitations associated with the
streamline approach  \cite{VERBAARSCHOT}, are somehow irrelevant.
Analytical and numerical  calculations with such an ensemble have led
to a satisfactory phenomenology \cite{SHU,DYAKONOV,NOWAK}.

The purpose of this paper is to show that the analytically derived results
using either resummation techniques \cite{DYAKONOV} or bosonisation techniques
\cite{DYAKONOV1,NOWAK} for two and three flavours in momentum space, are
consistent with the recent four-dimensional simulations \cite{SHU} as
well as with the cooled and quenched lattice simulations \cite{NEGELE}
up to a distance of 1.5 fm. At larger distances, the lack of confinement 
shows up in the form of spurious oscillations.
The physics of a screened gas of instantons and  antiinstantons is
well described by simple mean field arguments  \cite{DYAKONOV1,NOWAK}.
In section 2, we recall the effective action for a random instanton gas using
an approximate bosonisation scheme. In section 3, we discuss the structure of 
the massive quark propagator both in momentum and coordinate  spaces, and 
comment on heavy-light correlators. We note
that in the long wavelength limit, the quark propagator becomes  tachyonic for
all current quark masses. In section 4, we outline the result for the quark
condensate in the random instanton gas. In section 5, we give a brief account
of the various mesonic  correlators, including the scalars. We discuss 
issues related to the mixing between the scalars and the fluctuations in the
instanton scalar density through the scale anomaly. The mixing between the
pseudoscalar singlet and the fluctuations in the instanton pseudoscalar density
yields naturally to a resolution of the U$_{\rm A}$(1) problem.
 Issues related to the 
$\eta -\eta'$ mixing are also discussed. The
p-space results are discussed in detail for scalars, pseudoscalars and vectors,
and compared with the x-space simulations. While the analysis of the
p-space pseudoscalar correlator 
shows clear evidence of poles, the vector correlators
simply exhibit  two constituent quarks.  In section 6, we briefly discuss
nonstrange baryons in x-space. In section 7, scalar and pseudoscalar
 gluonic correlation functions are discussed.
 In section 8, quark and gluon form factors of the constituent quark are 
discussed. To leading order in $1/N_c$, they saturate the nucleon form factor 
following from the point-to-point correlator in terms of Ioffe's current.
 Our conclusions and recommendations are summarised in 
section 9.                                        

The details of  the bosonisation techniques are given in Appendix A. In
Appendix B, we  provide a direct calculation of the quark condensate. In
Appendix C, the  necessary elements for a Gaussian approximation are presented. 
In Appendix D, the effective action for the singlet and octet pseudoscalars is
explicitly worked out. 
In Appendix E, an extension bosonization scheme is presented.
In Appendix F, the various expressions  entering the
unconnected parts of the mesonic correlators are  summarized. Some of the
difficulties related with the expansion of  the mesonic vertices involving
the strange quark mass are discussed in Appendix G. In Appendix H,
 we outline the essentials of our numerical procedures. 

\section{Model}

$\bullet\,\,${\underline {Effective action}}

t'Hooft has shown that at scales larger
than a typical instanton size $\rho$ (fixed throughout this paper), 
instantons induce flavour mixing between the light $u,d$ and $s$ quarks in the
form of determinantal interactions (t'Hooft determinants) \cite{THOOFT}
\be
{\rm {det}}_{\pm} = \frac 1{N_f!}{\rm {det}}_{fg}
\left( m^{fg}\rho -\frac{\rho}{2i}
\langle\int\psi^{\dagger f}S_0^{-1}\phi^{\pm}\int\phi^{\pm \dagger }S_0^{-1}\psi^g\rangle\right)
\label{det}
\ee
where $m^{fg} ={\rm diag} (m,m,m_s)$ is the current mass matrix for  $(u,d,s)$
quarks,  $\phi_{\pm}$ are the instanton-antinstanton zero modes, $\psi$ is the
fermion  field in the long wavelength limit, and  $S^{-1}_0 = -\left(i\partial
\!\!\!/ +im\right) $ is the free fermion propagator. The averaging implied by
$\langle\cdots\rangle$  is over the instanton and antiinstanton color
orientations.

A random system of instantons and antiinstantons that is compatible with the 
U$_{\rm A}$(1) and scale anomaly yields the generating functional
\cite{THOOFT,ALKOFER,NOWAK1}
\be
Z[\eta , \eta^{\dagger}] = \int dn_+ dn_- {\cal D}\psi {\cal D}\psi^{\dagger}
\,\, \mu (n_+, n_-)\,\,e^{-\int d^4z {\cal L} [\eta , \eta^{\dagger}, n_+, n_-]}
\label{action}
\ee
where
\be
{\cal L} [\eta , \eta^{\dagger},n_+,n_-]=
\psi^{\dagger}S_0^{-1} \psi 
- n_{+}\log {\rm {det}}_{+} -n_{-} \log {\rm {det}}_{-} 
- \psi^{\dagger}\eta -\eta^{\dagger}\psi 
\label{aaction}
\ee
in (\ref{action}) at the saddle points. Throughout this paper, the generating 
functional will be used to carry calculations to leading order in $1/N_c$, where
$N_c$ counts the number of colors. The counting will be understood just as 
a convenient way of organizing the calculation, with $N_c=3$. In the presence 
of instantons, conventional $N_c$ arguments have to be amended \cite{WITTEN}
(see also below).

We are using a coarse grained action for the description of the instantons and 
antiinstantons, as discussed in \cite{NOWAK1}. The relation to the 
uncoarse-grained approach, follows from the identification
\be
n^{\pm} (z) = \pm \sum^{N_{\pm}}_{i=1} \delta^4 (z-z_i)
\ee
at the saddle points ($N_c=3 >>1$).  
The coarse grained version highlights the
role of the scalar and pseudoscalar glueball fields, and 
their mixing to the quark-antiquark excitations. 
The measure $\mu (n_+,n_-)$ refers to the distribution of instantons
and antiinstantons in the vacuum without the quarks (quenched approximation). 
Its form is generic, and follows solely from the U$_{\rm A}$(1) 
and scale anomalies
\cite{NOWAK1}
\be
\mu (n_+, n_-) = {\rm exp} \bigg(&-&
\frac n{\sigma_*^2}  \int d^4z (n^+ (z) + n^- (z) ) \bigg( 
{\rm log} \frac {n^+ (z) +n^- (z)}{n} \,\,-1\bigg)\nonumber\\
&-&\frac 1{2\chi_*} \int d^4z \,\, (n^+ (z) -n^-(z))^2 \bigg)
\label{measure}
\ee
where $n=N/V_4\sim N_c$ is the mean instanton and antiinstanton density in the 
thermodynamical limit, $\chi^* =n \sim N_c$ \cite{DYAKONOV2} the
quenched topological susceptibility 
\be
\chi_* = <\bigg( \int d^4z (n_+ -n_-) (z) \bigg)^2 >_{N_f=0} 
\label{susceptibility}
\ee
and $\sigma_*$ the quenched compressibility  ($a=1,2,...$)
\be
\bigg( \frac {\sigma_*^2}{n} \bigg)^{a-1} = \frac 1N
<\bigg( \int d^4z  \, (n^+ + n^- - n) (z) \bigg)^a >_{N_f=0}
\label{relations}
\ee

\vskip .5cm
$\bullet\,\,${\underline {Anomalies}}

The coarse grained effective action (\ref{aaction}) along with the measure 
(\ref{measure}) satisfies both the axial U(1) and scale anomaly \cite{NOWAK1}.
 Indeed, in the chiral limit the determinants in (\ref{action}) acquire a 
phase under a U(1) axial rotation, hence a non-conserved axial-singlet 
current,
\be
\partial_{\mu} j_{\mu 5} (z) =  2 N_f ( n_+ -n_-) (z) 
+ 2i\ {\rm Tr}_{f} \ m\ \psi ^{\dagger}(z)\gamma_{5}\psi (z)
\label{U1}
\ee
Also, in the quenched approximation, the measure (\ref{measure}) is not
scale invariant. As a result, the divergence of the dilatational
current (trace of the energy momentum tensor $\Theta_{\mu\nu}$), is not 
conserved,
\be
\Theta_{\mu\mu} (z) = \frac {4n}{\sigma_*^2}\, \, (n^+ (z) + n^- (z)) 
+ \frac {2}{\chi_*} (n_+ (z) - n_- (z) )^2 +{\cal O} (N_f)
\label{atrace}
\ee
Comparison with the QCD form of the trace anomaly \cite{QCDANOMALY}
gives $\sigma_*^2/ n\sim  {12}/{11N_c} \,\, \sim 1/N_c$.

In the fermionic part, we note that for $n_+ =n_-$, the generating functional 
(\ref{action}) involves  only the combination
(${\rm {det}}_+\,{\rm {det}}_-$) and is invariant under $U_L(3)\times
U_R(3)$. The fluctuations in $(n^++n^-)$ involve mixing between
the isosinglet scalar and the scalar ``glueballs", through
\be
-\frac 12 \int d^4z (n^++n^-) (z) \,\, {\rm ln} \bigg( {\rm det_+}\times {\rm 
det_-}\bigg) 
\ee
For $n_+ \neq n_-$, the combination (${\rm{det}}_+{\rm{/det}}_-$) is
also allowed. The latter dynamically breaks the axial U(1) symmetry through
\be
-\frac 12 \int d^4z (n^++n^-) (z) \,\, {\rm ln} \bigg( \frac{\rm det_+}{\rm 
det_-}\bigg) 
\ee
as originally suggested by t'Hooft.
The fluctuations in $(n^+- n^-)$ will mix with the isosinglet pseudoscalar,
thereby resolving the U$_{\rm A}$(1) problem (see below).

\vskip .5cm
$\bullet\,\,${\underline {Bosonization}}

In vacuum, the packing fraction is given by the dimensionless combination
\be
n_* \rho^4 = \frac{n}{2N_c} \rho^4 \sim 10^{-3}
\label{factor}
\ee
Since the density $n\sim N_c$, the packing fraction is of order $N_c^0$. 
The value (\ref{factor}) is small, and an expansion in the density is  
justified except in the presence of infrared singularities, as we will
specify below. In this spirit, 
the generating functional (\ref{action}) can be bosonized approximately
by inserting the identity \footnote{In what follows, we use the
shorthand notation $dk=d^{4}k/(2\pi)^{4}$ and $dz=d^{4}z$ when
integrating.}
 
\be
{\bf 1}\ =\ \int \! {\cal D}\pi^{\pm} {\cal D}P^{\pm}
\exp \bigg({\rm Tr}_{f}\int \!dk dl\ P^{\pm}(k,l)
\bigg(\pi^{\pm}(k,l) - \theta ^{\pm}(k,l)
\bigg)\bigg)
\label{identity}
\ee
in the partition function (\ref{action}), 
where $\pi^{\pm}$ and $P^{\pm}$ are bilocal auxiliary fields 
 and $N_{f}\times N_f$ valued such that 
$P^{\pm}(k,l)=P^{\pm}(k-l)$ and similarly for $\pi^{\pm}(k,l)$.
Also (see appendix A)
\be
\theta ^{\pm}(k,l)=
\langle \psi^{\dagger }(k)S_0^{-1}\phi^{\pm}(k)
\phi^{\pm \dagger }(l)S_0^{-1}\psi (l)
\rangle 
\ee
The trace over flavor indices is understood in the exponent (\ref{identity}). 
The auxiliary fields $\pi^{\pm}$ can be eliminated by 
using the saddle point approximation. From Appendix A, we have
\be
Z[\eta , \eta^{\dagger }] =
\int {\cal D}P^{\pm} 
\,{\rm e}^{-\eta^{\dagger }{\bf S}[P^+,P^-]\eta}
\,\,\,{\rm e}^{-{\rm S_{eff}(P^{\pm})}}
\label{partition}
\ee
where the effective action is given by
\begin{eqnarray}
S_{{\rm eff}}(P^{\pm})=
&-&N_c{\rm Tr}(\log {\rm {\bf S}}^{-1}[P^{+}, P^{-}]) +\frac {n}{2}
\int dz \,
\bigg({\rm Tr \,\,ln} \frac{4}{n \rho} P^{+}(z) + {\rm Tr\,\,ln} 
\frac{4}{n \rho}P^{-} (z)\bigg)
\nonumber \\
&-&2 \int \!dz {\rm Tr}_{f} m(P^{+}(z) +P^{-}(z))
\label{boson}
\end{eqnarray}
The trace ${\rm Tr }$ is over flavour and Dirac indices as well as four
momenta, the ${\rm Tr}_{f}$ over flavour indices and the ${\rm det}$ is over
flavour indices as well position space. We have explicitly used 
\footnote{This constraint will be relaxed below to address the $\eta '$ mass 
and the gluon correlators.}
$n_+=n_-=n/2$ and defined the operator momentum dependent inverse propagator
\be
{\bf S}^{-1}[P^+, P^-]= \hat {k}\!\!\!/ - im -
i\sqrt{M_{k}}
\left(1-i\frac{m\hat {k}\!\!\!/}{k^2}\right)\left(P^+\gamma_5^{+}+P^-\gamma_5^-\right)
\left(1-i\frac{m\hat {k}\!\!\!/}{k^2}\right)\sqrt{M_{k}} 
\label{propagator}
\ee
Here, $\hat{k}$ refers to the momentum operator and
$M_k$ is the induced momentum dependent screening quark mass.  
The screening mass arises 
from averaging over the instanton zero modes \cite{DYAKONOV}.
From Appendix C, we have after rescaling \\
\be
M_k = M_k (m) = \lambda (m) \frac {n}{N_c} k^2\phi'^2 =
                 \lambda (0) \frac {n}{N_c} k^2\left(\pi \rho^2
\frac{d}{d\tau} \left(I_0K_0 -I_1K_1\right)\right)^{2}
\label {mass}
\ee
\\
where $\tau =k\rho/2$ is the argument of the McDonald functions $I$ and $K$.
From (\ref{boson}) and (\ref{propagator}) it follows that in the 
long wavelength limit quarks
in the instanton vacuum interact via the exchange of
effective bosonic fields $P^{\pm}$. The latter are 
$N_f\times N_f$ valued and may be parametrized as 
\be
P^{\pm} = e^{\pm \frac 12 i \kappa} \sigma e^{\pm \frac 12 i\kappa}
\label{para}
\ee
Other parametrizations are also possible \cite{DIASPRIN}.
We note, however, that to the order we will discuss the correlation 
functions below (gaussian approximation), the results are 
parametrization independent. 

\vskip .5cm
$\bullet\,\,${\underline {Gap Equation}}

The matrices $\kappa$ and $\sigma$ are $N_f\times N_f$ valued and hermitean. 
The $\kappa$ variables can be identified as the pseudoscalar Goldstone modes,
except for $\kappa_0$. The matrix $\sigma$ contains the 
massive scalar-isoscalar and scalar-isovector excitations. 
The non zero value of $\sigma$ in the vacuum follows from the saddle point 
approximation to (\ref{boson}) by setting $\kappa =0$, and switching off the
sources. The result is an integral equation for each flavour
\\
\be
1-2m \lambda (m) =4\frac {N_c}n\int \frac{d^4k}{(2\pi)^4}
\frac{(k^2+m^2)M^2_k(m) -mM_k(m)k^2}{k^4-2mM_k(m)k^2 +(k^2+m^2)M_k^2(m)}
\label{gap}
\ee
\\
where we have set $M_k(m)/M_k(0) =\lambda (m)/\lambda (0)$. For an instanton 
density $n=1~{\rm fm}^{-4}$ and size $\rho=0.33~{\rm fm}$, 
the behaviour of the constituent quark mass $M_k (m)$ versus the dimensionless 
combination $z=k\rho/2$ is shown in Fig. 1 with 
current quark masses $m=0$ (dashed), $5$ MeV (dotted) and 10 MeV (solid),
respectively.
For $k\rightarrow 0$, $M_0(m)\rightarrow
\lambda \left(m\right) (n/N_c)(2\pi\rho)^2$,while for $k \gg 1$, $M_k(m)$ falls
off like $1/k^6$.  The width at half maximum is of order $1/\rho$.  The result
(\ref{gap}) was also obtained in \cite{ARMENIAN} using different arguments.

\section{Quark Propagator}

In a random instanton gas, quarks are ``screened". 
The light fermion propagator acquires a momentum dependent mass
\footnote{In general, the quark propagator is gauge dependent. Our case is no 
exception, and the present discussion should be understood as the evaluation
of the quark propagator in a random and classical background of instantons and 
antiinstantons in a singular gauge. }.
For one flavour, the results of Appendix A give (unless specified, we
denote $M_{k}(m)$ by $M_{k}$)
\\
\be
S(k,m) = \frac{1}{k\!\!\!/-im}-
\frac{1}{k\!\!\!/-i (m-k^2/M_{k})}
\label{prop}
\ee
\\
In the massless case \cite{DYAKONOV}
\be
S(k,0) =\frac{1}{k\!\!\!/-iM_k(0)}
\label{prop1}
\ee
We note, however, 
that at low momentum, $M_k (m)$ acquires a non-analytical contribution
($k\rightarrow 0$)
\\
\be
M_{k}(m) \sim M_{0}(m)\left(1+3z^{2}\log\frac{z}{2}{\rm e}^{c+\frac{1}{2}}\right)
_{z=k\rho /2}
\label{tachyon}
\ee
\\
where $c=0.577$ is Euler's constant. 
The screening mass $M_0 (m)$ does not show up as a simple pole.
What this means is that, as $k\rightarrow 0$, the screened quarks become 
tachyonic. To the extent that 
long wavelength quarks are unphysical, this should be of no real concern.
However, since the instanton model does not provide for 
confinement, these ``unphysical" effects will contaminate all 
large distance behaviours. 
The instanton simulations \cite{SHU}, or cooled lattice calculations
\cite{NEGELE} have not probed large distances.

With this in mind, we now proceed to x-space with the decomposition
\be
S(x,m)=S_{0}(x,m)+S_{1}(x,m)
\label{prop2}
\ee
where $ S(x, m)$ is the Fourier transform of (\ref{prop}) and
\\
\be
S_{0}(x,m)=\frac{im^{2}}{4\pi ^{2}}\left(\frac{x\!\!\!/}{x^{2}}
K_{2}(mx)+\frac{1}{x}K_{1}(mx)\right)
\ee
\\
is the free propagator of quark of mass m. $S_1 (x, m)$ follows from 
(\ref{prop2}) and will be understood as
\be
S_{1}(x,m)=i\left(x\!\!\!/
S_{1}^{odd}-S_{1}^{even}\right) (x, m)
\ee
\\
where
\\
\be
S_{1}^{odd}(x,m)=\frac{1}{x}\frac{\del }{\del x}\left(\frac{1}{4\pi ^{2}x}
\int _{0}^{+\infty}\!dk\frac{k^{2}}{k^{2}+(m-k^{2}/M_{k})^{2}}J_{1}(kx)\right)
\label{propodd}
\ee
\\
and
\\
\be
S_{1}^{even}(x,m)=\frac{1}{4\pi ^{2}x}
\int _{0}^{+\infty}\!dk \frac{k^{2}(m-k^{2}/M_{k})}{k^{2}+(m-k^{2}/M_{k})^{2}}
J_{1}(kx)
\label{propeven}
\ee
\\
Figs. 2a and 2b show  the behaviour of
$m{\rm Tr}S(x,m)/{\rm Tr}S_{0}(x,m=0)$ (chirality 
flip) and ${\rm Tr}\gamma _{4}S(x, m)/{\rm Tr}\gamma _{4}S_{0}(x,m=0)$ 
(chirality non-flip) versus  $x$ up to 2 fm,
for  quark masses of 5 MeV (lower curve) and 10 MeV (upper curve), 
respectively. 
The squares refer to the results of simulations of Ref. \cite{SHU} using 
128 instantons and 128 antiinstantons in a periodic Euclidean box
of $3.36^3\times 6.72$ fm$^4$. 
These simulations were carried out with equal 
$u$ and $d$ quark masses of $m=10$ MeV. The small discrepancy in the chirality 
flip part of the propagator may be due to the fact that the instanton 
simulations make explicit use of the single-instanton distorted propagator 
for the nonzero mode part, while the bosonized constructions presented above
make use of the undistorted light quark propagator for the nonzero mode part.

Figs. 2c and 2d show the chirality flip and non-flip part of the quark 
propagator for $m_u=10$ MeV and $m_s=140$ MeV over a wider range of $x$. The 
larger the quark mass, the larger the oscillation in the quark propagator 
at large distances. These spurious oscillations are due to the appearance of 
the tachyonic mass (\ref{tachyon}) and the occurence of the combination
$(m-k^2/M_k)$ in the quark propagator, and  will
cause most correlators to lack scaling at large distances (typically of the 
order of 2.5 fm and larger) as we will discuss below. We have checked that 
these oscillations persist in the massless case. In fact for $m=0$ Fig. 2c
is almost unchanged.

At this stage, we should point out  
that our treatment of the current masses is only approximate,
given our definitions (\ref{det}) and (\ref{action}). We will check below that 
the linear effects in the current mass do reproduce known results, while the
non-linear effects cancel out at large distances, leaving us with the expected
masses for the strange pseudoscalars. Similar observations apply to the 
instanton simulations in \cite{SHU}, although the handling of the 
current masses is not necessarily the same as the one discussed here.

Finally, we note that the naive interpretation that 
the x-space version of ${\rm Tr} (S(x,m) (1\pm \gamma_4)/2)$ as the
correlator of a light quark in the field of an infinitely heavy quark
\cite{SHU,RAD} overlooks the issue of binding. As it stands, the
non-relativistic projection of the heavy-light-propagator without the Wilson
line (Coulomb  field) for the heavy particle reflects solely on a screened
light quark.  In a heavy-light system like a D or B meson, the light quark  is
expected to bind to the heavy source,  causing the 
spectral function to develop a pole instead of a cut. A detailed analysis   of
systems with few heavy and light quarks in a random instanton gas has been 
given in Ref. \cite{CZ}. In the Coulomb field of a heavy quark, the light
quarks bind with a binding energy of  the order of a quarter of the screening
mass \cite{CZ}.

\section{Quark Condensate}

The formation of a quark condensate in the instanton vacuum follows from the
random nature of the system. From our bosonized construction, the 
quark  condensate is obtained from the effective action (\ref{boson}) through

\be
<\overline{\psi}\psi>
=\frac{1}{V_4}\frac{\del S_{{\rm_{eff}}}[0,0]}{\del m}
\ee
Since the present treatment is semi-classical, all the ambiguities associated
with the current mass singularities are ignored. 
At the saddle point, a straightforward calculation in the $m\rightarrow 0$
limit  gives 
\\
\be
<\overline{\psi}\psi> = -4N_{c} \left(\frac{n}{2N_{c}}\lambda (0)- \int \!dk \frac{M_k}{(k^2+M_k^2)}  \right) 
 - \frac{\lambda^{\prime}(0)}{\lambda(0)}
\left(4N_{c}\int\!dk\frac{M_{k}^{2}}{k^{2}+M_{k}^{2}}-n\right)
\ee
\\
Using the mass gap equation for zero current mass, the term  in brackets 
multiplying $\lambda^{\prime}(0)$ vanishes and we are left with 
\be
<\overline{\psi}\psi>=-4N_{c} \left(\frac{n}{2N_{c}}\lambda (0) - 
\int \frac{d^4k}{(2\pi)^4} \frac{M_k}{(k^2+M_k^2)}\right)  
\label{condensate}
\ee
As a check, we show in Appendix B how this result can be recovered 
from the original definition in the saddle point approximation,
prior to the bosonisation procedure. Numerically\footnote{With our choice of 
parameters, the discrepancy is $(10)$ MeV$^3$.},
\be
\frac n{2N_c}\lambda (0) = 2  
\int \frac{d^4k}{(2\pi)^4} \frac{M_k}{(k^2+M_k^2)}
\ee
so that
\be
<\overline{\psi}\psi>=-4N_{c} 
\int \frac{d^4k}{(2\pi)^4} \frac{M_k}{(k^2+M_k^2)}
=-\langle S(0, m\rightarrow 0_+)\rangle
\label{conexp}
\ee
which is the expected result to leading order in $1/N_c$.

\section{Mesonic Correlators}

To leading order in $1/N_c$, 
the mesonic correlation functions follow from (\ref{boson}) by differentiation 
with respect to the external sources in the presence of the auxiliary 
bosonic fields $P^{\pm}$. Generically,
\be
C_{\gamma} (x) = \langle T^* \psi^+\gamma\psi (x) 
\,\,\psi^+\gamma\psi (0)\rangle
\label{correlator}
\ee
with $\gamma = (1,\gamma_5, \gamma_{\mu}, \gamma_5\gamma_{\mu}, 
\sigma_{\mu,\nu})\otimes (1, T^a)$. From (\ref{boson}) we have
$C_{\gamma} =C_{\gamma}^0+C_{\gamma}^1$, 
where the connected part of the correlator is given by
\\
\be
C_{\gamma}^{0} (x) =-\frac{1}{Z[0,0]}\int \!{\it D}P^{\pm} 
{\rm Tr}\left({\bf S}[x,0; P]\gamma {\bf S}[0,x; P]\gamma \right)
e^{-S_{{\rm eff }}[P^{\pm}]}
\label{concorl}
\ee
\\
and the unconnected part is given by
\\
\be
C_{\gamma}^{1} (x) =\frac{1}{Z[0,0]}\int \!{\it D}P^{\pm}  
{\rm Tr}\left({\bf S}[x,x; P]\gamma \right)
{\rm Tr}\left({\bf S}[0,0; P]\gamma \right)
e^{-S_{{\rm eff }}[P^{\pm}]}
\label{unconcorl}
\ee
\\
Typical diagrams contributing to (\ref{correlator})
are shown in Figs. 3a and 3b. Only the  diagrams in  Figs. 3c and 3d are 
dominant.  They will be the only ones discussed here. 
In p-space, the contribution of Fig. 3c is
\\
\be
C_{\gamma}^0 (p) = -N_c \int \frac {d^4k}{(2\pi)^4}
{\rm Tr}
\left( S(1) \gamma S(2) \gamma\right)
\label{free}
\ee
\\
where $S(1,2)= S(k\pm p/2, m_{1,2})$ for two arbitrary flavors.
This contribution in the long wavelength limit reflects on the lack of 
confinement in the model. The contribution of Fig. 3d is
\\
\be
C_{\gamma}^1 (p) = \frac {N_c}2\sum_{\pm}
\frac{\left(R_{\gamma}^{\pm} (p)\pm R_{\gamma}^{\mp}(p)\right)\left(R_{\gamma}^{\pm}(-p)\pm R_{\gamma}^{\mp}(-p)\right)}
{\Delta_{\pm}(p)}
\label{corr}
\ee
\\
The extraction of $\Delta_{\pm}$ and $R_{\pm}$ from (\ref{boson})
is performed in Appendices B and C, respectively. With the above
approximation in mind, the total correlation function reduces to the sum 
of (\ref{free}) and (\ref{corr}), $i.e.$ 
$C_{\gamma} =C_{\gamma}^0+C_{\gamma}^1$. The results (\ref{free}-\ref{corr})
were first derived by Dyakonov and Petrov for two massless flavours using
detailed resummation procedures \cite{DYAKONOV}.

At this stage, it is interesting to compare the
expression we have for the mesonic correlator in the instanton model with the 
one derived in planar QCD$_2$. In the large $N_c$ limit,
the two-fermion cut in QCD$_2$ 
is infrared sensitive and cancels exactly against the infrared 
sensitive one-gluon exchange graph \cite{EINHORN}. 
This cancellation makes explicit use of 
Ward identities in Feynman graphs. It is essentially quantum and thus absent 
from the present semiclassical argument. The lack of confinement in our case
will have dramatic consequences on the large distance behavior of the various
correlation functions as we will discuss below.

The expressions used to generate the various correlators 
in p-space are tabulated in Appendix D.  In Figs. 4, we show 
the behaviour of the connected (minus the vacuum) correlators in the various 
channels versus the momentum $p$, for the Axial- (A), Vector- (V), 
Pseudoscalar- (P), Scalar- (S) and Tensor-channel (T),  without
strangeness (Fig. 4a) and with strangeness (Fig. 4b). Similar correlators are 
shown in Figs. 5 for the unconnected part. By about $p\sim 10$ fm$^{-1}$ the 
correlations are totally washed out. The plots are for $u$ and $d$
quark masses of 10 MeV  and a strange quark mass of 140 MeV. Although 
unconventional, this choice of the current masses allow for a comparison 
with the numerical simulations of Ref. \cite{SHU}.

\subsection{Gell-Mann-Oakes-Renner Relation}

In the pion channel, a pole is produced by the unconnected part 
of the correlator that lies well below the two-constituent quark cut.
This is a good example of an infrared sensitive channel, where a
simple expansion in the instanton density fails. The presence of small 
denominators through zero modes forces the resummation of an infinite string 
of terms of increasing powers in the instanton density, causing the correlation
function to develop a pole. 
Using  the small momentum expansion (see Appendix F), we have in
the pseudoscalar channel 
\\
\be
\Delta_{-} (p) = \frac {f^2}{4N_c} \left( M_{-}^2 + p^2 + {\cal O}(m^2,p^2)
\right)
\label{taylor}
\ee
\\
to leading order in the current quark mass $m$. Above, the decay constant $f$
satisfies
\\
\be
f^2 =  4N_c \int \frac{d^4k}{(2\pi)^4} \frac{M_k^2-\frac k2 M_k' +\frac{k^2}4M_k'^2}{(k^2+M_k^2)^2}
\ee
\\
and the pseudoscalar mass $M_-$ is given by
\\
\be
\frac{f^2}{4N_{c}}M_-^2 =2m\left(\frac{n}{2N_{c}}\lambda\left(m\right)-
 \int \frac{d^4k}{(2\pi)^4} \frac{M_k}{k^2+M_k^2}\right)
\label{mass}
\ee
\\
To this order, the quark condensate is current mass independent and is 
given by (\ref{condensate}). Thus,
\be
f_{\pi}^{2}m_{\pi}^{2}=-2m
<\overline{\psi}\psi>
\ee
which is the current algebra result derived by Gell-Mann, Oakes and Renner 
(GOR) \cite{GOR}. 
For equal 
$u$ and $d$ quark masses with $m=5$ MeV, we obtain $m_{\pi}=158$ MeV, 
$<\overline{\psi}\psi>=-(249\ {\rm MeV})^3$ and to leading order in the 
current mass $f_{\pi}= 88 $ MeV. Similar results can be derived for $K$
and $\eta$, although the small momentum expansion is no longer valid for the 
unconnected part of the correlation function with a large strange quark mass.
This point is further discussed in Appendix F.  

In the expansion discussed above, the consistency of the GOR
result can be further checked by noting that the unconnected part in the pion 
channel reads
\be
C_{\pi}^{1}(p\sim 0)=\frac{1}{2}\left(\frac{4N_{c}}{f\pi}R_{\pm}^{\gamma
 _{5}}\left(0\right)\right)^{2} \frac{1}{p^2+m_{\pi}^{2}}
\ee
where
\be
R_{\pm}^{\gamma_{5}}\left(0\right)=-2\int \frac{d^4k}{(2\pi)^4} \frac{M_k}{k^2+M_k^2}
\ee
The term in brackets in the expression for $C_{\pi}^{1}$ can be 
identified with the usual pseudoscalar strength $g_{\pi}$. 
From (\ref{condensate}-\ref{conexp}), 
it follows that $g_{\pi}\sim 2<\overline{\psi}\psi>/f_{\pi}$.

\subsection{Pseudoscalars}

\vskip 0.5cm
$\bullet$ {\underline {$\,\,\,\pi$ and $K$}}

Figs. 6 and 7 show the behaviour of the pion and kaon 
correlators versus $x$, respectively, as they follow 
from (\ref{free}) and (
\ref{corr}) by Fourier transforms. The upper curve is for 
 $m=5$ MeV, while the lower curve is for $m=10$ MeV.
  The squares are the results of simulations using 128 instantons and 128
 antiinstantons in a ($3.36^3\times 6.72$) fm $^{4}$ periodic box.
 The dotted circles are the results from cooled and quenched 
lattice gauge calculations on a $16^3\times 24$ lattice with a physical lattice 
spacing of $0.17$ fm.

The momentum dependent parts display a low-lying spurious cut at about 627 MeV,
as well as a pole in the scalar and pseudoscalar channels. 
In Fig. 8, we display these two separate contributions  to the pion channel
for an average quark mass of 10 MeV. 

The asymptotic form of the correlation function in x-space is strongly
influenced by the position of the pole in most channels. Indeed, the large 
distance behaviour produced by the pole is approximately of the form 
\\
\be
C_{\gamma}^1 (x\rightarrow \infty )\simeq
 \frac{(gM_{\pm})^2}{4}
\frac{e^{-M_{\pm} x}}{\left(2\pi M_{\pm}x\right)^{3/2}}
\label{pole}
\ee
\\
which is to be compared with the contribution of two ``regular"
(not tachyonic) screened quarks: 
\\
\be
C_{\gamma}^{0}(x\rightarrow \infty)\simeq
\frac{N_{c}}{4}M_{0}^{6}\frac{{\rm e}^{-2M_{0}x}}{(2 \pi M_{0}x)^3}
\left({\rm Tr}(\gamma \gamma)-\frac{{\rm Tr}(x\!\!\!/\gamma x\!\!\!/\gamma)}{x^2}\right)
\label{cuttwo}
\ee
\\
with $M_0 = M_0 (m)$. Beyond 2 fm, the running mass $(m-k^2/M_k)$
causes  (\ref{cuttwo}) to oscillate as shown in Figs. 9 and 10.
 These oscillations, however, are overpowered in the pion channel
given the very large signal caused 
by the pion pole compared to the spurious cut (about 100 : 1). We 
note that for $m=10$ MeV, the screening mass for the two screened quarks is 
about 627 MeV. In (\ref{pole}), the pseudoscalar mass squared $M^{2}_{-}$
follows  from the GOR relation          
\\
\be
M_-^{2} = - (m_1 +m_2) \frac {<\overline{\psi}\psi>
}{f^2}
\ee
\\
In the case of the pion, we plot in Fig. 11 the total correlator times
$x^{3/2}$ for quark masses of $5$ and $10$ MeV, respectively. The pion
mass sets in at about 2.5 fm.
From the asymptote, we read a slope of $m_{\pi} \sim 157$ and
$215$ MeV,  respectively. The agreement of the slopes with the GOR result
provides consistency checks on the various Fourier transforms performed.  We
stress that to read  the masses through slopes requires a proper identification
of the preexponent  power (here $x^{-3/2}$). A raw plot of the total correlator
versus x does not  show any scaling up to 10 fm ! 
                                     
A similar analysis for the kaon channel is shown in Fig. 12, where 
only the rescaled and unconnected part $C^{1}(x)$ is shown. The
connected part oscillates at distances of the order of 2.5 fm and larger, 
as shown in Fig. 9b, for two strange quarks. In contrast to the pion channel, 
the ratio of 
the connected to unconnected parts in this case is about 5:1. 
The linear fall off in Fig. 12 sets in between 2 and 3 fm. From the slope, we 
read $m_K= 490$ MeV, for $m=5$ MeV and $m_s=140$ MeV. We note that all the 
non-linearities in the strange quark mass cancel out to give a kaon mass that
is compatible with the mass obtained by a naive use of the GOR relation,
as indicated above.

\vskip .5cm
$\bullet$ {\underline {$\,\,\,\eta$ and $\eta '$ }}

In the $\eta$ and $\eta '$ channels, the situation is a bit more subtle 
because of mixing and the anomaly. First, let us follow the nonet 
decomposition used in Appendix D, for the singlet ($\kappa_0$) and
 the octet ($\kappa_8$) excitations.
 The connected part of the correlator in the  (00,08,88)
channels reads
\be
C^{0}(p)=-\int\!dk{\rm Tr}\left(\gamma _{5} \lambda_{\eta}S(k_{1},m)\gamma _{5} \lambda_{\eta}S(k_{2},m)\right)
\ee
with $\lambda _{\eta}$ any of the singlet or octet $U(3)$ generator. 
Specifically,
\\
\be
C_{0}^{0}(x)=&&\frac{4}{3}C^{0}(x,m_{u})+\frac{2}{3}C^{0}(x,m_{s})\nonumber \\ 
C_{8}^{0}(x)= &&\frac{2}{3}C^{0}(x,m_{u})+\frac{4}{3}C^{0}(x,m_{s})\nonumber \\
C_{08}^{0}(x)= &&\frac{2\sqrt{2}}{3}\left(C^{0}(x,m_{u})-C^{0}(x,m_{s})\right)
\label{corr0}
\ee
\\
with $C^{0}(x,m_{u})$ and $C^{0}(x,m_{s})$ the correlators of two screened 
$uu$ and $ss$ quarks, with $m_u = m_d =10$ MeV and $m_s=140$ MeV. The behaviour
of (\ref{corr0}) is shown in Fig. 13 versus $x$. The oscillations seen in all 
channels  beyond 2.5 fm are due to the spurious quark modes.

The unconnected  part of the correlators in the singlet, octet and mixed 
channels follow from the results of Appendix D. Since the $\eta_0$ and 
$\eta_8$ fields are integrated over, we can desentangle them by a unitary
rotation of angle $\theta$ ($\Delta =\Delta _{-}(k,m,m)$, $\Delta
_{s} = \Delta _{-}(k,m_{s},m_{s})$ and $z=\chi _{*}N_{f}/N_{c}$)
\be
\sin 2\theta (p)=\frac{4\sqrt{2}}{3}\frac{\Delta -\Delta _{s}}
{\lambda _{+}(k) -\lambda_{-}(k)}
\label{angle}
\ee
where
\be
\lambda _{\pm}(k)=\Delta +\Delta _{s}+\frac{z}{2}\pm
\bigg((\Delta - \Delta_{s} +\frac{z}{2})^{2}
+\frac{2z}{3}(\Delta_{s}-\Delta)\bigg)^{1/2}
\label{eigen}
\ee
at the expense of rotating the vertices (sources) as well. The result is
\\
\be
C^{1}_{0}(p)=2N_{c}\left(\frac{(R_{0}\cos\theta+R_{08}\sin\theta)^{2}}
{\lambda _{+}(k)}+
\frac{(-R_{0}\sin\theta+R_{08}\cos\theta)^{2}}{\lambda _{-}(k)}\right)
\ee
\\
\be
C^{1}_{8}(p)=2N_{c}\left(\frac{(R_{08}\cos\theta+R_{8}\sin\theta)^{2}}
{\lambda _{+}(k)}+\frac{(-R_{08}
\sin\theta+R_{8}\cos\theta)^{2}}{\lambda _{-}(k)}\right)
\ee
\\
\be
C^{1}_{08}(p)&=&2N_{c}\frac{(R_{0}\cos\theta+R_{08}\sin \theta)(R_{08}
\cos\theta+R_{8}\sin\theta)}{\lambda _{+}(k)}
\nonumber \\ 
\nonumber \\
&+&2N_{c}\frac{(-R_{0}\sin\theta+R_{08}\cos
\theta)(-R_{08}\sin\theta+R_{8}\cos\theta)}{\lambda _{-}(k)}
\ee
\\
The poles in the unconnected parts are just the $\eta$ and $\eta'$ masses, 
since we have rewritten the singlet and octet correlators in the 
$\eta$ and $\eta'$ basis
\footnote{Since the diagonalization is momentum dependent, it is 
not possible to devise a local source that would trigger precisely the $\eta$
or $\eta'$ quantum numbers without mixing.}. 
At low momentum $\theta \sim -13.1 ^{\circ }$, 
 which is to be compared with $\theta \sim -11.5 ^{\circ }$ in 
\cite{ALKOFER}. From the effective action 
of Appendix D, we conclude that to leading order in the current masses
\be
\lambda _{\pm}(k\sim 0)=&-&\frac{m+m_{s}}{2N_{c}} <\overline \psi \psi> 
+\frac{z}{2}
\nonumber \\
&\pm&\bigg((\frac{m_{s}-m}{2N_{c}} <\overline \psi \psi>
+\frac{z}{2} )^{2}+\frac{2z}{3}\frac{m_{s}-m}{2N_{c}} <\overline \psi
\psi>\bigg)^{1/2}
\label{eigen0}
\ee
Then,
\be
f^{2}m_{\eta^{\prime}}^{2}=2N_{c}\ \lambda _{+}(k\sim 0 )
\label{goretaprime}
\ee
and
\be
f^{2}m_{\eta}^{2}=2N_{c}\ \lambda _{-}(k\sim 0 )
\label{goreta}
\ee
The above relations give $m_{\eta'} = 1163$ MeV and $m_{\eta} =557$ MeV. These
values are to  be compared with
 $m_{\eta ^{\prime}}=1172$ MeV and $m_{\eta}=527$ MeV  for
 $<\overline\psi \psi> =$ ($-255$ MeV)$^{3}$ and $f=91$ MeV as
used in Ref.  \cite{ALKOFER}.
From (\ref{eigen0}), (\ref{goretaprime}) and (\ref{goreta}) we have
\be
f^{2}(m_{\eta ^{\prime}}^{2}+m_{\eta}^{2}-2m_{K}^{2})=2N_{f}\chi _{*}
\ee
which is the Veneziano-Witten formula \cite{{WITTEN}, {VENEZIANO}}.

Figs. 14a and 14b show the behaviour of the rescaled and unconnected
parts of the correlators versus $x$.
At about 3 fm, the asymptotic slopes set in. 
The large distance behaviour being dictated by the smaller pole, 
we obtain a slope of $220$ MeV  (essentially the pion mass) when the term 
$\chi_*N_f/N_c$ is switched off, and a slope of 480 MeV (essentially the $\eta$ mass)
when it is not. The contribution due to the large $\eta '$ mass dies off too
rapidly, as seen from the asymptotic behavior.  In this sense,  it is very hard
to measure the $\eta'$ characteristics from an x-space analysis of the
correlation functions. The x-space analysis of the topological 
susceptibilities offers a  better probe \cite{ZAHED}, 
although on the lattice there may be
subtleties related to the definition of gluonic sources.  
Finally, we note that in the
presence of the  connected parts of the correlator, no asymptote sets in
within 5 fm due again to the spurious oscillations discussed above.

We note that since $\chi_*=n\sim N_c$, the instanton-induced shift in the
$\eta$' mass $2\chi_* N_f/f^2\sim N_c^0$, at variance with Witten's argument
\cite{WITTEN}. This is not totally surprising, if we recall that the original
instanton gauge-configuration $A\sim 1/g\sim \sqrt{N_c}$. Also for $n\sim N_c$,
we have a fixed compressibility $\sigma_*\sim N_c^0$. However, when the density 
$n$ grows, the instanton and antiinstanton system is no longer dilute, and one
would a priori expect a phase change \cite{DYAKONOV2}, whence a breakdown of the 
conventional large $N_c$ arguments. The academic case of 
$n\sim N_c^0$ yields zero compressibility, with the quantum fluctuations
dwarfing the instanton effects.

\subsection{Scalars}

In Fig. 15 we plot the (normalized) connected $C^{0}(x)$ and
unconnected $C^{1}(x)$ parts of the correlator in short and long
dashed lines respectively.
 The solid line represents the sum of these two. As seen from Fig. 3d
the contribution from the first three diagrams is non vanishing in 
the scalar channel. If we were to repeat the calculation leading to
the unconnected part of the correlator in x-space we would obtain the 
additional term
\be
N_{c}\int \! {\rm Tr}\ S(k,m)
\left(N_{c}\int \! {\rm Tr}\ S(k,m) -2 \int \! \frac{M_{k}}{\Delta_{+}(k-l)}
{\rm Tr}\ C^{2}(k,m)\left( B(l,m)l\!\!\!/+iA(l,m)\right)\right)
\ee
If we recall the definition for the unconnected correlator we expect
this additional term to be amenable to the square of the condensate
$<\overline{\psi}\psi>$. The decay constant and the sigma meson mass
follow from the last diagram of Fig. 3d. They can be evaluated
through the use of a similar expansion of $\Delta _{-}$ to 
$\Delta _{+}$. Specifically
\\
\begin{eqnarray}
f^2 &=&  4N_c \int \frac{d^4k}{(2\pi)^4} \frac{M_k^2}{4{\cal D}^{2}}
\nonumber \\
&\times&
\left(
(\frac{ M_{k}^{\prime}}{k}+M_{k}^{\prime \prime})
(1-\frac{2M_{k}^{2}}{{\cal D}})
-\frac{M_{k}}{2{\cal D}}
\left(1+\frac{ M_{k}^{\prime 2}}{4}-\frac{k^{2}}{2{\cal D}}
(1+\frac{M_{k}M_{k}^{\prime }}{k})
^{2}
\right)\right)
\end{eqnarray}
\\
and the scalar mass $M_+$ is given by
\\
\be
\frac{f^2}{4N_{c}}M_+^2 =2m\left(\frac{n}{2N_{c}}\lambda\left(m\right)-
 \int \frac{d^4k}{(2\pi)^4} \frac{M_k}{\cal D}\right)
(1-\frac{8M_{k}^{2}k^{2}}{{\cal D}})
+\int \frac{1}{8\pi^{2}}\frac{4M_{k}K^{2}}{{\cal D}^{2}}
\label{sigmass}
\ee
\\
where ${\cal D}=k^2+M_{k}^{2}$. Numerically,  we obtain 
 $m_{\sigma}$=640 MeV and $f$=109 MeV. The scalar mass is about twice the 
constituent mass of $2\times 310 =620$ MeV. This is generic of all bosonized 
interactions at the mean-field level ($e.g.$ Nambu-Jona-Lasinio model). The 
nearness of the quark-antiquark threshold is expected to yield a large 
width for the scalar-isoscalar.

\subsection{Vectors}

\indent{$\bullet$ {\underline {$\,\,\, \rho$, $K^{*}$ and $\phi$} }}

To leading order in the instanton density, the vector 
correlation functions for both the $\rho$ and $\phi$ do not acquire
 any unconnected part.  (The vector 
correlation functions are just the correlation functions of two screened 
quarks.) This is expected, since the instanton-antiinstanton 
interaction acts primarily in the spin-isospin zero channel. Figs. 16
and 17 show respectively, the behaviour of the $\rho$- , and 
$\phi$-correlators versus $x$ up to 2 fm, for a light quark mass of $5$
MeV and a strange quark mass of $140$ MeV. The squares in these figures 
correspond to the instanton simulations,  while the filled circles
in Fig. 15 correspond to the cooled and quenched lattice simulations. 

The failure to produce correlations in the vector channel, while obvious in 
the p-space analysis, is implausible from the x-space analysis. In general,
simple spectral guesses as used in the instanton simulations or lattice 
calculations for an x-space analysis within 1 to 2 fms may be misleading.
They cannot differentiate between cuts and poles within 1.5 fm. At these 
distances it is difficult to reliably differentiate between poles and cuts 
(the pion-channel being an exception). A resolution of the two requires a 
careful analysis of the preexponents and the asymptotics, as we have discussed.

In the case of the $K^*$, it is clear that a contribution due to the mixing
between the up (down) and strange sectors occurs in the unconnected
part of the correlator. This is evident from appendix E, where we see that 
the coupling of the $\rho$ and $\phi$ to the quarks vanishes
identically, whereas in the case of the $K^*$ a contribution in 
${\cal O}(m_{s}-m_{u})$ arises. The possibility of the unconnected part
of the $K^*$-correlator 
being contaminated by the excitations of its scalar partner (in flavour
space) $\pi_{s}^{K}$ is allowed. 
Having said this we display in Fig. 18 the behaviour of the 
$K^*$-correlator versus $x$ up to 2 fm, for a light quark mass of both $5$
MeV (upper curve) and $10$ MeV (lower curve), and a strange quark mass
 of $140$ MeV. Again, the squares in this figure 
correspond to the instanton simulation.

\vskip .5cm
{$\bullet$ {\underline {$\,\,\, A_{1}$ and  $K_{1}$.} }}

Similar conclusions apply to the axial-vector correlators, although the latter
are contaminated by pion and kaon excitations through 
their longitudinal parts. Generically, the nonstrange axial-vector correlator
can be decomposed along the tranverse
and longitudinal directions that consist of the $A_{1}$ and $\pi$,
respectively: 
\be
C_{\mu \nu}(p)=(\delta_{\mu \nu}-\hat{p}_{\mu}\hat{p}_{\nu})C^{T}(p)+\hat{p}_{\mu}\hat{p}_{\nu}C^{L}(p)
\ee
From the p-space analysis, each contribution is well separated. 
$C^T$ contains solely 
a cut, while $C^L$ displays only a pole. Similar remarks apply to the strange
axial-vector correlator $K_1$. Figs. 19 and 20 show the behaviour of the 
combination $3C^T + C^L$ versus $x$ in the $A_{1}$ and $K_{1}$ channel, 
respectively. The squares refer to the results of simulations using instantons.

Since the longitudinal pole reflects on the pion pole, consistency with the 
pseudoscalar correlators requires that the pion properties (mass and decay 
constant) should be the same. The explicit form of the longitudinal part of the 
axial correlator reads
\be
C_{\mu \nu}^{L}(p)=\hat{p}_{\mu}\hat{p}_{\nu}2N_{c}
\frac{\left(R^{+}_{\gamma _{5} \gamma _{\mu}}(p)\right)^2}{\Delta _{-}(p)}
\ee
where at zero momentum
\be
R^{+}_{\gamma _{5} \gamma _{\mu}}(p=0)=2\int\frac{d^4k}{(2\pi)^4}\frac{M_{k}^2-kM_{k}M_{k}^{\prime}/2}{(k^2+M_{k}^2)^2}
\ee
which is just $f_{\pi}^2/2N_{c}$. Numerically, we obtain from the 
axial-correlator $f_{\pi}=76$ MeV, which is about $10\ \%$ off from the value
of $f_{\pi}=88$ MeV extracted from the pseudoscalar-correlator. This point 
illustrates some of the systematic uncertainties
introduced by the use of undistorted scattering states for 
the nonzero mode states around a single instanton or antiinstanton
\cite{DYAKONOV}.

\section{Baryon Correlators}

In the large $N_c$ limit, a baryon is made out of $N_c$ quarks, and is believed 
to be a soliton \cite{WITTEN}. In our case, we will think of a nucleon as made 
of $N_c=3 >>1$ quarks. To leading order in $1/N_c$, the nucleon is just three 
free streaming constituent quarks. In contrast to the meson case, the induced 
instanton (or gluon interaction) interaction between diquarks is subleading in 
$1/N_c$. We note that the soliton case in this model was considered in 
\cite{DYAKONOV2}.

Generically, the baryonic correlators will be defined to be
\be
R (x) = i\ \langle T^* J_{B}(x) \,\,J_{B}(0) \rangle
\ee
where we use for the nucleon and delta currents
\be
J^{N}(x)=&&\epsilon _{abc}\left(u^{a}(x)C\gamma_{\mu}u^{b}(x)\right)
\gamma ^{\mu}\gamma _{5}d^{c}(x)\nonumber \\
J^{\Delta}_{\mu}(x)=&&\epsilon _{abc}\left(u^{a}(x)C\gamma_{\mu}u^{b}(x)\right)
u^{c}(x)
\label{ioffe}
\ee
respectively.
Using Wick's theorem, we can reduce the nucleon and delta correlators into
(Minkowski)
\be
 R^{N}(x)=&&2 \epsilon _{abc} \epsilon _{a^{\prime}b^{\prime}c^{\prime}}
\gamma ^{\mu}\gamma_{5}
S^{cc^{\prime}}(x)
\gamma ^{\nu}\gamma_{5}
{\rm Tr}\left(\gamma_{\mu}S^{bb^{\prime}}(x)
\gamma_{\nu}S^{aa^{\prime}}(-x)\right)\nonumber \\
 R^{\Delta}(x)=&&3 \epsilon _{abc} \epsilon _{a^{\prime}b^{\prime}c^{\prime}}
S^{cc^{\prime}}(x)
{\rm Tr}\left(\gamma_{\mu}S^{bb^{\prime}}(x)
\gamma^{\nu}S^{aa^{\prime}}(-x)\right)
\label{ioffe1}
\ee
In the free case, (\ref{ioffe1}) reduces to
\be
i\frac{24x\!\!\!/}{\pi^6x^{10}}\ \ \ and \ \ \ -i\frac{18x\!\!\!/}{\pi^6x^{10}}
\ee 
respectively. 

Fig. 21 shows the behaviour of three constituent quarks versus $x$.
The two solid lines are for 5 and 10 MeV, respectively, 
the open circles are the results 
of instanton simulations and the full circles are those of quenched and cooled 
lattice simulations. Clearly, both simulations show attraction in the nucleon 
channel, which is very likely due to the fact that in the instanton model, the 
instanton induced interaction in a spin-zero isospin-zero diquark 
configuration  (qq)$_{I=0}^{J=0}$ is attractive. This follows from the large
attraction already observed in the spin-zero isospin-zero
quark-anti-quark configuration ($\overline q q$)$_{I=0}^{J=0}$ by crossing.
This attraction is, however, an order of magnitude 
smaller than the attraction in the pion channel. 
Whether these interactions can result in a pole remains an open question and 
requires a more detailed analysis. Indeed, the screened quarks
amount to a mass of about 940 MeV, which is close to the empirical value of the
nucleon  mass.

Fig. 22 shows the results  in the delta channel. From this, we 
conclude that the present simulations cannot distinguish between a cut 
and a pole in this channel. In fact, it is very unlikely that a dilute 
instanton gas can yield binding in decuplet channels, since the 
instanton induced interaction is usually non-existent in these channels.

\section {Gluonic Correlators}

The present construction allows for a convenient analysis of correlation 
functions involving $F\cdot F$ and $F\cdot \tilde F$ reflecting on the 
scalar and pseudoscalar glueballs in the model \cite{ZAHED,ALKOFER}. 
In the quenched approximation, these correlators are ultralocal and given 
by our choice of the measure (\ref{measure}). Through the identification
\be
\frac 1{32\pi^2} F\cdot F (x) = (n^+ + n^- ) (x) 
\ee
the scalar gluon correlator reads
\be
{\cal C}_{FF} (x-y) = &&
<T^* \frac 1{32\pi^2} F^2 (x) \,\, \frac 1{32\pi^2} F^2 (y)>_{\rm conn., N_f=0}
\nonumber\\ 
=&&< T^*\bigg((n^+ + n^- ) (x) - n\bigg)\bigg((n^+ + n^- ) (y) - n\bigg) >_{N_f=0}
\nonumber\\
=&& {\sigma_*^2} \,\, \delta^4 (x-y)
\label{sglue}
\ee
Also, through the identification
\be
\frac 1{32\pi^2} F\cdot\tilde F (x) = (n^+ - n^- ) (x) 
\ee
the pseudoscalar gluon correlator reads
\be
{\cal C}_{F\tilde F} (x-y) = &&
<T^* \frac 1{32\pi^2} F\tilde F (x) \,\, 
\frac 1{32\pi^2} F\tilde F (y)>_{N_f=0}
\nonumber\\ 
=&&<T^* (n^+ -n^- ) (x) \,\, (n^+ - n^- ) (y) >_{N_f=0}
\nonumber\\=&& \chi_*  \,\, \delta^4 (x-y)
\label{pglue}
\ee
In (\ref{action}),the glueballs in the quenched  approximation carry infinite
mass and zero size. They act as heavy sources.

In the presence of quarks, the glueball sources mix. 
The mixing is of order $1/N_c$. In the scalar channel,
\be
{\cal C}_{FF} (x-y) = &&
< T^* \bigg((n^+ + n^- ) (x) - n\bigg)\bigg((n^+ + n^- ) (y) - n\bigg) >
\nonumber\\
= && {\sigma_*^2} \bigg( \,\, \delta^4 (x-y) +  2N_f{\sigma_*^2}
<T^* \pi_0 (x) \pi_0 (y)> \bigg)  
\label{sglue1}
\ee
where the unconnected correlator in the $\lambda _{0}$ scalar channel is
\\
\be
<\pi_0 (x) \pi_0 (y)> =\frac{1}{2N_c }\int \! dk \ {\rm e}^{ik(x-y)}
\bigg(\frac{2/3}{2 \Delta _{+}(k,m,m)}+\frac{1/3}{2 \Delta
_{+}(k,m_{s},m_{s})}\bigg)
\label{eta0}
\ee
\\
The large separation behaviour of the above result follows from 
section 5 with the pole 
$m_0$ = 640 MeV as the mass of the scalar-isoscalar.
Because of the mixing, the fall-off is dictated by the
scalar-isoscalar masses. 
Fig. 23 shows the plot of the scalar correlator ${\cal C}_{FF} (x)$ 
(minus the ultralocal term). From (\ref{sglue1}) and (\ref{eta0}), the 
 compressibility takes the form

\be
\sigma ^{2}= \frac {1}{V_4} <\bigg( \int d^4 z (n_+ +n_--n) (z)\bigg)^2 >
\simeq \frac {4n}{b}
\label{gell1}
\ee
where $b$ is given by
\be
b=\frac{11 N_{c}}{3}-\frac{2N_{f}}{3}\alpha_{+}
\label{gell2}
\ee
with
\be
\alpha_{+}=\frac{n}{N_{c}} \sum _{f} \frac{1}{\Delta_{+}(k=0,m_{f},m_{f})}
\label{alphaplus}
\ee
Numerically we find $\alpha_+ =1.22$, which is to be compared with $\alpha_+=1$
in the QCD trace anomaly. This is only suggestive, however, since the two 
calculations are totally different in spirit. Ours is classical, while in QCD 
it is quantum.

In a similar way, we have in the pseudoscalar channel
\be
{\cal C}_{F\tilde F} (x-y) = &&
< T^*(n^+ -n^- ) (x) \,\, (n^+ - n^- ) (y) >
\nonumber\\
= && {\chi_*} \bigg( \,\, \delta^4 (x-y) -  2N_f{\chi_*}
<T^* \eta_0 (x) \eta_0 (y)> \bigg)  
\label{pglue}
\ee
where the unconnected correlator in the $\lambda _0 $ pseudoscalar
channel is
\\
\be
<\eta_0 (x) \eta_0 (y)> =\frac{1}{2N_c }\int \! dk \  {\rm e}^{ik(x-y)}
\bigg(\frac{\cos ^{2}\theta (k)}{ \lambda _{+}(k)}
+\frac{\sin ^{2}\theta (k)}{ \lambda _{-}(k)}\bigg)
\label{etacorl}
\ee
\\
 The rotation angle $\theta (k)$ along with $\lambda _{\pm}$ are
defined in (\ref{angle}) and (\ref{eigen}), respectively. The
fluctuations in the pseudoscalar gluonic source fall off with a rate
that is given by the lightest mass (the $\eta$ in our case) 
$m_{\eta}$ = 557 MeV.
Fig. 26b shows the plot of the scalar correlator ${\cal C}_{F\tilde F} (x)$ 
(minus the ultralocal term).
 Let us now evaluate ${\cal C}_{F\tilde F} (x-y) $ using the
pseudoscalar field decomposition, in which the quadratic part of the
action is diagonal. We obtain
\\
\be
{\cal C}_{F\tilde F} (x-y) 
= \int \!dk \ {\rm e}^{ik(x-y)} \bigg( \frac {1}{\chi_*} +
\frac{1}{2N_c }  \sum_{i=1}^{N_f} 
\frac {1}{\Delta _{-}(k,m_{i},m_{i})}\bigg)^{-1}
\label{ppglue}
\ee
\\
One should not be alarmed by the two different expressions for the 
pseudoscalar gluonic correlators (\ref{pglue}) and (\ref{ppglue}).
Using the two relations (denoting $\Delta =\Delta _{-}(k,m,m) $ and
$\Delta _{s}=\Delta _{-}(k,m_{s},m_{s}) $ )
\\
\be
\lambda _{+}(k) \lambda _{-}(k)=
4\Delta \Delta _{s}+\frac{2}{3}\frac{\chi_{*}N_{f}}{N_{c}}
(\Delta +2\Delta _{s})
\ee
and
\be
\lambda _{+}(k) \sin ^{2}\theta(k)+\lambda _{-}(k) \cos ^{2}\theta(k)=
\frac{2}{3}(\Delta+2\Delta _{s})
\ee
\\
we can easily rewrite (\ref{ppglue}) to (\ref{pglue}).
 The mixing causes the topological susceptibility to decrease. From 
(\ref{ppglue}) we have
\be
\chi = \frac {1}{V_4} <\bigg( \int d^4 z (n_+ -n_-) (z)\bigg)^2 >
= \frac {\chi_*}{1- \sum_{i=1}^{N_f} \frac {\chi_*}{m_i
 <\overline\psi \psi>}}
\label{topscree}
\ee
and vanishes for any quark mass going to zero. The topological charge
is totally screened in the chiral limit.

\section{Nucleon form factor}
All hadrons are characterized by various form factors, each of which carry 
information on the various charge and current distributions. In this part, we 
show how various nucleon form factors can be analyzed in $1/N_c$, thinking
of $N_c=3>>1$. In this section, we will 
distinguish between purely gluonic form factors ${\bf G}(x) \sim F^2 (x)\,\,, 
F\tilde F (x), \,\, \sigma_{\mu\nu} F^{\mu} F^{\nu}\,\, , ...$ and fermionic 
form factors ${\bf F}(x) = \psi^{\dagger} \, {\bf \Lambda} \psi$, where ${\bf 
\Lambda} =\gamma\otimes T$ is a spin-flavour matrix. Mixed form factors
${\bf M}(x) = \psi^{\dagger} \sigma_{\mu\nu} F_{\mu\nu} \psi, ...$ can be 
obtained in a similar way, although they will not be discussed here.

\vskip .4cm
${\bullet}$ {\underline{ Gluonic Form Factor of a Constituent Quark}}
\\
Since the model lacks confinement, the nucleon form factor receives 
contribution from the unconfined constituent quark states. This is
represented in Figs. 26(a) and (b) as $P^{\pm}$ insertions.
To leading order in $1/N_c$, these contributions are either direct as
shown in Fig. 26(a), or meson  mediated (Figs. 26(c) and (d)). The constituent
quark gluonic form factor is defined (when x $\rightarrow \infty$) as                            
\be
{\bf F}_G (k^2)  
\,\,\langle T^* \psi (\frac{ x}{2} ) \psi^{\dagger} (-\frac{ x}{2} )
\rangle
=\langle T^* \psi (\frac{x}{2} ){\bf G} (k) 
\psi ^{\dagger}(-\frac{x}{2}) \rangle _{\rm con.}
\label{form1}
\ee
with
\be
{\bf G} (k) = \int \, dy\, {\rm e}^{ik\cdot y}{\bf G} (y)
\label{form2}
\ee
Throughout, we will think of ${\bf F}_G$ as matrix valued (here in spin
space), so that various components of the form factor can be extracted by 
proper tracing.
To leading order in $1/N_c$, the form can be readily evaluated using the 
bosonization results developed in appendices C and D. The result is
($x\rightarrow \,\infty$)
\be
{\bf F}_{FF} (k^2) \,\, S(x, m) \,\, = 
i\int \!dp\ e^{ip\cdot x} \sqrt{M_{p_-}M_{p_+}} \,C(p_-)
\ 32 \pi^{2}\ \langle \pi_s (-k) \sigma (k) \rangle C(p_+) 
\label{form3}
\ee
for a scalar gluon insertion, and
\be
{\bf F}_{F\tilde F} (k^2) \,\, S(x, m) \,\, =
i\int \! \ dp e^{ip\cdot x} \sqrt{M_{p_-}M_{p_+}} \,C(p_-)
\ i 32\pi^2 \gamma _{5} \langle \pi_{ps} (-k) \chi (k) \rangle  C(p_+) 
\label{form4}
\ee
for a pseudoscalar gluon insertion. Here $p_{\pm} = p\pm k/2$.
The mixed spin-gluon matrix element can be
discussed using similar arguments. The expectations in (\ref{form3}-\ref{form4}) 
involve a Gaussian integral over the effective bosonic fields, with quadratic 
actions as discussed in Appendix C. After integration, the results are
(x$\rightarrow\,\infty$)
\be
{\bf F}_{FF} (k^2) \,\, S(x,m)
 = &+&i
\ \int \! \ dp e^{ip\cdot x} \sqrt{M_{p_-}M_{p_+}} \,C(p_-)\,C(p_+)
\nonumber \\
&\times& \frac{32\pi^{2}}
{2N_{c}\Delta _{+}(k,m,m)}
\bigg(\frac{n}{n_*\sigma_{*}^2}
+\frac{1}{2N_{c}}\sum_{f}\frac{1}{\Delta_{+}(k,m_{f},m_{f})}\bigg)
^{-1}
\label{form5}
\ee
and
\be
{\bf F}_{F{\tilde F}} (k^2) \,\, S(x,m) 
 = &+& i
\ \int \! \ dp e^{ip\cdot x} \sqrt{M_{p_-}M_{p_+}} \,C(p_-)\,\gamma
_{5}\,C(p_+)
\nonumber \\
&\times& \frac{32\pi^{2}}
{2N_{c}\Delta _{-}(k,m,m)}
\bigg(\frac{1}{\chi_{*}}
+\frac{1}{2N_{c}}\sum_{f}\frac{1}{\Delta_{-}(k,m_{f},m_{f})}\bigg)
^{-1}
\label{form6}
\ee

From our numerical analysis of section 3, the constituent quark propagator 
$S(x, m)$ shows a rough scaling in the window $0<x<2.5$ fm, with $M_0\sim 
300-400$ MeV, but then oscillates for $x>2.5$ fm, due to non-analyticities. In
the  window $0-2.5$ fm,                    
\be
S(x, m) \sim \frac {iM_0^2}{4\pi^2 x} \sqrt{\frac{\pi}{2M_0 x}} e^{-M_0 x}
(\hat{\rlap/x} +1 )
\label{form7}
\ee
It would be interesting to see how the present form factors 
(\ref{form5}-\ref{form6}) with (\ref{form7}) compare with simulations in the 
range $0<x<2.5$ fm. This is only indicative, since the channel is contaminated 
by spurious oscillations for $x>2.5$ fm.

The large $x$ separation provides for a way to select the constituent quark 
on its ``mass-shel"\footnote{This is, of course, suggestive in Euclidean space.},
hence the analogy with the Minkowski definition of the form factor. We can also 
define a totally ``off-shell" form factor by considering 
(\ref{form5}-\ref{form6}) for finite $x$ and integrating $x$ over $V_4$. In 
this way, one obtains off-mass shell form factors with zero-momentum 
constituent quarks. For $k=0$, the results are
\be
{\bf F}^*_{FF} (0)=\frac 1{2N_{c}\Delta _{-}(0,m,m)}\
\bigg(\frac{n}{n_*\sigma_{*}^2}
+\frac{1}{2N_{c}}\sum_{f}\frac{1}{\Delta_{+}(0,m_{f},m_{f})}\bigg)
^{-1}
\label{form8}
\ee
and
\be
{\bf F}^*_{F\tilde F} (0)=\frac 1{2N_{c}\Delta _{-}(0,m,m)}\
\bigg(\frac{1}{\chi_{*}}
+\frac{1}{2N_{c}}\sum_{f}\frac{1}{\Delta_{-}(0,m_{f},m_{f})}\bigg)
^{-1}
\label{form9}
\ee
To leading order in $1/N_c$, the ``off-shell" scalar form factor reduces to
\be
{\bf F}^*_{FF} (0)=\frac 3{11N_{c}} \alpha
\label{form10}
\ee
where $\alpha =2\alpha_+ \sim 2.45$. We note that (\ref{form10})
differs by almost a factor of 2 from its ``on-shell" analogue with
$\alpha=1$, as argued from a QCD low-energy theorem based on the trace anomaly
\cite{QCDANOMALY,WEISS}. For the pseudoscalar form factor (\ref{form9}) 
the result is ${\bf F}^*_{F\tilde F} (0)$=0.44, which is to be compared with 
the gluonic part of the ``on-shell" value of the axial-singlet 
form factor $g_A^0$, as determined from the U(1) anomaly (\ref{U1}) in the 
constituent quark state
\be
&g_A^{0} (0)&<T^{*} \psi^{\dagger}
(\frac{x}{2}) \gamma_{5} 
\psi (-\frac{x}{2})>
=
\nonumber \\
& &
<T^{*} \psi^{\dagger}(\frac{x}{2})\bigg(\int \!dz \
\frac{F\tilde F (z)} {32\pi^2} 
+\frac{i}{N_{f}}
\int \! dz\ {\rm Tr}_{f}\ m\psi^{\dagger}\gamma_{5}\psi (z)\bigg) 
\psi (-\frac x2)>_{conn.}
\label{form11}
\ee
The mass term in (\ref{form11}) involves the U$_{{\rm A}}$(1)
 form factor in the constituent quark state. 
In the last few years, efforts have been made to understand the data
from the European Muon Collaboration (EMC) \cite{VENEZIANO1}--
\cite{CHENG}.  One of its remarkable
results has been to yield a small value for the singlet axial coupling
constant $g_{{\rm A}}^{0} =0.13\pm 0.24$. The result obtained above ``off-mass 
shell" seems to be close to this value. The approach described here, provides 
some insights from a instanton vacuum model, to the effective approach 
discussed by many \cite{VENEZIANO1,SCHECHTER,CHENG}. In fact, the modified 
bosonisation scheme discussed in Appendix E, is very close in spirit to
these models. A comprehensive discussion of all these issues goes beyond the 
scope of this work.

\vskip .4cm
${\bullet}$ {\underline{ Fermionic Form Factor of a Constituent Quark}}
\\
The fermionic form factors can be analyzed in the same way as the gluonic form 
factors. The mechanism consisting of $P^{\pm}$ insertion is shown in 
Fig. 27. In Fig. 28, we show the leading contributions to the mesonic form 
factor to order $1/N_c$. Fig. 28(a) counts the bare charge, while Fig. 28(b)
involves a typical meson-exchange with non-local form factors. Generically
(x $\rightarrow\, \infty$), 
\be
{\bf F}_{\Lambda} (k^2)\,\, 
\langle T^* \psi (\frac{x}{2}) \psi^{\dagger} (-\frac{x}{2} ) \rangle
=
\langle T^* \psi (\frac{x}{2} )\,\,\psi^{\dagger}
{\bf \Lambda} \psi (k)\,\, \psi ^{\dagger}
 (-\frac{x}{2} ) \rangle_{\rm conn}
\label{form12}
\ee
Parametrizing all meson fields by $\pi^A=\gamma^A\pi^A$, where $\gamma^A =
(1,\gamma_5)\otimes T$, yields to leading order in $1/N_c$
\be
{\bf F}_{\Lambda} (k^2) \,\, S (x, m_{f}) =&-& \int \! dp\,dq\,
{\rm  e}^{i(p+k/2)\cdot x} \, \, 
 \sqrt{M_{p}}
\,C(p,m_{f}) 
\gamma
C (q+k,m_{g})\, \sqrt{M_{q+k}}\, \Lambda
\nonumber \\
&\times&
\sqrt{M_{q}}\, C (q,m_{g})\,\gamma \,C(k-p,m_{f})\, \sqrt{M_{k-p}}
\langle \pi^{fg} (p-q-k) \, \pi^{gf} (q+k-p) \rangle
\nonumber \\
&+&\int \! dp \,  {\rm e}^{i(p-k/2)\cdot x}
\, \sqrt{M_{p}}
\,C(p,m_{f}) 
\gamma C(p-k,m_{g})
\,\sqrt{M_{p-k}} 
\langle \pi^{fg} (k) \, \pi^{gf} (-k) \rangle
\nonumber \\
&\times&{\rm Tr} \left(\Lambda \sqrt{M_{q}}
C(q,m_{g}) \gamma C (q+k,m_{f})
\sqrt{M_{q+k}}
 \right)
\label{form13}
\ee
where summation over flavour $g$ is understood and $f$ is the flavour
of the quark being probed. $\gamma$ is 1 $(\gamma _{5})$ for 
the scalar (pseudoscalar) sector.
The first and second terms of (\ref{form13}) are
displayed in Fig. 28(a) and 28(b), respectively. 
The expectation value involves a Gaussian integration over the measure derived 
in Appendix C and can be evaluated for arbitrary momentum $q$. In the 
scalar sector
\be
\langle \pi_{s}^{fg} (q) \, \pi_{s}^{gf} (-q) \rangle
=&&\!\!\!\frac{1}{2N_{c}\Delta_{+}(q,m_{f},m_{f})}
\,\bigg(1+\frac{1}{2N_{c}\Delta_{+}(q,m_{f},m_{f})}\bigg)
\nonumber \\
&\times&
\bigg(\frac{n}{n_{*}\sigma_{*}^{2}}
+\sum _{g}\frac{1}{2N_{c}\Delta_{+}(q,m_{g},m_{g})}\bigg)^{-1}
\label{form14}
\ee
For the pseudoscalar case, we replace $\Delta _{+}$ by $\Delta _{-}$
and $n/(n_{*}\sigma_{*}^{2})$ by $1/\chi$. The case where $\gamma=\gamma_{\mu}$
(vector form factors) and $\gamma =\sigma_{\mu\nu}$ (tensor form factors) can 
be analyzed similarly.

It is interesting to note at this stage that most of these form 
factors may be used to assess the strength of the meson-constituent quark
interaction in some constituent quark models, as recently discussed by
Glozman and Riska  \cite{RISKA}. When couched in the $1/N_c$ framework, the 
present analysis provides some rationale for their successful phenomenology.

As in the gluonic case, we can investigate the ``off-shell" limit of the form 
factor at $k=0$. Using (\ref{form13}) for fixed $x$, integrating numerator and 
denominator over the entire $V_4$, and taking the $k=0$ limit, yields
\be
{\bf F}_{\Lambda} (0) \,\,S (0, m) =& -&
M_{0}^{2}\, C(0,m)\gamma\,\int\!dq\ M_{q}
\nonumber \\
&\times&
\bigg(C(q,m)\,\Lambda\,  C(q,m)\gamma\,-\,{\rm Tr}\left(
\Lambda \,  C(q,m)\gamma\, C(q,m)\right)\bigg)
\, C(0,m)
\label{limit1}
\ee
where all momenta are taken to be zero.
Numerically, the meson-meson expectation value
$\langle \pi_{s} (0) \, \pi_{s} (0) \rangle$
 in the scalar sector is
0.69 fm $^{-4}$ and 0.92 fm $^{-4}$, for the up (down) and strange 
quark, respectively. The same applies for the pseudoscalar
sector, where $\langle \pi_{ps} (0) \, \pi_{ps} (0) \rangle$
is 4.68 fm $^{-4}$ and 6.69 fm $^{-4}$, for the up (down)
and strange quark, respectively.
In short, formula (\ref{limit1}) along with the numerical values
for the meson-meson expectation value, will serve us as a check
point when numerically generating the values of
${\bf F}_{\Lambda} (k^{2}) $ using (\ref{form13}).

\vskip .4cm
${\bullet}$ {\underline{ Form Factors from Ioffe's Currents}}
\\
If we were to think about the nucleon as made out of three constituent quarks, 
then the nucleon form factor follows from the additive 
constituent quark picture. When simulations are performed, however, it is 
customary to use Ioffe's currents (\ref{ioffe}) for the nucleon. This results
in some non-trivial combinatorics and folding of the single constituent quark 
propagators, as we now explain. Let $J_N^{\alpha} (x)$ be Ioffe's current 
(\ref{ioffe}). Then, the nucleon form factor reads ($x\rightarrow\infty$)
\be
{\bf F}_N (k^2 )\, \
\langle T^* J_N^{\alpha} (\frac x2 ) 
\overline{J}_N^{\beta} (-\frac x2 ) \rangle
 =\langle T^* J_N^{\alpha} (\frac x2 ) {\bf O} (k) 
\overline{J}_N^{\beta} (-\frac x2 ) \rangle
\label{form15}
\ee
where ${\bf O} = {\bf G}, {\bf F}$, which are short for the gluonic and 
mesonic insertions discussed above. 
Typical diagrams for mesonic insertions are displayed in Fig. 29.
The term (L.H.S.)
multiplying ${\bf F}_N
(k^2 )$ in the left hand side of (\ref{form15}) can be
 readily reduced to give (\ref{ioffe1}). The right hand side (R.H.S.)
takes the form 
\be
{\rm R.H.S.} =&&+ 6\left( \gamma_{\mu} \gamma_5 {\cal OS} (x;k;m) 
\gamma_{\rho}\gamma_5\right)_{\alpha\beta} 
{\rm Tr}_s \left( \gamma_{\mu} S (x, m) \gamma_{\rho} S(-x, m) 
\right)\nonumber\\
&&+ 6\left( \gamma_{\mu} \gamma_5 S(x,m) 
\gamma_{\rho}\gamma_5\right)_{\alpha\beta} 
{\rm Tr}_s \left( \gamma_{\mu} {\cal OS}(x;k;m) \gamma_{\rho} S(-x, m) 
\right)\nonumber\\
&&+ 6\left( \gamma_{\mu} \gamma_5 S(x,m) 
\gamma_{\rho}\gamma_5\right)_{\alpha\beta} 
{\rm Tr}_s \left( \gamma_{\mu} S (x, m) \gamma_{\rho} {\cal OS}(-x,k,m) 
\right)
\label{form16}
\ee
where ${\cal OS} (x;k;m)$ follows from the right-hand side of (\ref{form3})
and (\ref{form4}) for the gluonic insertions, and, (\ref{form13}) for the 
mesonic insertions. 

It would be interesting to see how (\ref{form16})
compares to actual simulations. As noted above, the actual constituent quark 
propagator oscillates at distances larger than 2.5 fm. Hence, a true 
asymptotic 
form factor may not be reached in this model for the nucleon. In the region 
$0<x<2.5$ fm, the constituent quark propagator seems to be damped following the
behaviour described in (\ref{form7}). Using this behaviour, the left
hand side term in (\ref{form15}) reduces to
\be
{\rm L.H.S.} = &&\left( \frac {iM_0^2}{4\pi^2 x}\sqrt{\frac{\pi}{2M_0 
x}}e^{-M_0x}\right)^3\nonumber\\&&\times
6\left( \gamma_{\mu} \gamma_5 {\cal OS} (\hat{\rlap/x} +1 ) 
\gamma_{\rho}\gamma_5\right)_{\alpha\beta} 
{\rm Tr}_s \left( \gamma_{\mu} (\hat{\rlap/x} +1 )
\gamma_{\rho} (-\hat{\rlap/x} +1 ) \right)
\label{form17}
\ee
while the right hand side reduces to
\be
{\rm R.H.S.} = &6&\left( \frac {iM_0^2}{4\pi^2 x}\sqrt{\frac{\pi}{2M_0 
x}}e^{-M_0x}\right)^2
\nonumber \\
&\times&
\,\bigg(\left( \gamma_{\mu} \gamma_5 {\cal OS} (x;k;m)
\gamma_{\rho}\gamma_5\right)_{\alpha\beta} 
{\rm Tr}_s \left( \gamma_{\mu} (\hat{\rlap/x} +1 )
\gamma_{\rho} (-\hat{\rlap/x} +1 ) \right)
\nonumber\\
&+& \left(\gamma_{\mu} \gamma_5 (\hat{\rlap/x} +1 )
\gamma_{\rho}\gamma_5\right)_{\alpha\beta} 
{\rm Tr}_s \left( \gamma_{\mu} {\cal OS} (x;k;m)
\gamma_{\rho} (-\hat{\rlap/x} +1 )\right)
\nonumber \\
&+& \left(\gamma_{\mu} \gamma_5 (\hat{\rlap/x} +1 )
\gamma_{\rho}\gamma_5\right)_{\alpha\beta} 
{\rm Tr}_s \left( \gamma_{\mu} (\hat{\rlap/x} +1 )
\gamma_{\rho} {\cal OS} (-x;k;m) \right) \bigg)
\label{form18}
\ee
Numerical results for the resulting form factors will be given elsewhere.

\section{Discussion}

We have analysed the mesonic correlators  in a random instanton gas
in momentum space using bosonization techniques, and, in coordinate 
space by performing direct Fourier transforms. Our starting point  
was a grand-canonical ensemble of instantons and antiinstantons, where
the t' Hooft vertices play the role of ``fugacities".
The momentum space results are in agreement 
with the original analysis in both the massless \cite{DYAKONOV} and massive
cases \cite{NOWAK}. Following 't Hooft's 
suggestion, the resolution of the $\eta'$ problem follows  
by assuming that the topological charge is screened
\cite{NOWAK,ZAHED}, with a finite screening length (non-zero topological 
susceptibity). This effect is leading in $1/N_c$ counting  
 and results in a contribution of order $N_c^0$ to the $\eta'$ mass.
Without this effect, the $\eta'$ would be degenerate
with the $\eta$. 

We remark that a non-vanishing topological susceptibility should not be
taken for granted \cite{PROBLEM}. In the present 
case, it follows directly from the use
of instantons and antiinstantons in a singular gauge. A check would be to 
repeat the analysis using instantons and antiinstantons in a regular 
(non-singular) gauge, or, carry out cooled lattice simulations with free
boundary conditions.

Our x-space translation  of the p-space correlators shows that the results
of simulations 
using either a large sample of instantons and antiinstantons in four 
dimensions, or quenched and cooled lattice gauge configurations, are in 
agreement with the Fourier transformed analytical calculations within
the reported  range of (0-1.5) fm. The recent analysis carried out in Ref. 
\cite{HUTTER} for two flavours
differs from the bosonized results \cite{NOWAK} \footnote{Eq. (58) in
 Ref. \cite{HUTTER} relies on a resummation of the quark
propagator Eq.(57) which is valid only for zero quark mass.}, hence our
analysis.
                                  
We have shown that the running quark mass causes the quark propagator to 
oscillate at large x. The oscillations are larger for larger quark 
masses and affect most of the correlation functions at large distances. 
These effects are spurious and reflect on the lack of confinement in the model.
They are easily subtractable in a p-space analysis. They are harder to track
down in an x-space formulation. The extent to which these spurious modes impact
on the subtracted results is presently unclear.

We have shown that, while the asymptotics of suitably subtracted
correlators yield pseudoscalar  masses that are
accurate to within a few per cent,   the non-asymptotic readings could be as 
inaccurate as 100 \%. From our calculations, the subtracted and 
rescaled correlators show good
asymptotics between 2 and 3 fm. The non-rescaled correlators do not show any
reasonable asymptotics even up to 10 fm.  This point merits further scrutiny 
in lattice calculations.

The bosonized results show that while it is possible to infer the 
existence of light pseudoscalars in a dilute instanton gas, they do not seem 
to support the appearance of bound vectors. We have explicitly shown that the 
results of simulations are consistent with the presence of just screened
quarks in these channels. We have noted that the use  of schematic poles 
and cuts to analyze the x-space correlators in these channels would have 
implied otherwise. Due to mixing between the octet and singlet pseudoscalars,
we have found it difficult to extract the $\eta$ and $\eta'$ masses from 
the x-space analysis. The extraction is straightforward in the p-space 
analysis.

We have presented a simple analysis of the baryonic correlators in both the 
nucleon and the delta channels. The attraction seen in the nucleon channel is 
expected from general arguments. In this channel, however, it appears to be
difficult to identify a nucleon mass without going to the asymptotics,  since
three screened quarks already yield a mass of the order of 940 MeV.  This may
cause the nucleon to unbind, although soliton-inspired 
calculations with constituent
quarks seem to suggest otherwise \cite{PETROV}. In this respect, it would be 
interesting to repeat our analysis by including diquark fields. The results 
of simulations in the delta channel  are also consistent with three 
constituent quarks. A dilute instanton gas  does not induce correlations 
in the decuplet channels.

Using Ioffe's current for the nucleon, we have worked out various gluonic 
and mesonic form factors "on- and off-mass" shell, to leading order in $1/N_c$. 
The form factors are sensitive to the the three constituent quark 
cut. Moreover, the appearance of spurious oscillations in the single 
constituent quark propagators causes the form factors to be ill-defined for
point-to-point separations that are larger than 2.5 fm. In the region $0<x<2.5$ 
fm, some estimations have been made that would be of some interest for 
future simulations. The analysis of the nucleon form factor presented 
in this work could also be extended
to other mesonic and baryonic channels. It also provides 
insights in to some recently used constituent quark models \cite{RISKA}.

The fluctuations in the number sum and difference of the instantons and 
antiinstantons relate directly to the scalar and pseudoscalar glueball
correlation functions. In the quenched approximation, the glueballs are
infinitely heavy and stable. In the unquenched approximation, they mix 
with their scalar and pseudoscalar counterparts and decay. The mixing and
decay are of order $1/N_c$.

The overall agreement between the instanton simulations and the present
analysis within 1.5 fm shows that a random set of instantons and 
antiinstantons that is suitably stabilized in the infrared is well 
described by gaussian fluctuations over a mean field solution. The mean 
field solution follows from a simple bosonisation scheme. It also shows that 
constituent quark models with dynamically generated masses, $e.g.$ 
Nambu-Jona-Lasinio model, are also likely to give similar results provided 
that chiral symmetry is dynamically broken. In all these models, however, 
the subtle issue is that of confinement with its impact on large distance 
asymptotics and form factors.

\vskip 2cm
{\bf \noindent  Acknowledgements } \\ 
\noi 
This work was supported in part by the 
US Department of Energy under Grant No. DE-FG-88ER40388.

\newpage
\section{Appendices}
\vskip 2cm
\subsection{ Appendix A : Generating Functional}

In this Appendix, we provide the necessary details for the derivation of the 
generating functional (\ref{partition}) discussed in the text. 
Although these calculations were extensively used in establishing the
results of Refs. \cite{NOWAK}, they were never published.
We start by evaluating the color averages occuring in the 't Hooft determinants
(\ref{det}) for $N_f=1$. For convenience, we will use the shorthand notation 
$d^{4}k/(2\pi ) ^{4} \rightarrow \ dk$ and $d^{4}x \rightarrow \ dx$ when 
integrating out. If we denote by
\\
\be
\theta ^{\pm}(z)=\langle\int \!dx\psi^{\dagger}S_0^{-1}\phi^{\pm}(x-z)\int \! dy \phi^{\pm \dagger }(y-z)S_0^{-1}\psi (y)\rangle_{U\rho}
\label{average}
\ee
\\
then its Fourier transform reads
\\
\be
\theta ^{\pm}(z)=
\int \!dk\ dl\ e^{-i\left(k-l\right)z}\ \theta ^{\pm}(k,l)
\ee
with
\\
\be
\theta ^{\pm}(k,l)=
\psi^{\dagger}_{i,\alpha}(k)\left(k\!\!\!/-im\right)_{ij}
\langle\phi_{j,\alpha}^{\pm}(k)\phi_{k,\beta}^{\pm \dagger}(l)\rangle _{U\rho}
\left(l\!\!\!/-im\right)_{kl}
\psi _{l,\beta}(l)
\ee
\\
Averaging over the color group, we obtain \cite{NOWAK,NOWAK1}
\\
\be
\theta ^{\pm}(k,l)=
\frac{k\phi^{\prime}\left(k\right)l\phi^{\prime}\left(l\right)}{N_{c}}
 \psi^{\dagger}_{i,\alpha}(k)\left(\left(1-\frac{imk\!\!\!/}{k^2}\right)
\gamma _{5}^{\mp}
\left(1-\frac{iml\!\!\!/}{l^2}\right)\right)_{ij}
\psi _{j,\alpha}(l)
\ee
\\
where $\phi ^{\prime}(k)$ is the Fourier transform of the fermion zero mode
profile, and is given by
\\
\be
\phi ^{\prime}(k)=\pi \rho ^{2} \frac{\del}{\del z}(I_{0}(z)K_{0}(z)-I_{1}(z)K_{1}(z))_{z=k\rho/2}
\ee
\\
With the use of (\ref{identity}), the partition function (\ref{action}) 
takes the form 
\\
\begin{eqnarray}
Z[\eta,\eta ^{\dagger }] & = &\int {\cal D}\psi ^{\dagger}  {\cal D}\psi 
{\cal D}P^{\pm} {\cal D}\pi ^{\pm}\ (-2im)^{N}\ 
{\rm e}^{-\int \psi ^{\dagger} S_{0}^{-1} \psi -\psi ^{\dagger}\eta-\eta^{\dagger}\psi } \nonumber \\
& \times & 
\ \exp\ \frac{n}{2}\int \!dz\log 
\left(1 -\frac{1}{2im}\theta ^{\pm}(z)\right)
\nonumber \\
& \times & 
\ \exp \  i\int\!dk\,dl \ P^{\pm}(k,l)
\left(\pi^{\pm}(k,l)-\theta ^{\pm}(k,l)\right)
\ee
\\
where the integral in the last exponent is performed in both variables
$k$ and $l$ of the bilocal auxiliary fields.
 The field $\pi ^{\pm}$ is eliminated using the mean field equation 
\\
\be
-iP ^{\pm}(k,l)=\frac{n}{2}\int \! dz\ \ \frac{1}{1-\frac{1}{2im}\pi ^{\pm}(z)}\ \frac{{\rm e}^{-i(k-l)z}}{2im}
\label{MFEQ}
\ee
\\
For $N_f >1$, the auxillary fields $\pi ^{\pm}$ and $P ^{\pm}$ are 
$N_f\times N_f$ valued along with the average 
$\theta ^{\pm}$ (\ref{average}) entering the 't Hooft determinants 
(\ref{det}). As a result, 
additional traces over flavor indices will be needed. With this in mind, the 
previous results can be generalized in a straightforward way. For
$m={\rm diag}(m_{1},...,m_{N_{f}}) $, the result is (after absorbing in 
the measure a term in the size $\rho$ to have a 
dimensionless argument in the $\log $)
\\
\begin{eqnarray}
Z[\eta,\eta ^{\dagger }]& = &\int {\cal D}\psi ^{\dagger} {\cal D}\psi 
{\cal D}P^{\pm} \,
{\rm e}^{-\int \psi ^{\dagger} {\bf S}^{-1} [P^+, P^-]
\psi -\psi ^{\dagger}\eta-\eta^{\dagger}\psi } \nonumber \\
& \times &\ {\rm e}^{ -\frac{n}{2}\int \!dz{\rm Tr}_{f}\log\left(
\frac{4}{n\rho}P^{+}(z) \frac{4}{n\rho}P^{-}(z) 
\right)}
{\rm e}^{2\int \! dz {\rm Tr}_{f}m \left(P^{+}(z)+P^{-}(z)\right)}
\label{A1}
\ee
\\
where ${\bf S}^{-1}[P^+,P^-]$ is given in the text (\ref{propagator}). 
At the saddle point $P^{\pm}=P$, and ${\bf S}$ is the quark propagator 
in the external background $P$ such that in momentum space
\be
{\bf S}(k,l)=\delta ^{4}(k-l)\ S(k,m)
\ee
 with $S(k,m)$ written down in (\ref{prop}). From
(\ref{A1}), the partition function (\ref{partition}) follows after
 integration over the fermionic fields.

\subsection{Appendix B : Quark Condensate}

In this Appendix, we will show that (\ref{condensate}) follows from an exact
 derivation using the standard definition prior to the bosonization
procedure. Following the method used in Ref. \cite{NOWAK}, the
partition function (\ref{action}) can easily be written as (ignoring
fluctuations in the density and switching off the sources)
\begin{equation}
Z=\langle \int \!{\cal D}\psi ^{\dagger}{\cal D}\psi  {\rm e}^{ -\int \psi ^{\dagger}S^{-1}\psi}
\rangle
\label{partition2}
\end{equation}
where in the one flavour case
\be
S^{-1}=S^{-1}_{0}+\frac{1}{2im} S^{-1}_{0}\phi _{I}\phi _{I}^{\dagger}
S^{-1}_{0}
\label{inverse}
\ee
The (Euclidian) quark condensate follows as
\be
<\! \psi ^{\dagger} \psi\! >=
\frac{1}{V_{4}Z} \langle \int \!{\cal D}\psi ^{\dagger} {\cal D}\psi \int \psi ^{\dagger} \psi {\rm e}^{ -\int \psi ^{\dagger} S^{-1} \psi}\rangle
\ee
where averaging over all pseudoparticles is understood. Specifically,
\\
\be
<\!\psi ^{\dagger} \psi\!>=
-\frac{\langle{\rm Tr}S(0,m)\ det (-S^{-1})\rangle}{\langle det(-S^{-1})\rangle }
\label{cond}
\ee
\\
Introducing a set of Grassman variables for the pseudoparticle
ensemble, the partition function (\ref{partition2}) reads
(sum over $I,J$ understood)
\begin{equation}
Z=\langle \int \!{\cal D}\psi ^{\dagger}{\cal D}\psi {\cal D}\chi {\cal D}\chi ^{\dagger}  {\rm e}^{ -\int \psi ^{\dagger}S_{0}^{-1}\psi}{\rm e}^{\chi ^{\dagger}_{I}(T-im)_{IJ}\chi _{J}}
\rangle
\end{equation}
Here, $T$ is the kinetic part of the overlap matrix \cite{DYAKONOV,NOWAK} and
 the integration is over fermionic fields $\psi ,\psi^{\dagger}$ and
 Grassman variables $\chi_I, \chi_I^{\dagger}$, where $I$ is an integer that 
runs over all the instantons and antiinstantons in the ensemble.
Similarly, the condensate

\be
<\!\psi ^{\dagger} \psi\!>=\frac{1}{V_4Z}\langle 
\int \!{\cal D}\psi ^{\dagger}{\cal D}\psi {\cal D}\chi {\cal D}
\chi ^{\dagger}  
\left(\int \psi^{\dagger}\psi
-\chi _{I}^{\dagger}\chi _{I}\right)
{\rm e}^{- \int \psi ^{\dagger} S_{0}^{-1}\psi}{\rm e}^{\chi ^{\dagger}_{I}(T-im)_{IJ}\chi _{J}}
\label{conden}
\rangle
\ee
From the two formulas above, it is easy to show that
\be
<\! \overline{\psi}\psi>=-i\ <\! \psi^{\dagger}\psi>
=-\frac{1}{V_{4}}\frac{\del \log Z}{\del m}
\ee
We now shift the fermion fields according to $\psi\rightarrow \psi 
+i\phi _{I} \chi _{I}$ and 
$ \psi ^{\dagger}\rightarrow  \psi ^{\dagger}+
i\chi ^{\dagger} _{I}\phi_{I}^{\dagger}$ (sum over $I$ understood). 
We then expand $\chi$ and $\chi ^{\dagger}$ around the respective classical 
solution  of the shifted action.
The remaining integral (\ref{conden}) has now a Gaussian form in $\chi ^{\dagger} \chi$ 
and can be performed. We obtain
\\
\begin{eqnarray}
<\! \psi ^{\dagger} \psi \!>&=&\frac{1}{V_4Z} \langle \int \!{\cal D}\psi ^{\dagger}{\cal D}\psi \  (-2im)^{N}\ {\rm e}^{ -\int \psi ^{\dagger}S^{-1}\psi } \nonumber \\ \nonumber
 & \times & \left(\int \psi^{\dagger} \left(-1+S_{0}S^{-1}+S^{-1}S_{0}+S_{0}^{-1}\frac{\phi _{I} \phi _{I}^{\dagger}}{2(im)^{2}}S_{0}^{-1}\right)\psi +\frac{N}{im} \right) \rangle
\end{eqnarray}
Rewriting the pseudoparticle sum in the exponent as a product over $I$, and 
noting that only the first two terms in the Taylor expansion contribute, 
we can easily perform the color group average to yield
\\
\begin{eqnarray}
<\! \psi ^{\dagger} \psi \!>&=&\frac{1}{V_{4}Z}  \int \!{\cal D}\psi ^{\dagger}{\cal D}\psi (-2im)^{N}{\rm e} ^{-\int \psi ^{\dagger}S _{0}^{-1}\psi} \nonumber \\
&\times&\left(\int \psi ^{\dagger} \psi +\frac{N}{im}-i\frac{\del }{\del m}\right) \prod _{I}\int \!dz_{I}\left(1-\frac{1}{2im}\theta ^{\pm}(z_{I})\right)
\end{eqnarray}
\\
where $\theta ^{\pm}(z)$ is given in appendix A. As in \cite{NVZ1}, we 
assume a sufficient amount of coarse graining so as to rewrite the 
product over $I$  with the result 
\\
\begin{eqnarray}
<\! \psi ^{\dagger} \psi \!>&=&\frac{1}{V_4Z} \int \!{\cal D}\psi ^{\dagger}{\cal D}\psi (-2im)^{N}{\rm e} ^{ -\int \psi ^{\dagger} S _{0}^{-1}\psi } 
\nonumber \\
 &\times& \left(\int \psi ^{\dagger} \psi +\frac{n}{2im}\int \! dz \frac{1-\dot{\theta}^{\pm}(z)/2i}{1-\theta^{\pm}(z)/2im}\right) {\rm e}^{\frac{n}{2}\int \!dz\log(1-\frac{1}{2im}\theta ^{\pm}(z))}
\end{eqnarray}
\\
where the dot on $\theta ^{\pm}(z) $ indicates the derivative with respect to $m$. The functional integral above can be evaluated exploiting the same bosonisation scheme used for the partition function $Z$
\\
\begin{eqnarray}
<\! \psi ^{\dagger} \psi \!>&=&\frac{1}{V_4Z} \int \!{\cal D}\psi
^{\dagger}{\cal D}\psi {\cal D} P^{\pm}(-2im)^{N}{\rm e} ^{-\int \psi
^{\dagger} {\bf S }^{-1}[P^{+},P^{-}]\psi } \nonumber \\
&\times&
{\rm e}^{ -\frac{n}{2}\int \!dz\log\left(\frac{4m}{n}P ^{\pm}(z)\right)}
\nonumber \\ 
 &\times&
 {\rm e}^{ \left(-N+2m\int \! dz P^{\pm}(z)\right)} \left(\int \psi ^{\dagger}\psi+\int \! dz \left(\dot{\theta}^{\pm}(z)-2i\right)P^{\pm}(z)\right)
\end{eqnarray}
\\
where ${\bf S}^{-1}[P^{+},P^{-}]$ is given in the main text (\ref{propagator}). 
At the saddle point 
\\
\begin{eqnarray}
<\! \psi ^{\dagger} \psi \!>&=&\frac{1}{V_4Z} 
\int \!{\cal D}\psi ^{\dagger}{\cal D}\psi (\frac{n}{2iP})^{N}
{\rm e}^{-\int \psi ^{\dagger}S ^{-1}\psi } {\rm e}^{ -N \left( -N+4mP\right)}
\nonumber \\ 
 &\times&\left(\int \psi ^{\dagger}\left(-1+2S_{0}S^{-1}\right)\psi-4iVP\right) 
\end{eqnarray}
\\
where (aside from rescaling $M_{k}(m)$) the quark propagator $S$ is
written down in momentum space (\ref{prop}).
After performing the integral and properly rescaling $P$, we recover 
the expression (\ref{condensate}) quoted in the main 
text for the condensate (in the chiral limit). This result is expected, since 
to leading order in $1/N_c$, the determinants in (\ref{cond}) cancel out, after
factorization (quenched approximation).

\subsection{ Appendix C : Gaussian Approximation}

In what follows, we give details leading to the Gaussian
approximation in the partition function. We can repeat the steps
performed in appendix A with the constraint $n_{+}=n_{-}=n/2$ now
relaxed and the parametrization
\be
n_{\pm}(z)&=&\frac{n^{*}}{2}+\frac{\sigma(z)\pm\chi(z)}{2}
\nonumber \\
P^{\pm}(z)&=&P+\tilde\pi^{\pm}(z)
\ee
A few comments are in order.  In the equations above, $\sigma(z)$ and
$\chi(z)$ respectively represent the scalar and pseudoscalar glueball
sources.
The field $\tilde\pi^{\pm}$ contains pseudoscalar and scalar excitations
and will be discussed further below.
Following appendix A, the auxilliary field $\pi^{\pm}$ is eliminated using 
the mean field equation
\be
-iP^{\pm}(k,l)\ =\ \frac{n^{*}}{2}\int \!dz\ \frac{{\rm e}^{-i(k-l)z}}{2im}\
 \frac{1}{1-\pi^{\pm}(z)/2im}
\ee
Along with the contribution from the measure $\mu (n_{+},n_{-})$
(\ref{measure}) 
the bosonized effective action reads
\\
\be
S_{\rm eff}
=&-& N_{c}{\rm Tr} \log{\bf S}^{-1} [P^+, P^-]
\ -\ 2\int \! dz \ {\rm Tr}_{f}m \left(P^{+}(z)+P^{-}(z)\right)
\nonumber \\
&+&\int \!dz\ n_{\pm}(z)\ 
{\rm Tr}_{f}\log \left(\frac{4P^{\pm}(z)}{n^{*}\rho}\right)
+ 
\int \!dz\ n_{\pm}(z)\ \log N_{\! f}{\rm !}
\ +\ {\tilde S}_{G}
\label{eff}
\ee
where
\be
{\tilde S} _{G}=
&+&\frac{1}{2\chi_{*}}\int \! dz\ (n_{+}(z)-n_{-}(z))^{2}
\nonumber \\
&+&\frac{n}{\sigma_{*}^{2}}\int \!dz \ \left(n_{+}(z)+n_{-}(z)\right)
\bigg(\log\frac{n_{+}(z)+n_{-}(z)}{n}-1\bigg)
\label{glue0}
\ee
The trace (${\rm Tr}_{f}$) is in flavour space and the 
trace (${\rm Tr}$) is over flavour and Dirac indices with an 
integration over momentum. 
\vskip 0.5cm
$\bullet$ {\underline {Gluonic contribution}}

Let us first turn our attention to the last three terms of the effective
action (\ref{eff}).
\\
\be
S_{G}\ [P^{\pm},n_{\pm}]=
&+&\int \!dz\ n_{\pm}(z)\ 
{\rm Tr}_{f}\log \left(\frac{4P^{\pm}(z)}{n^{*}\rho}\right)
\nonumber \\
&+& 
\int \!dz\ n_{\pm}(z)\ \log N_{\! f}{\rm !}
\nonumber \\
&+&\frac{1}{2\chi_{*}}\int \! dz\ (n_{+}(z)-n_{-}(z))^{2}
\nonumber \\
&+&\frac{n}{\sigma_{*}^{2}}\int \!dz \ \left(n_{+}(z)+n_{-}(z)\right)
\bigg(\log\frac{n_{+}(z)+n_{-}(z)}{n}-1\bigg)
\ee
\\
Using the saddle approximation in the scalar glueball source $\sigma
(z)$ fluctuations, we obtain
\be 
n_{*}=n\exp\bigg(-\frac{\sigma _{*}^{2}}{n-N_{f}/\sigma_{*}^{2}}
\log N_{f}{\rm !}\prod _{f}
\frac{4P}{n\rho}\bigg)
\ee
As first discussed in \cite{DYAKONOV2} and later in \cite{NVZ1}, the
distribution of the fluctuations in the number densities $n_{\pm}(z)$
is Gaussian (exact) in $\chi (z)$ with a width given by 
(\ref{susceptibility}). The distribution is logarithmic in the sum
$\sigma (z)$ and Gaussian (approximate) in the large $N_{c}$
limit with a dispersion relation given by (\ref{relations}).
 Along with the saddle point decomposition of the bilocal auxilliary field
$P^{\pm}=P{\rm e}^{\pm i\pi_{ps}/2}(1+\pi_{s}){\rm e}^{\pm i\pi_{ps}/2}$
, we obtain \footnote{This
parametrization is reminiscent of the action being invariant (for
massless quarks) under global axial transformation with the subscript
{\it s} and {\it ps} respectively standing for the scalar and pseudoscalar
mesonic excitations.}
\\
\be
S_{G}\ [P^{\pm},n_{\pm}]\ =\ -\frac{Nn}{\sigma_{*}^{2}}
+\ S_{G}^{(1)}\ [\pi_{s,ps}]\ 
+S_{G}^{(2)}\ [\pi_{s,ps},\sigma, \chi]\
\label{gluonaction}
\ee
\\
Adopting the nonet decompositions 
$\pi _{ps}=\lambda _{0}\eta_{0}+\sum \lambda _{a} \pi _{ps}^{a}$ and
$\pi _{s}=\lambda _{0}\pi_{s,0}+\sum \lambda _{a} \pi _{s}^{a}$, the 
term $\ S_{G}^{(1)}\ [\pi_{s,ps}]\ $ contains mesonic fluctuations only and
reads
\\
\be
\ S_{G}^{(1)}\ [\pi_{s,ps}]\ 
=-\frac{n_{*}}{2}\int \!dz\ {\rm Tr}_{f}\bigg(\pi_{s}^{2}(z)
-\pi_{ps}^{2}(z)\bigg)
+\int \!dz\
N_{f}\bigg(\chi_{*}\eta_{0}^{2}(z)-\sigma_{*}^{2}\pi_{s,0}^{2}(z)\bigg)
\ee
\\
We point out that the term $\ S_{G}^{(1)}\ [\pi_{s,ps}]$ should be put
 in concert with the first two terms of 
$S_{\rm {eff}}[P^{\pm},n_{\pm}]$ in order to obtain
 the total mesonic contribution to (\ref{eff}).

The last term $S_{G}^{(2)}\ [\pi_{s,ps},\sigma, \chi]$ involves mixing on the 
one hand between the isosinglet scalar and the scalar glueballs, and,  between
the isosinglet pseudoscalar and the pseudoscalar glueballs,  on the other hand. 
\\
\be
S_{G}^{(2)}\ [\pi_{s,ps},\sigma, \chi]\ =
&+&\int \!dz\ \frac{1}{2\chi_{*}}\bigg(\chi (z)+i\chi
_{*}\sqrt{2N_{f}}\eta _{0}(z)\bigg)^{2}
\nonumber \\
&+&\int \!dz\ \frac{1}{2\sigma _{*}^{2}}\bigg(\sigma (z)+\sigma
_{*}^{2}\sqrt{2N_{f}}\pi  _{s,0}(z)\bigg)^{2}
\ee

\vskip 0.5cm
$\bullet$ {\underline {Mesonic contribution}}

Performing  a Taylor expansion of $P^{\pm}$ around the saddle point
$P$ in the first two terms of $S_{{\rm eff}}[P^{\pm},n_{\pm}]$ 
(\ref{eff}) along with $S_{G}^{(1)}\ [\pi_{s,ps}]$ the total mesonic
 contribution reads 
\footnote{The sum over flavour indices $f$ and $g$ is understood.}
\be
S_{{\rm meson}}\ [\pi_{s,ps}]\ =
&-&N_{c}{\rm Tr}\log S^{-1}(P)-4V\ m_{f}P(m_{f})
\nonumber \\
&-&N_{c}\int\!dk\ \pi _{s}^{fg}(k) \Delta _{+}(k,m_{f},m_{g})
\pi_{s}^{gf}(-k)
\nonumber \\
&+&N_{c}\int\!dk\ \pi _{ps}^{fg}(k) \Delta _{-}(k,m_{f},m_{g})
\pi_{ps}^{gf}(-k)
\nonumber \\
&+&\int \!dk\ \bigg(\eta_{0}(k)\frac{\chi_{*}N_{f}}{N_{c}}\eta_{0}(-k)
-\pi_{s0}(k)\frac{\sigma_{*}N_{f}}{N_{c}}\pi_{s0}(-k)\bigg)
\label{mesonaction}
\ee
\\
where the saddle point approximation leads to an integral (gap)
equation in $P(m_{f})$ for each flavour $f$
\\
\be
\frac {4N_c}{n}\int\! dk A(k;M_{k}P(m_{f});m_{f})\ =\ 
1\ - \ 2m_{f}\frac{2 P(m_{f})}{n} 
\ee
\\
After rescaling the constituent mass according to 
$M_{k}P\rightarrow M_{k}$ with $P(m_{f})=n\lambda (m_{f})/2$, we obtain
 the gap equation (\ref{gap}) in the text. 
 We define below the various quantities introduced in the mesonic
action (\ref{mesonaction}).
In momentum space, we write the inverse quark propagator in the background of 
instantons and antiinstantons as
\be
\langle k | S^{-1}(P) |l\rangle =\delta (k-l) S^{-1}\left(k,m\right)
\ee
and
\be
S^{-1} \left(k,m\right) = \frac{-iM_{k}P}{k^2}
\left(k\!\!\!/-im\right)
\left(k\!\!\!/-i\left(\frac{k^{2}}{M_{k}P}-m\right)\right)
\ee
\\
The coefficient $A\left(k;M_{k};m\right)$ appearing in the gap
equation is given by
\\
\be
A\left(k;M_{k};m\right)
=\frac{k^2}{M_{k}}\frac{M_{k}-m+M_{k}m^2/k^2}
{k^2+\left(m-k^2/M_{k}\right)^2}
\ee
\\
Except for the isosinglet scalar and pseudoscalar, the inverse meson 
propagator in the background of instantons and antiinstantons,
apart from the factor $f^2/4N_{c}$, 
 can be identified with $\Delta _{\pm}(k)$ 
appearing in the quadratic
part of (\ref{mesonaction}) and reads 
\\
\be
\Delta _{\pm}(k,m_{1},m_{2})=\frac{n}{2N_{c}}-2\int \! dq\left(A_{1}A_{2}\mp\left(q_{1}.q_{2}\right)B_{1}B_{2}\right)
\ee
\\
where we have
set $q_{1,2}\!=\!q\pm k/2$,
$M_{1}\!=\!M_{q_{1}}(m_{1})$, $ A_{1}\!=\!A\left(q_{1};M_{1};m_{1}\right)$, 
$ B_{1}=B\left(q_{1};M_{1};m_{1}\right) $, 
$m_{1}$  being one of the quark masses in $SU(3)$ flavour space and $B$ is
given by
\\
\be
B\left(k;M_{k};m\right)=\frac{k^2}{M_{k}}\frac{1}{k^2+\left(m-k^2/M_{k}\right)^2}
\ee
\\
In what follows, we will always consider the rescaled constituent mass 
$M_{k}(m_{f})$.

\vskip 0.5cm
$\bullet$ {\underline {Bosonized partition function}}
\\
To be thorough, let us exhibit the bosonized partition function
utilized in evaluating the various (mesonic, baryonic and gluonic)
 correlators. To this end, 
\\
\be
Z\ =\ \int\! {\cal D}\pi _{s,ps}\ {\cal D}\sigma \ {\cal D}\chi\ 
{\rm e}^{-S_{{\rm eff}}[\pi_{s,ps},\sigma,\chi]}
\label{gaussian}
\ee
\\
where the bosonized action follows from regrouping terms in
$S_{G}$ (\ref{gluonaction}) and $S_{{\rm meson}}$ (\ref{mesonaction}).
\\
\be
S_{{\rm eff}}[\pi_{s,ps},\sigma,\chi]
\ =\ 
S_{{\rm eff}}^{(0)}[0,\frac{n}{2},0]
+S_{{\rm eff}}[\pi_{s,ps}]
+S_{{\rm eff}}[\sigma,\chi]
\label{bosonizedaction}
\ee
with
\be
S_{{\rm eff}}^{(0)}[0,\frac{n}{2},0]
\ =\ -N_{c}{\rm Tr}logS^{-1}(P)-\frac{Nn}{\sigma_{*}^{2}}
+N {\rm Tr}_{f}\log \frac{P}{n \rho}-4Vm_{f}P(m_{f})
\ee
\\
along with the mesonic part of the effective action
\\
\be
S_{{\rm eff}}[\pi_{s,ps}]\ =
&-&N_{c}\int\!dk\ \pi _{s}^{fg}(k) \Delta _{+}(k,m_{f},m_{g})
\pi_{s}^{gf}(-k)
\nonumber \\
&+&N_{c}\int\!dk\ \pi _{ps}^{fg}(k) \Delta _{-}(k,m_{f},m_{g})
\pi_{ps}^{gf}(-k)
\nonumber \\
&+&\int \!dk\ \bigg(\eta_{0}(k)\frac{\chi_{*}N_{f}}{N_{c}}\eta_{0}(-k)
-\pi_{s0}(k)\frac{\sigma_{*}N_{f}}{N_{c}}\pi_{s0}(-k)\bigg)
\label{effmeson}
\ee
\\
and the gluonic part of the effective action
\\
\be
S_{{\rm eff}}[\sigma,\chi]\ =
&+&\int \!dz\ \frac{1}{2\chi_{*}}\bigg(\chi (z)+i\chi
_{*}\sqrt{2N_{f}}\eta _{0}(z)\bigg)^{2}
\nonumber \\
&+&\int \!dz\ \frac{1}{2\sigma _{*}^{2}}\bigg(\sigma (z)+\sigma
_{*}^{2}\sqrt{2N_{f}}\pi  _{s,0}(z)\bigg)^{2}
\label{effglue}
\ee
\\

We are now in a position to evaluate the correlation functions of
interest.

\vskip 0.5cm
$\bullet$ {\underline {Connected meson correlator}}
\\
From the expression of $C_{\gamma}^{0}(x)$ (\ref{concorl}) in the text
along with the bosonized partition function (\ref{gaussian}), we
easily find that to leading order in $N_{c}$ (the trace being over
flavour as well as Dirac indices)
\\
\be
C_{\gamma}^{0}(x)
=-N_{c}{\rm Tr}\bigg(\ S(x,m)\ \gamma \ S(-x,m)\ \gamma \ \bigg)
\ee
\\
along with its p-space version $C_{\gamma}^{0}(p)$ quoted in
(\ref{free}) of the main text.

\vskip 0.5cm
$\bullet$ {\underline {Unconnected meson correlator}}
\\
From the expression of $C_{\gamma}^{1}(x)$ (\ref{unconcorl}) in the
text, we need to examine the term ${\rm Tr}\ \gamma S(x,x,P^{\pm})$ 
in the integrand (the trace being in flavour, color and Dirac space).
With the shorthand notation
\be
\tilde \pi=\pi _{s}-\frac{\pi_{ps}^{2}}{2}+i\gamma _{5}(\pi _{ps}
+\frac{1}{2}\pi_{s}\pi_{ps}+\frac{1}{2}\pi_{ps}\pi_{s})
\ee
we can write
\be
{\rm Tr}\ \gamma S(x,x,P^{\pm})
=\int \!dk\ dl\ {\rm e}^{i(k-l)x}\ {\rm Tr}\ \gamma  S(k,l,P^{\pm})
\ee
where the relevant term in the large $N_{c}$ limit is given according
to
\\
\be
S(k,l,P^{\pm})&=&S(k,m)\ \delta (k-l)\ + \ i\sqrt{M_{k}}C(k,m)\tilde
\pi (k-l) C(l,m) \sqrt{M_{l}}
\nonumber \\
\nonumber \\
&-&\sqrt{M_{k}}C(k,m)\ \bigg(\int \!dq\ \tilde \pi (k-q)  
\sqrt{M_{q}}
\bigg(1-\frac{imq\!\!\!/}{q^{2}}\bigg)
\nonumber \\
\nonumber \\
&\times&
S(q,m)
\bigg(1-\frac{imq\!\!\!/}{q^{2}}\bigg)
\sqrt{M_{q}}
\tilde \pi (q-l)\bigg) \ C(l,m) \sqrt{M_{l}}
\ee
\\
The coefficient $C(k,m)$ is given further below. Defining
\be
C_{\gamma}^{1}(p,q)=\delta(p-q)\ \bigg( {\tilde C}_{\gamma}^{1}(p) +
C_{\gamma}^{1\prime }(p) \bigg)
\ee 
where
\be
{\tilde C}_{\gamma}^{1}(p)
&=&-\frac{\delta(p)}{Z} \int \!{\cal D}\pi
_{s,ps} \ {\cal D}\sigma \ {\cal D}\chi\ \ 
\exp \big(-S_{{\rm eff}}[\pi_{s,ps},\sigma,\chi]\big)
\nonumber \\
\nonumber \\
&\times&
{\rm Tr}\ \gamma S
\bigg( {\rm Tr}\ \gamma S
-2 {\rm Tr}\ \gamma \sqrt{M_{k}} 
C\ {\tilde \pi}\ 
(B{\hat k}\!\!\!/+iA)\ {\tilde \pi}\ C\ \sqrt{M_{k}} \bigg)
\ee
\\
Here, the trace carries an integral over momentum. The meson field
${\tilde \pi}$ is left inside the trace so as to reflect its bilocal
character in the momentum.
It is clear that ${\tilde C}_{\gamma}^{1}(p)$ vanishes identically
except in the isoscalar singlet $\pi_{s,0}$. From a diagram approach,
it has two unconnected closed fermion loops with possibly the
$\pi_{s,0}$ being emitted within one loop (Fig. 3c). Further
discussion will be presented in the text regarding this term.

We are therefore left with the second term contribution in 
$C_{\gamma }^{1}(p)$. This term amounts to a propagating meson
from one closed fermion loop to the other (Fig. 3d) and reads
\\
\be
C_{\gamma}^{1}(p)&=&
-\frac{N_{c}}{Z} \int \!{\cal D}\pi
_{s,ps} \ {\cal D}\sigma \ {\cal D}\chi\ 
\ R_{\gamma}(p,m_{f},m_{g})\ R_{\gamma}(-p,m_{f},m_{g}) 
\nonumber \\ 
&\times&  \exp \bigg(-S_{{\rm eff}}[\pi_{s,ps},\sigma,\chi]\bigg)
\ee 
\\ 
where 
\be
R_{\gamma}(p,m_{f},m_{g})&=&\int\!dk\ \sqrt{M_{1}M_{2}}{\rm Tr}\bigg(\gamma
C(k_{1},m_{f}){\tilde \pi}(p)\ C(k_{2},m_{g})\bigg)
\label{vertex} 
\ee
\\
with $k_{1,2}=k\pm p/2$, $M_{1}=M_{k_{1}}(m_{f})$ and
\be
C(k,m)=S(k,m)\bigg(1-\frac{imk\!\!\!/}{k^{2}}\bigg)
\ee
If we redefine $R_{\gamma}$ so as to extract the meson field, we have 
\be 
R_{\gamma}(p,m_{f},m_{g})=R_{\gamma}^{\pm}
(p,m_{f},m_{g})\pm R_{\gamma}^{\mp}(p,m_{f},m_{g}) 
\ee 
with 
\be
R_{\gamma}^{\pm}
\left(p,m_{1,2}\right)=\int \!
\frac{d^4k}{(2\pi)^4}\sqrt{M_{1}M_{2}} {\rm Tr}\left(\gamma C_{1}
\frac{1\pm\gamma _{5}}{2} C_{2}\right) \ee and \be
C_{1,2}=\frac{k\!\!\!/_{1,2}}{k_{1,2}^2}\left(1-A_{1,2}\right)+iB_{1,2} 
\ee
Performing the functional integral, we obtain (\ref{corr}).
As an example, we exhibit the case of the mixing singlet $\eta _{0}$
and octet $\eta _{8}$. The specific flavour character of
 $R_{\gamma}^{\pm}$  follows from
 the effective action (\ref{effmeson}) for the nonet
decomposition.
 For instance, the pertinent terms $R_{\gamma _{5}}$ (\ref{vertex})
for the  $\eta _{0}$ 
and $\eta_{8}$ correlators are given by
\\ 
\be
\int\!dk\sqrt{M_{1}M_{2}}{\rm Tr}_{f}\left(\gamma _{5}\lambda _{0,8}C(k_{1},m)\gamma _{5}C(k_{2},m)\kappa (p)\right)
\ee
\\
Explicitly, for the unconnected parts of the 
$\eta _{0}$ and $\eta _{8}$ correlators, we obtain 
\\
\be
C^{1}_{\eta _{0,8}}(p)&=-\frac{N_{c}}{Z}
\int \!{\cal D}\eta _{0,8}\ 
{\bf \tilde{R}}_{0,8}(-p) {\bf \eta}(-p) {\bf \tilde{\eta}} (p)
 {\bf R}_{0,8}(p) 
{\rm e}^{N_{c}\int {\bf \tilde{\eta}} [{\bf \Delta}] {\bf \eta}}
\ee
\\
where the partition function $Z$ in the denominator contains only
$\eta_{0}$ and $\eta_{8}$.
For the $\eta _{0}$ we have
\\
\be
{\bf R}_{0}(p)=\left(\begin{array}{c}
\frac{4}{3}R(p)+\frac{2}{3}R_{s}(p)
\\

\\
\frac{2\sqrt{2}}{3}(R(p)-R_{s}(p))
\end{array}
\right)
\ee
\\
and for the $\eta _{8}$
\\
\be
{\bf R}_{8}(p)=\left(\begin{array}{c}
\frac{2\sqrt{2}}{3}(R(p)-R_{s}(p))
\\

\\
\frac{2}{3}R(p)+\frac{4}{3}R_{s}(p)
\end{array}
\right)
\ee
\\
with $R(p)=R_{\gamma _{5}}(p,m)$ and similarly $R_{s}(p)=R_{\gamma _{5}}(p,m_{s})$ where $m$ and
$m_{s}$ are the up (down) and strange quark mass respectively (\ref{vertex}).
\\      
\be
C^{1}_{\eta _{0,8}}(p)
=\frac{N_{c}}{2}
{\bf \tilde{R}}_{0,8}(-p)
[{\bf \Delta }(p)]^{-1}
 {\bf R}_{0,8}(p) 
\ee
\\

\subsection{Appendix D : Nonet Decomposition}

Taking the partition function (\ref{gaussian}) derived for 
$N_{f}>1$ with the same decomposition for the meson 
fields $\pi_{s,ps}$, we have as the mesonic effective action
(\ref{effmeson})
\\
\be
S_{{\rm eff}}[\pi_{s,ps}]\ =
&-&N_{c}\int\!dk\ \pi _{s}^{fg}(k) \Delta _{+}(k,m_{f},m_{g})
\pi_{s}^{gf}(-k)
\nonumber \\
&+&N_{c}\int\!dk\ \pi _{ps}^{fg}(k) \Delta _{-}(k,m_{f},m_{g})
\pi_{ps}^{gf}(-k)
\nonumber \\
&+&\int \!dk\ \bigg(\eta_{0}(k)\frac{\chi_{*}N_{f}}{N_{c}}\eta_{0}(-k)
-\pi_{s0}(k)\frac{\sigma_{*}N_{f}}{N_{c}}\pi_{s0}(-k)\bigg)
\label {nonet}
\ee
\\
where $f$ and $g$ are flavour labels
(f is not to be confused with the pseudoscalar decay constant). 
Using the decomposition 
$\pi _{ps}=\sum _{k=0}^{8}\kappa _{k} \ \lambda _{k}$, 
the $\kappa _{0}$ and $\kappa _{8}$ excitations contribute to $ S_{\rm 
eff}$ in the form
\\
\be
S_{\rm eff}[\eta_{0},\eta_{8}]
\ =\ N_{c}\int \!dk\ 
{\bf \tilde{\eta}} (k)\ [{\bf \Delta} (k)]\ {\bf \eta}(-k)
\ee
\\
where ${\bf \tilde{\eta}}(p)=(\eta _{0}(p),\eta _{8}(p))
=f(\kappa_0 (p), \kappa_8 (p) )$ and
\\
\\
\be
[{\bf \Delta} (k)]=\left( \begin{array}{cc}
\frac{2}{3}\left(2\Delta +\Delta _{s}\right)+\frac{\chi_*N_{f}}{N_{c}}
&
\frac{2\sqrt{2}}{3}\left(\Delta -\Delta _{s}\right)
\\
&
\\
\frac{2\sqrt{2}}{3}\left(\Delta -\Delta _{s}\right)
&
\frac{2}{3}\left(\Delta +2 \Delta _{s}\right)
\end{array}
\right)
\ee
\\
\\
with the shorthand notation $\Delta =\Delta (k,m,m)$ 
$\Delta _{s}=\Delta (k,m_s,m_s)$ with the strange quark mass inserted. The
contribution $\chi_*N_f/N_c$ follows from the singlet mixing with the topological
fluctuations through the measure (\ref{measure}).
At low energy, and to leading order in the current mass 
\\
\be
S_{\rm eff}[\eta_{0},\eta_{8}]\ =&+&
 \int \!dk\, \frac{1}{2}f^{2}k^{2}(\eta _{0}^{2}+\eta _{8}^{2})
+\int \frac 12 ({2N_F\chi_*}) \,\eta_0^2
\nonumber \\
&-&\ \int \!dk \frac{1}{2}\langle \overline\psi \psi \rangle
\left(\eta _{0}^{2}(\frac{4}{3}m+\frac{2}{3}m_{s})
+\eta _{8}^{2}(\frac{2}{3}m+\frac{4}{3}m_{s})\right)
\nonumber \\
&-&\ \int \!dk \frac{1}{2}\langle \overline\psi \psi \rangle
\eta _{0}\eta _{8}\frac{4\sqrt{2}}{3}(m_s-m)
\label{etaaction}
\ee
Note that $S_{\rm eff}[\eta_{0},\eta_{8}] \sim N_c$. 
The above result yields the GOR relations for the singlet and the octet, 
if we were to drop $\frac{nN_{f}}{N_{c}}$. As is
well known, the GOR result is badly violated in the singlet channel by
the axial U(1) anomaly. The latter is carried over by local
fluctuations in the topological charge, which results in a mixing with
the singlet quantum numbers as displayed in (\ref{etaaction}).

\subsection{ Appendix E : Extended Bosonization.}

The use of the mean-field equation 
(\ref{MFEQ}) in Appendix A has allowed for a bosonization scheme that 
is trouble free. Indeed, if we were to carry a gaussian analysis around all the 
fields $including$ the auxillary field $\pi^{\pm}$ (and hence expand 
(\ref{MFEQ}), then instabilities show up along the scalar $(\pi_s$) and 
pseudoscalar ($\pi_{ps}$) directions. This, however, can be easily fixed 
through a generalization of (\ref{identity}) to include fluctuations around
the instanton densities. We start with a modification in (\ref{identity})
\\
\be
{\bf 1}\ =\ \int \! {\cal D}\pi^{\pm} {\cal D}P^{\pm}  
\exp \bigg({\rm Tr}_{f}\int \!dz\ P^{\pm}(z)
\bigg(\pi^{\pm}(z) - \frac{N^{\pm}(z)}{n^* /2}\, \theta ^{\pm}(z)
\bigg)\bigg)
\label{identity1}
\ee
\\
that is inserted in the partition function (\ref{action}), 
where the term in the exponent
\\
\be
\frac{N^{\pm}(z)}{n^* /2}\ =\ 1\
+\frac{g_{s}\sigma(z)+g_{ps}\chi(z)}{n^*}
\ee
\\
clearly couples scalar (pseudoscalar) glueballs to the quarks with
strength $g_{s}$ \ $(g_{ps})$ (see (\ref{propglue}) below).

Following the steps of appendix C, the mean field equation in
$\pi^\pm$ reads
\\
\be
-iP^{\pm}(z)\ =\ \frac{n^{*}}{4im}\frac{n^{\pm}(z)}{N^{\pm}(z)}
\ \frac{1}{1-n^* \pi^{\pm}(z)/(4im\, N^{\pm}(z))}
\ee
\\
Along with the contribution from the measure $\mu (n_{+},n_{-})$
(\ref{measure}) 
the bosonized effective action reads
\be
S_{\rm eff}
=&-& N_{c}{\rm Tr} \log{\bf S}^{-1} [P^{\pm},\sigma ,\chi ]
\ -\ 2\int \! dz \ \frac{N^{\pm}(z)}{n^* /2}
{\rm Tr}_{f}m \ P^{\pm}(z)
\nonumber \\
&+&\int \!dz\ n_{\pm}(z)\ 
\log
\bigg( N_{\! f}{\rm !}
\prod _{f}\ \frac{4P^{\pm}(z)}{n^{*}\rho}\frac{N^{\pm}(z)}{n^{\pm}(z)}
\bigg)
\nonumber \\
&+& N_{f}\int \! dz\ \bigg(n^{+}(z)+n^{-}(z)\bigg)\ +\ {\tilde S}_{G}
\label{eff1}
\ee
where 
${\tilde S} _{G}$ is the gluonic contribution (\ref{glue0}). The
inverse
quark propagator ${\bf S}^{-1} [P^{\pm},\sigma ,\chi ]$ in
(\ref{eff1}) is
\be
{\bf S}^{-1} [P^{\pm},\sigma ,\chi ]\ = &+&{\bf S}^{-1}[P^+,P^-]
\nonumber \\
&-&i\sqrt{M_{k}}
\bigg(1-\frac{im{\hat k}\!\!\!/}{k^2}\bigg)\ 
\frac{g_{s}\sigma(z)+g_{ps}\chi(z)}{n^*}
\nonumber \\
&\times&
\bigg(1+\pi_{s}+i\gamma_{5}\pi_{ps}\bigg)
\bigg(1-\frac{im{\hat k}\!\!\!/}{k^2}\bigg)
\sqrt{M_{k}}
\label{propglue}
\ee
and its second term clearly exhibits glueballs coupling to quarks.
The first term ${\bf S}^{-1}[P^+,P^-]$ in (\ref{propglue}) 
 is given in (\ref{propagator}).
Following the saddle point approximation used in appendix C, we
obtain the effective action as follows
\be
S_{{\rm eff}}[\pi_{s,ps},\sigma,\chi]
\ =\ 
&-&N_{c}{\rm Tr} \log S^{-1}(P)+n^{*}V\bigg(N_{f}-{\rm
Tr}_{f}\frac{4mP}{n^*}-\frac{n}{\sigma_{*}}\bigg)
\nonumber \\
&-&N_{c}\int\!dk\ \pi _{s}^{fg}(k) \Delta _{+}(k,m_{f},m_{g})
\pi_{s}^{gf}(-k)
\nonumber \\
&+&N_{c}\int\!dk\ \pi _{ps}^{fg}(k) \Delta _{-}(k,m_{f},m_{g})
\pi_{ps}^{gf}(-k)
\nonumber \\
&+&S_{{\rm eff}}[\sigma^2,\sigma\,\pi_{s}]
\nonumber \\
&+&S_{{\rm eff}}[\chi^2,\chi\,\pi_{ps}]
\label{bosonizedaction1}
\ee
In the last two terms of (\ref{bosonizedaction1}), we have lumped
terms in $\sigma^2$, $\sigma \pi_{s}$, $\chi ^2$ and $\chi \pi_{ps}$
where $\pi_{s}$ $(\pi_{ps})$ is decomposed in the scalar
(pseudoscalar) unphysical basis (in flavour space) 
$\pi_{s}^{fg}$ $(\pi_{ps}^{fg})$. Only diagonal terms $\pi_{s}^{ff}$
 $(\pi_{ps}^{ff})$ mix with scalar (pseudoscalar) glueballs 
according to
\be
S_{{\rm eff}}[\sigma^2,\sigma\,\pi_{s}]\ =
&+& \int \! dk\ 
\sigma (k)\ g_{s}^{\sigma }(k)\ \sigma (-k)
\nonumber \\
&+&\int \! dk \ \sigma (k)\  \sum_{f}\bigg(
1-g_{s}
\ \frac{2N_{c}}{n_{*}}\Delta_{+}(k,m_{f},m_{f})\bigg)\ \pi _{s}^{ff}(-k)
\label{gluescalar1}
\ee
where
\be
g_{s}^{ \sigma}(k)=\frac{n}{2n_{*}\sigma _{*}^{2}}
+\frac{g_{s}^{2}}{2n_{*}}\sum_{f}
\bigg(1-\frac{2N_{c}}{n_{*}N_{f}} \Delta_{+}(k,m_{f},m_{f})\bigg)\
-\ \frac{N_{f}}{2n_{*}}(g_{s}-1)^2 
\label{xnew1}
\ee
The point made at the start of this Appendix can now be appreciated. 
We see that if we set $g_s=0$
the result differs from what we had in Appendix C by the last term. 
The latter with its negative sign can cause an instability in the 
scalar glueball fluctuations. This is easily tamed by the use of the 
bosonization scheme discussed in Appendix C.

For the pseudoscalar part, we have 
\be
S_{{\rm eff}}[\chi^2,\chi\,\pi_{ps}]\ =
&+& \int \! dk\ 
\chi (k)\ g_{ps}^{\chi }(k)\ \chi (-k)
\nonumber \\
&+&\int \! dk \ \chi (k)\ \sum_{f} \bigg( 
1-g_{ps}\ \frac{2N_{c}}{n_{*}}\Delta_{-}(k,m_{f},m_{f})\bigg)\ 
i\pi _{ps}^{ff}(-k)
\label{gluepseudoscalar1}
\ee
where
\be
g_{ps}^{\chi}(k)=\frac{1}{2\chi _{*}} 
+\frac{g_{ps}^{2}}{2n_{*}}\sum_{f}
\bigg(1-\frac{2N_{c}}{n_{*}N_{f}} \Delta_{-}(k,m_{f},m_{f})\bigg)\
-\ \frac{N_{f}}{2n_{*}}(g_{ps}-1)^2 
\ee
Again, if $g_{ps}$=0 the latter term may cause the fluctuations in the 
pseudoscalar glueball direction to be unstable. This is easily tamed by
the bosonization scheme discussed in Appendix C.

As a check, we clearly see that in the absence of fermion we recover
the scalar (pseudoscalar) gluonic contribution $\tilde S _{G}$ for
$n_{\pm}=(n+\sigma\pm\chi)/2$.
The expression (\ref{gluescalar1}) in terms of physical scalar meson
field in $\lambda _{0}$ and $\lambda _{8}$ channels (flavour space)
reads
\be
S_{{\rm eff}}[\sigma^2,\sigma\,\pi_{s}]\ =
&+& \int \! dk\ 
\sigma (k)\ g_{s}^{\sigma }(k)\ \sigma (-k)
\nonumber \\
&+&\int \! dk \ \sigma (k)\  \sqrt{2N_{f}}\bigg(
g_{s}^{0}(k)
\ \pi _{s0}(-k)\ +\ g_{s}^{8}(k)\ \pi _{s8}(-k)\bigg) 
\ee
and similarly (\ref{gluepseudoscalar1}) for the pseudoscalar sector reads
\be
S_{{\rm eff}}[\chi^2,\chi\,\pi_{ps}]\ =
&+& \int \! dk\ 
\chi (k)\ g_{ps}^{\chi }(k)\ \chi (-k)
\nonumber \\
&+&\int \! dk \ \chi (k)\ i\sqrt{2N_{f}}\ \bigg( 
g_{ps}^{0}(k)\ \eta_{0}(-k)\ +\ g_{ps}^{8}(k)\ \eta_{8}(-k)\bigg) 
\ee
We have defined above for the scalar (pseudoscalar) part of the action
\\
\be
g_{s,ps}^{0}(k)=1-g_{s,ps} \frac{2N_{c}}{n_{*}N_{f}}
\bigg(2\Delta_{\pm}(k,m,m)+\Delta_{\pm}(k,m_{s},m_{s})\bigg)
\ee
\be
g_{s,ps}^{8}(k)=g_{s,ps} \frac{2N_{c}}{n_{*}N_{f}}\sqrt{2}\bigg(\Delta_{\pm}(k,m_{s},m_{s})-
\Delta_{\pm}(k,m,m)\bigg)
\ee
\\
The expression of $n_{*}$ and the mass gap equation of appendix C
remain unchanged.
\vskip 0.5cm
$\bullet$ {\underline {Scalar gluonic correlator $C_{FF}(x,y)$}
\\
The form of the scalar gluonic correlator is unchanged up to a
constant term
\be
{\cal C}_{F F} (x-y) 
= \int \!dk \ {\rm e}^{ik(x-y)}\ \bigg(
 \frac{n}{n_{*}\sigma_{*}^{2}}\
+\ \frac{1}{2N_{c} } 
 \sum_{f=1}^{N_{f}} \bigg(
\frac{1}{\Delta _{+}(k,m_{f},m_{f})}-\frac{2N_{c}}{n_{*}}
\bigg)\bigg)^{-1}
\label{spglue1}
\ee
\vskip 0.5cm
$\bullet$ {\underline {Pseudoscalar gluonic correlator $C_{FF}(x,y)$}
\\
Similarly, for the pseudoscalar gluonic correlator
\be
{\cal C}_{F\tilde F} (x-y) 
= \int \!dk \ {\rm e}^{ik(x-y)}\ \bigg(
  \frac{1}{\chi_{*}}\ +
\ \frac{1}{2N_c }  \sum_{f=1}^{N_f}\bigg( 
\frac{1}{\Delta _{-}(k,m_{f},m_{f})}-\frac{2N_{c}}{n_{*}}\bigg)\bigg)
^{-1}
\label{ppglue1}
\ee
$\bullet$ {\underline {Pseudoscalar form factor}}
\\
For the pseudoscalar case, we have
\be
F_{p}(0)=\frac{32\pi^{2}}{2N_{c}\Delta _{-}(0,m,m)}\
\bigg(\frac{1}{\chi_{*}}
+\frac{1}{2N_{c}}\sum_{f}\bigg(\frac{1}{\Delta_{-}(0,m_{f},m_{f})}
-\frac{2N_{c}}{n_{*}}\bigg)\bigg)
^{-1}
\ee
as indicated in section 8.

\vskip 0.5cm
$\bullet$ {\underline {Scalar form factor}}
\\
For the scalar case, we have
\be
F_{s}(0)=\frac{32\pi^{2}}{2N_{c}\Delta _{+}(0,m,m)}\
\bigg(\frac{n}{n_{*}\sigma^{2}_{*}}
+\frac{1}{2N_{c}}\sum_{f}\bigg(\frac{1}{\Delta_{+}(0,m_{f},m_{f})}-\frac
{2N_{c}}{n_{*}}\bigg)\bigg)^{-1}
\ee
as indicated in section 8.

\vskip 0.5cm
$\bullet$ {\underline {Pseudoscalar coupling constant $g_{ps}$}}
\\
In order to determine the strength $g_{ps}(k=0)$ of the pseudoscalar
glueballs coupling constant to the quarks, we fit the experimental
value of the mixing angle $\theta =-20^{\circ}$ when diagonalizing.
The result gives an estimate for $g_{ps}$ ($k$=0 is understood)
 according to
\be
2\, \cot 2\theta\ \bigg(\frac{2\sqrt{2}}{3}g_{ps}^{\chi}
(\Delta - \Delta_{s})
+\frac{N_{f}}{2}g_{ps}^{0}\, g_{ps}^{8}\bigg)=\frac{2}{3}g_{ps}^{\chi}(\Delta
-\Delta_{s})
+\frac{N_{f}}{2}(g_{ps}^{0\ 2}-g_{ps}^{8\ 2})
\ee
We obtain $g_{ps}=-7.025$.
\vskip 0.5cm

\subsection{ Appendix F : Unconnected Correlators}

We tabulate below the expressions for the unconnected correlator
$C_{\gamma}^{1}(p)$ in the various channels 
\vskip 1cm 
\begin{center}
\begin{tabular}{|c|c|c|}    
\hline & & \\ ${\bf \gamma} $&$ {\bf Tr\left(\gamma
C_{1}\frac{1+\gamma_{5}}{2}C_{2}\right)}$ & {\bf
$C_{\gamma}^{1}\left(p\right)/2N_{c}$}  \\ \hline  & & \\ 1 &
$2\left(\left(1-A_{1}\right)\left(1-A_{2}\right)\frac{k_{1}k_{2}}{k_{1}^2k_{2}^2}
-B_{1}B_{2}\right)$ & $\frac{\left(R_{1}^{+}\left(p\right)\right)^2}{\Delta
_{+}\left(p\right)}$  \\ \hline  & & \\ $\gamma _{5}$ &
$-2\left(\left(1-A_{1}\right)\left(1-A_{2}\right)\frac{k_{1}k_{2}}{k_{1}^2k_{2}^2}
+B_{1}B_{2}\right)$ & $\frac{\left(R_{\gamma
_{5}}^{+}\left(p\right)\right)^2}{\Delta _{-}\left(p\right)}$ \\ \hline  & & \\
$\gamma _{\mu}$ & $2\left(\frac{kx+p/2}{k_{1}^{2}}\left(1-A_{1}\right)B_{2}  
+\frac{kx-p/2}{k_{2}^{2}}\left(1-A_{2}\right)B_{1}\right)$ &
$\frac{\left(R_{\gamma _{\mu}}^{+}\left(p\right)\right)^2}{\Delta
_{+}\left(p\right)} $  \\ \hline  & & \\ $\gamma _{5}\gamma _{\mu}$ &
$2\left(\frac{kx+p/2}{k_{1}^{2}}\left(1-A_{1}\right)B_{2}  
-\frac{kx-p/2}{k_{2}^{2}}\left(1-A_{2}\right)B_{1}\right)$ &
$\frac{\left(R_{\gamma _{5}\gamma _{\mu}}^{+}\left(p\right)\right)^2}{\Delta
_{-}\left(p\right)} $  \\ \hline  & & \\ $\sigma _{\mu \nu}$ & $R_{\pm}^{\mu
\nu}\left(p\right)=0$ & 0  \\  \hline 
\end{tabular} 
\end{center} 
\vskip 1cm

\subsection{Appendix G : m$_s$ Expansion}

In this Appendix, we give the details leading to (\ref{taylor}).
Inserting the mass gap equation for small quark masses 
$m_{1}$ and $m_{2}$ in the first term of 
$\Delta {\pm}(p,m,m)$  with the following approximations
\be
A\simeq\frac{M^2-mM}{k^2+M^2-2mM}\,,
\ \ \ \ B\simeq\frac{M}{k^2+M^2-2mM}
\ee
 we obtain (denoting $\Delta_{\pm}(p)=\Delta_{\pm}(p,m,m)$)

\be
\Delta _{\pm}(p)&=&
\frac{n}{2N_{c}}\left(m_{1}\lambda_{1}+m_{2}\lambda_{2}\right)
+\int\!\frac{d^{4}k}{(2\pi)^4}\frac{\left(k_{1}M_{2}\pm k_{2}M_{1}\right)}
{\left(k_{1}^2+M_{1}^2\right)\left(k_{2}^2+M_{2}^2\right)}
\nonumber \\ \nonumber \\
&-&(m_{1}+m_{2})\int \!\frac{d^{4}k}{(2\pi)^4}\frac{M_{k}^4+k^4-2M_{k}^2k^2\mp4M_{k}^2k^2}
{\left(k^2+M_{k}^2\right)^3}
\ee
At small momentum, and using
$\left(k_{1}M_{2}-k_{2}M_{1}\right)^2=p^2M_{k}^2+\left(k.p\right)^2
\left(M_{k}^{\prime 2}-2M_{k}M_{k}^{\prime}/k\right)$, we obtain

\be
\Delta _{\pm}(p)&=&p^2\int\!\frac{d^{4}k}{(2\pi)^4}\frac{\left(M_{k}^{2}- kM_{k}M_{k}^{\prime}/2+k^2M_{k}^{\prime 2}/4\right)}{\left(k^2+M_{k}^2\right)^2} \nonumber \\ \\ \nonumber 
&+&\frac{n}{2N_{c}}\left(m_{1}\lambda_{1}+m_{2}\lambda_{2}\right)
-(m_{1}+m_{2})\int \!\frac{d^{4}k}{(2\pi)^4}\frac{M_{k}}{k^2+M_{k}^2}
\ee
To illustrate the fact that the approximations used for A and B 
fail in the case of a strange quark mass, we show the plots of A (Fig. 23)
and B (Fig. 24) for both the up (a)  and strange quark (b). The solid
line is the unexpanded result, while the dotted line is the expanded one.\\
 
\subsection{Appendix H. Outline of the Numerics}
In this appendix, we sketch how the numerical calculations were performed. 
First, we 
solve the integral equation (\ref{gap}) for $\lambda (m)$. With the constituent
mass $M_{k}(m)$ fully known (\ref{mass}), the propagator $S(x,m)$ follows.  The
decomposition (\ref{prop2}) of $S(x,m)$ lends itself to a straightforward
numerical integration of $S_{1}(x,m)$ (\ref{propodd},\ref{propeven}). The
singular behavior at $x=0$ is contained in $S_{0}(x,m)$. In p-space, each
correlator is the sum of an connected part (\ref{free}) and a unconnected part
(\ref{corr}). To speed up the convergence of the numerical integration of
(\ref{free}), the free bubble diagram is removed by hand and later added. The
evaluation of (\ref{corr}) is achieved in stages (numerator and denominator).
Because we will later on numerically Fourier transform the p-space version of
(\ref{corr}), great care is taken at low momenta since the reading of a meson
mass is done at large distance. 
In x-space, the connected part of each correlator is
directly evaluated from the x-space version of the propagator $S(x,m)$ as
\be
C_{\gamma}^{0}(x)=-N_{c}Tr(S(x,m)\gamma S(-x,m) \gamma) 
\ee
 For the unconnected correlator, we numerically Fourier transform its p-space
version. In the pion case, we sum the two parts of the correlator and read the
pion mass from the large distance behavior of $x^{3/2}$ times the correlator.
The kaon unconnected correlator is exponentially damped by a factor of $m_{\pi}-
m_{K}$ with its pion analog and turns out to be of about the same order as its
connected part. Therefore, we single out the unconnected part to read the kaon
mass.  

\vglue 1cm

\bibliographystyle{aip}
\bibliography{vacref}

\newpage
{\bf Figure captions}

\vskip 1cm
{\bf Fig. 1}. The constituent quark mass $M_{k}(m)$ versus 
$z=k\rho /2$ ($k$ being the momentum and $\rho $ the average
 size of the pseudoparticle) for current masses $m$=0 (dashed curve),
 $m$=5 MeV (dotted curve) and $m$=10 MeV (solid curve), respectively. 

{\bf Figs. 2}.
The chirality flip ({\bf 2a}) and non-flip ({\bf 2b}) parts of the quark
propagator (normalized to the free massless quark propagator) versus  x (fm)
for current masses of $5$ MeV (lower curve) and $10$ MeV (upper curve),
respectively. The squares are results of simulations  carried out in \cite{SHU}
for 128 instantons and 128 antiinstantons in a periodic box. The large
distance behaviours are also shown for the chirality  flip ({\bf 2c}) and  
chirality non-fliAp parts ({\bf 2d}).

{\bf Figs. 3}.
The connected ({\bf 3a}) and unconnected ({\bf 3b}) 
parts of the correlator with arbitrary quantum numbers.
The insertions correspond to the external 
background field $P^{\pm}$ as discussed in the text.
The resulting connected ({\bf 3c}) and unconnected ({\bf 3d}) parts 
to leading order in $1/N_c$ counting.

{\bf Figs. 4}. The connected part $C_{\gamma}^{0}(p)$ of the correlator
(normalized to the free and massless correlator). ({\bf 4a})
for the up (down)
quark ($m_{u}=m_{d}=10$ MeV), and  ({\bf 4b}) for
the strange quark  ($m_{s}$=140 MeV)
versus $p$ (fm$^{-1}$).  The channels shown are  scalar
(S), pseudoscalar (P), vector (V), tensor (T) and axial-vector (A),
respectively.

{\bf Figs. 5}.  The same as {\bf Figs. 4} but for the unconnected part 
$C_{\gamma}^{1}(p)$ of the correlator.
The tensor channel vanishes identically for the up and strange  quarks
as does the vector channel for the up quark.

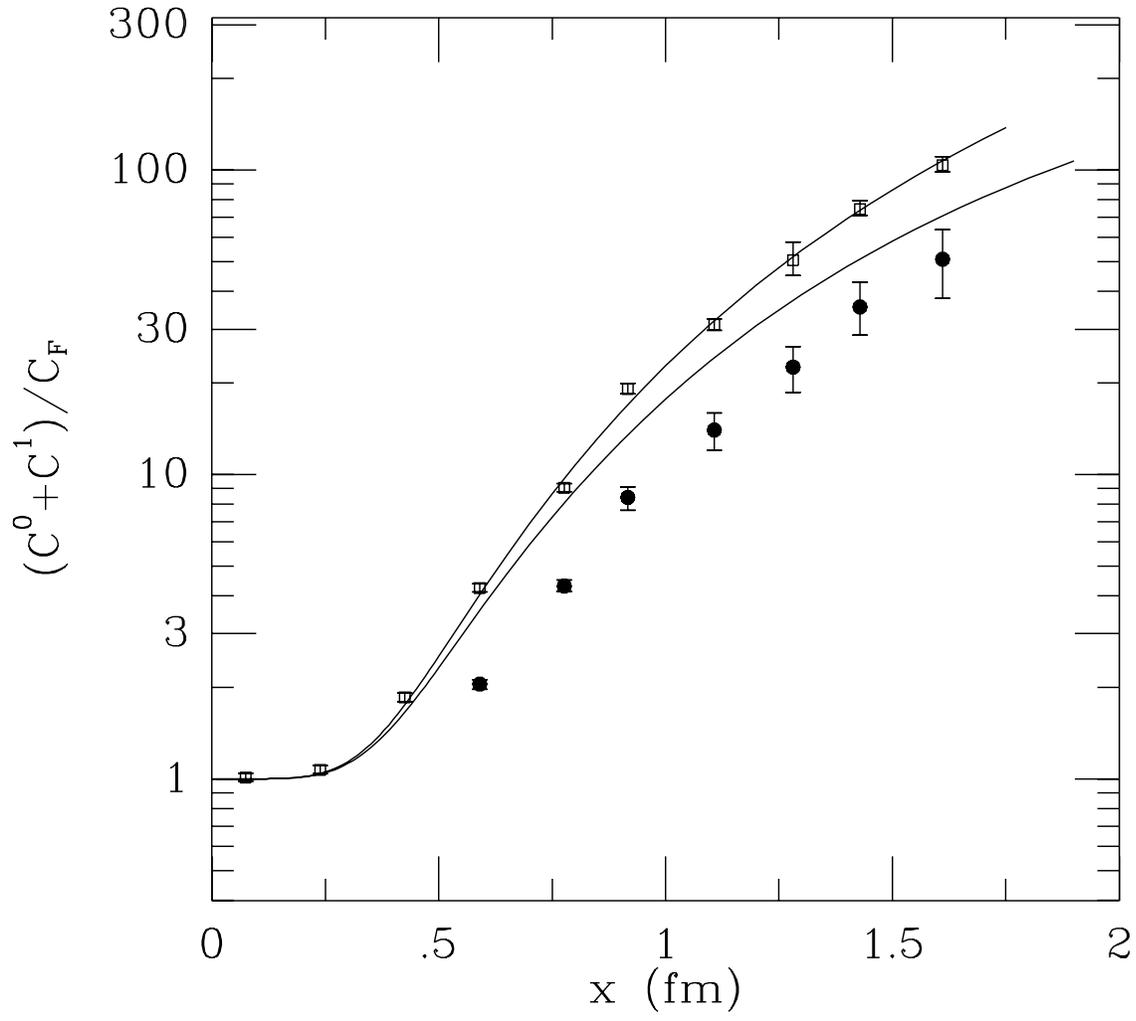
{\bf Fig. 6}. The pion-correlator (normalized to the free and massless 
correlator) versus  $x $ (fm) for $m_{u}=m_{d}=5$ MeV (upper curve) and  
$m_{u}=m_{d}= 10$ MeV (lower curve). 
The squares are the results obtained in \cite{SHU1}
 using 128 instantons and 128 antiinstantons in a periodic box.
  The circles are the results obtained in \cite{NEGELE} from cooled 
 lattice gauge calculations.

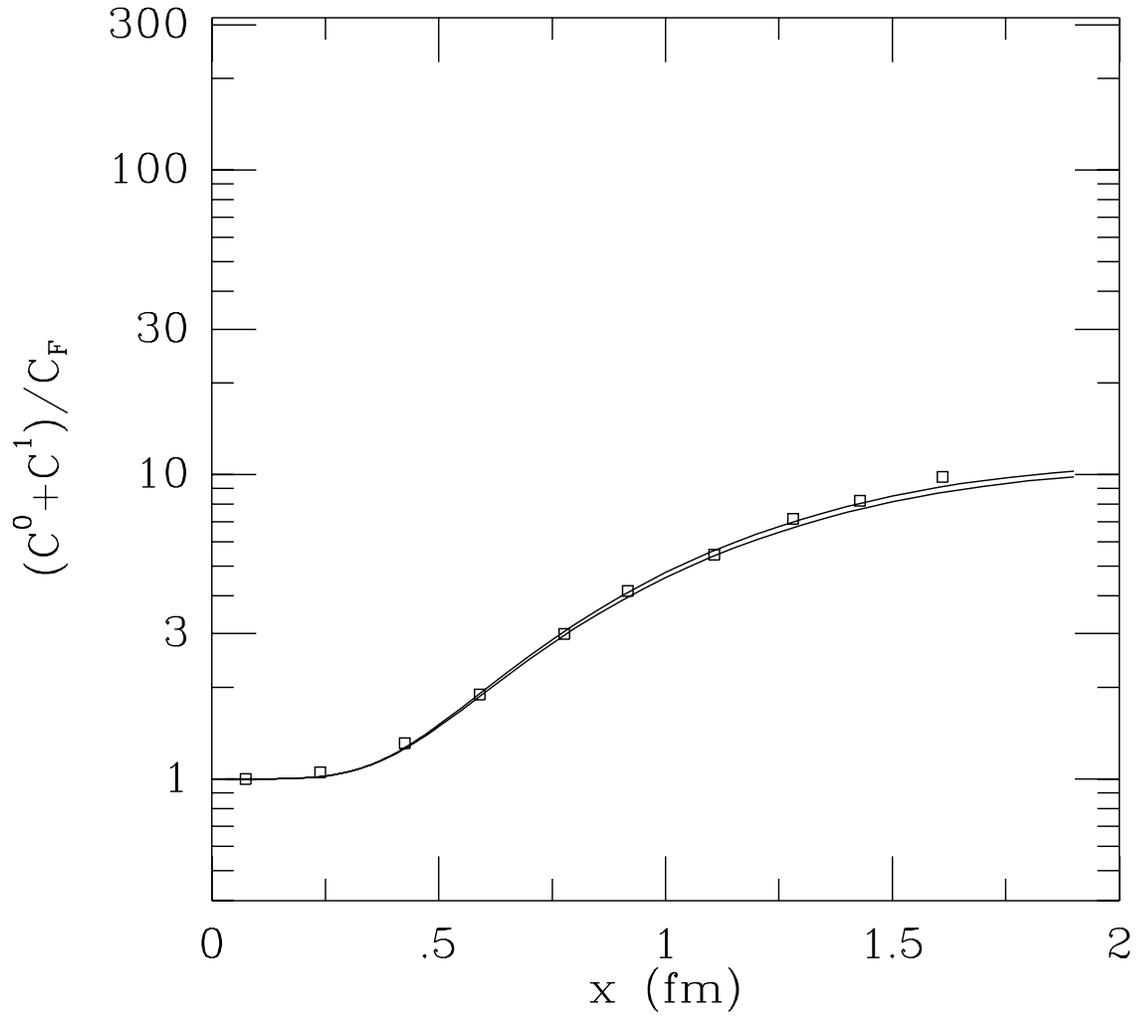
{\bf Fig. 7}. The kaon-correlator (normalized to the free and massless 
correlator) versus $x$  (fm) for $m_{u}=m_{d}=5$ MeV (upper curve) 
and for $m_{u}=m_{d}= 10$ MeV (lower curve). 
The squares are the results obtained in \cite{SHU1}
 using 128 instantons and 128 antiinstantons in a periodic box.

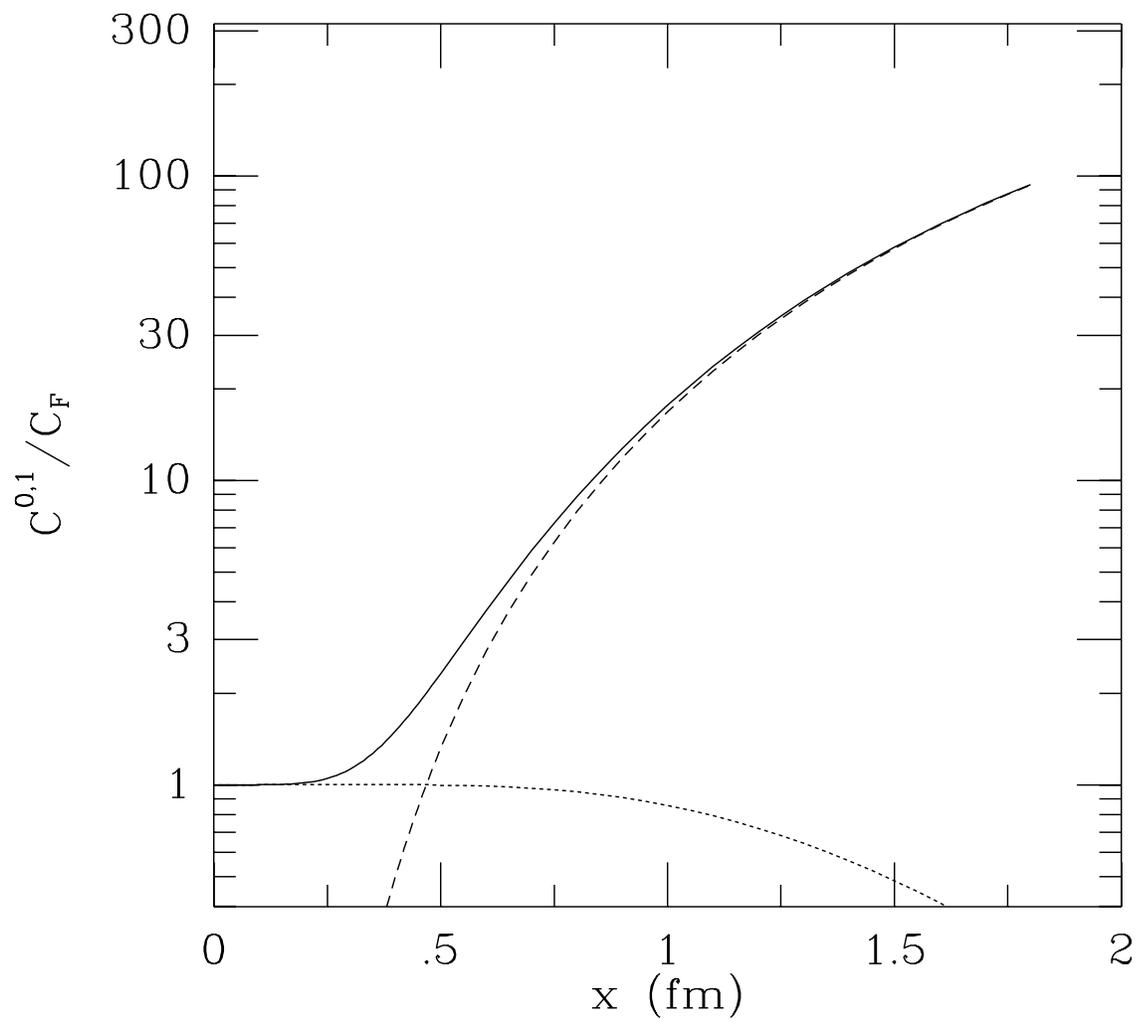
{\bf Fig. 8}. In the pion channel, the connected $C^{0}$($x,m_{u}$), unconnected
$C^{1}$($x,m_{u}$) and full correlators for $m_{u}$=10 MeV versus $x$ (fm) are
plotted  in dashed, dotted and solid lines, respectively.

{\bf Figs. 9}. Large distance behaviour of the connected and normalized
correlator $C^{0}(x,m)$. ({\bf 9a}) for up (down) and ({\bf 9b}) for
strange quarks.

{\bf Figs. 10}. Large distance behaviour of the connected
correlator times $x^{3}$. ({\bf 10a}) for up (down) quark, 
and  ({\bf 10b}) for strange quark, respectively.

{\bf Fig. 11}. In the pion channel, the total correlator times $x^{3/2}$ versus 
 $x$ (fm) is plotted for $m_{u}$=5 MeV 
 (upper curve) and $m_{u}$=10 MeV (lower curve), respectively.

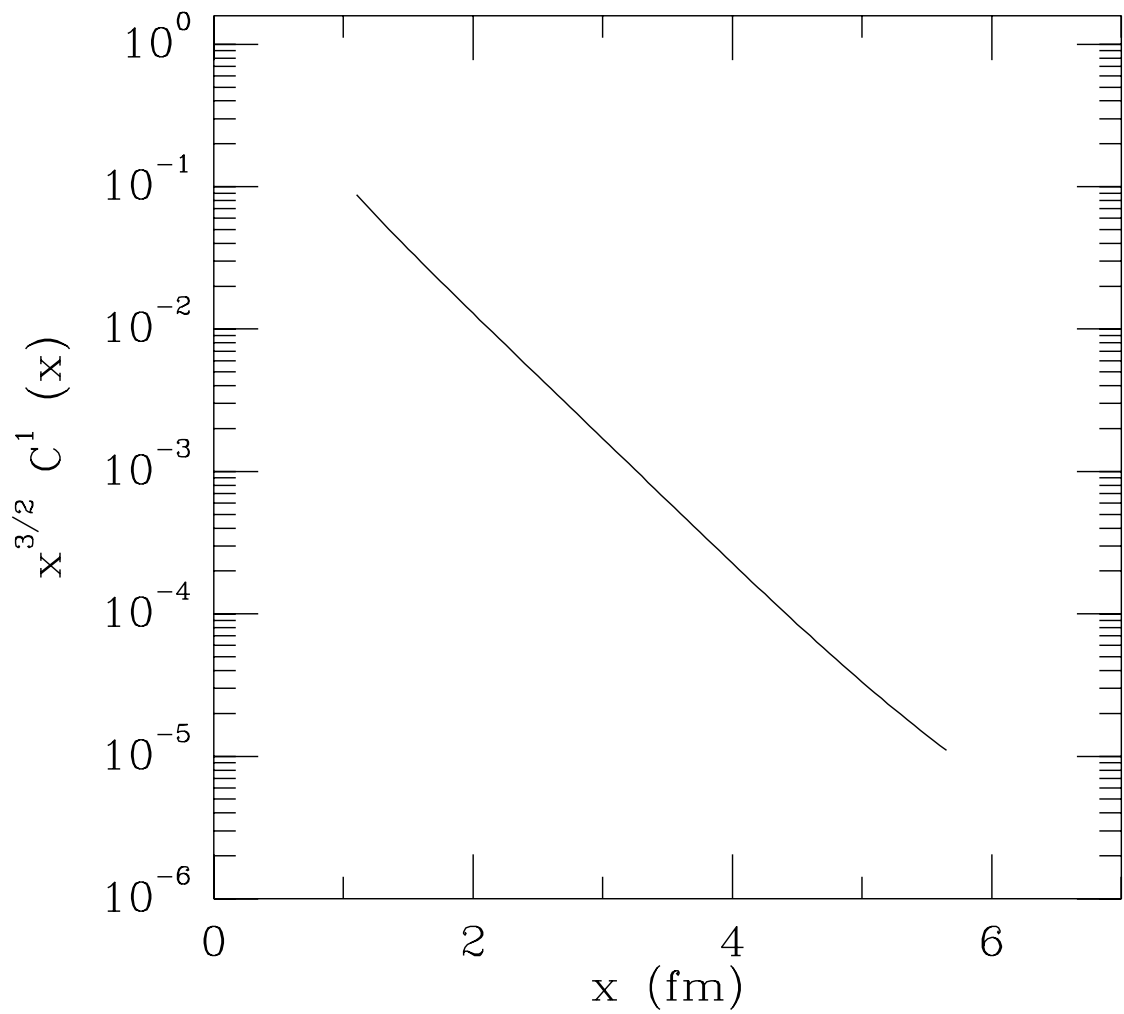
{\bf Fig. 12}. In the kaon channel, the connected correlator times $x^{3/2}$
 versus  $x$ (fm) is plotted for $m_{s}$=140 MeV.

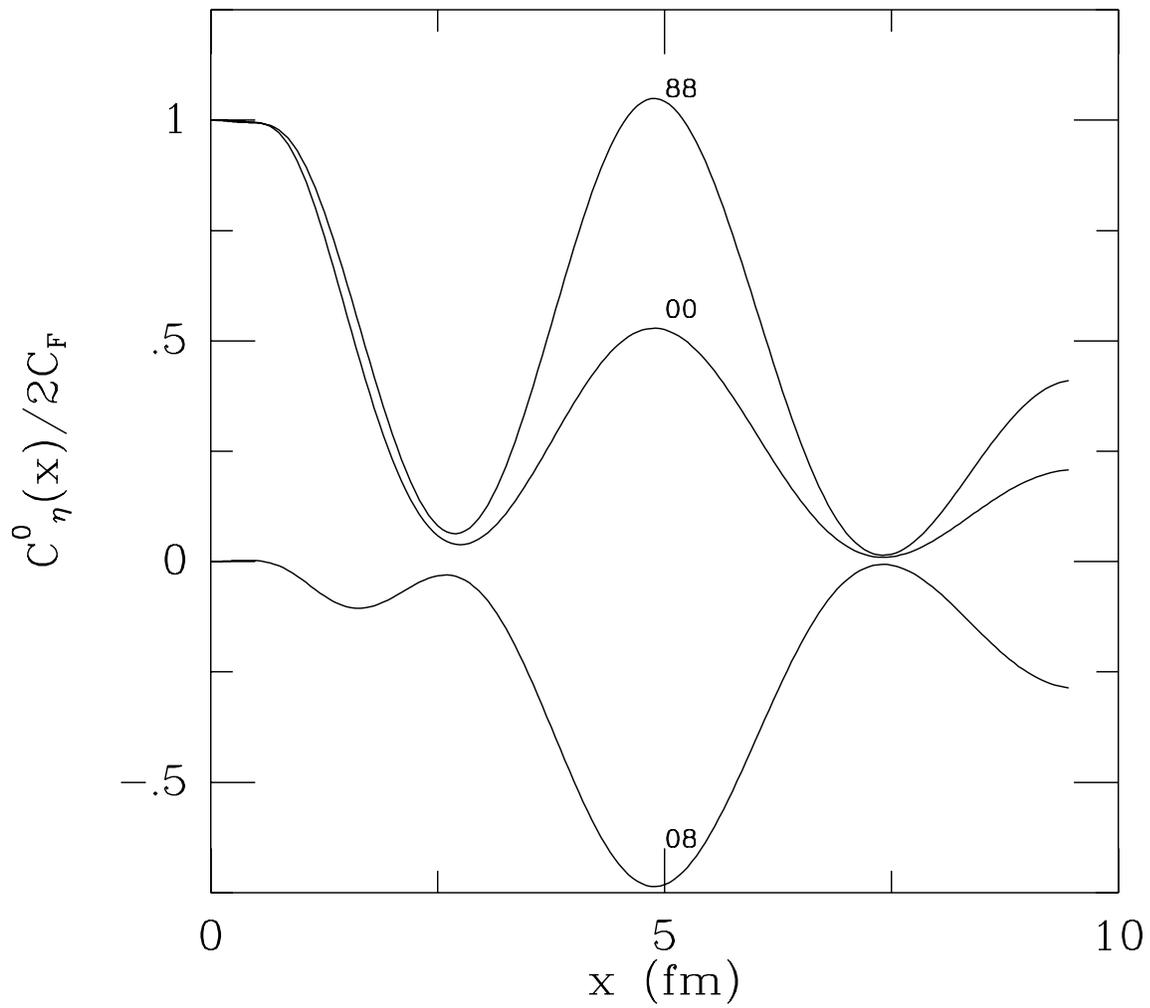
{\bf Fig. 13}. 
 The connected correlator (normalized to the free and massless 
correlator) for the $\eta$ (00,88,08) versus  $x$  (fm). 

{\bf Figs. 14}. The unconnected correlator times $x^{3/2}$ for the (00) channel
({\bf 14a}) and the (88) channel ({\bf 14b}) versus  $x$  (fm) is  
plotted with (upper curve) and without (lower curve) a topological 
susceptibility.

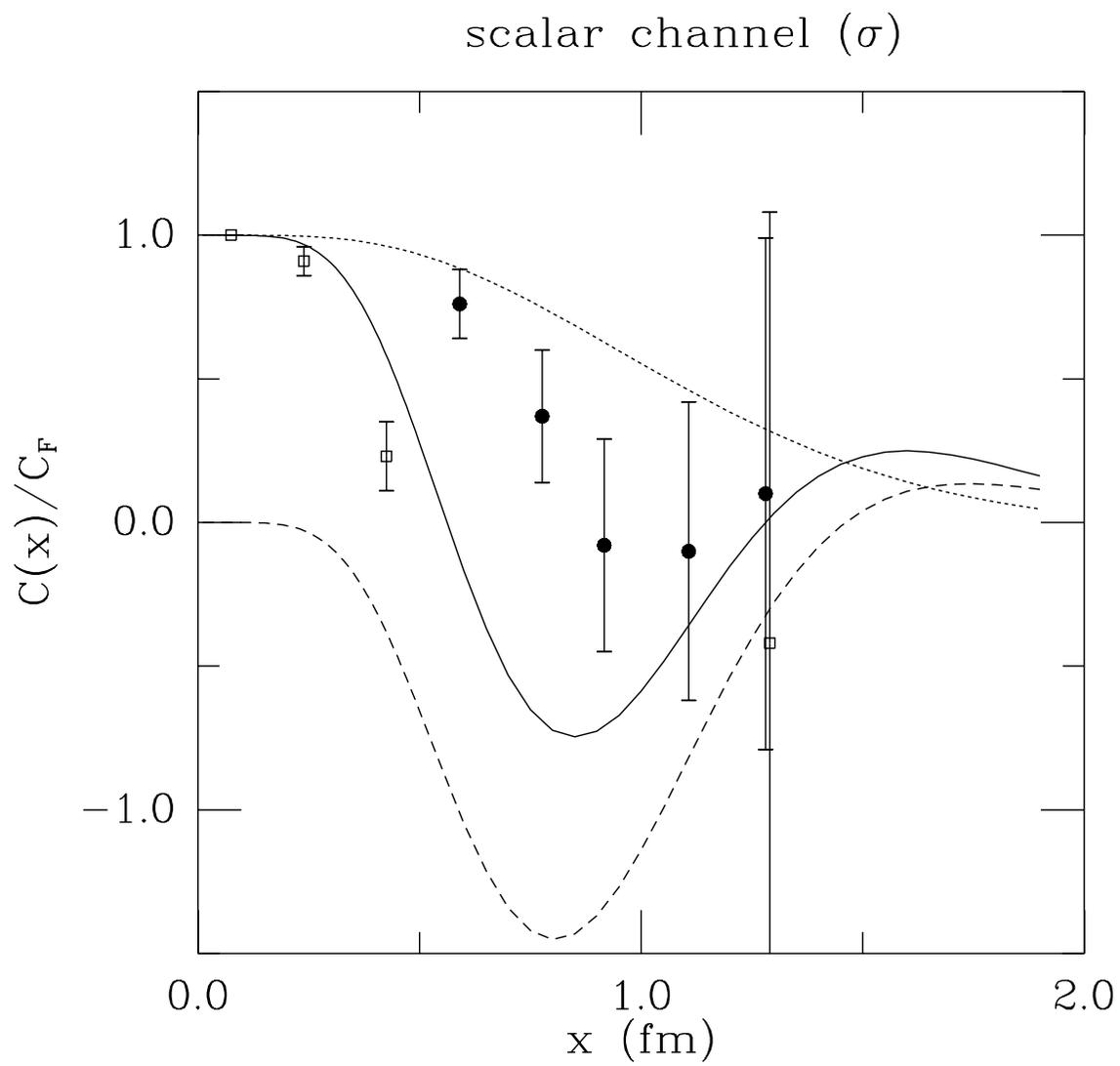
{\bf Fig. 15}. The  scalar connected and unconnected
sigma meson-correlator 
(normalized to the free and massless correlator)
 versus  $x $ (fm) for $m_{u}=m_{d}$= 10 MeV are respectively plotted
in short dashed and long dashed lines. The solid line represents their sum.  
The squares are the results obtained in \cite{SHU1}
 using 128 instantons and 128 antiinstantons in a periodic box.
  The circles are the results obtained in \cite{NEGELE} from cooled 
 lattice gauge calculations.

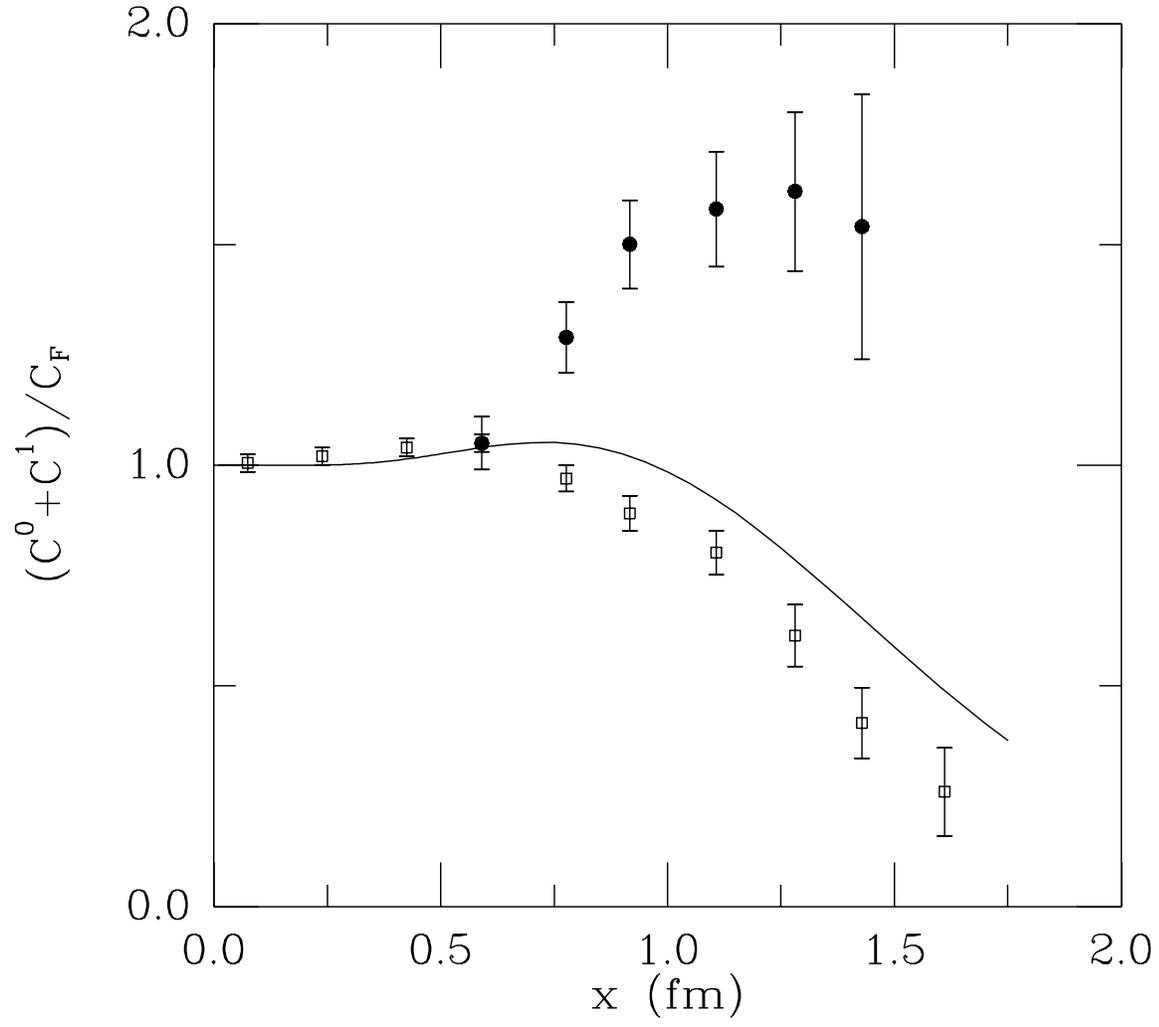
{\bf Fig. 16}.  The total correlator (normalized to the free and massless 
correlator) in the $\rho$ meson channel versus  $x$ (fm), 
for $m_{u}=m_{d}=10$ MeV. The squares are the results of
\cite{SHU1} using 128 instantons and 128 antiinstantons in a periodic box.

{\bf Fig. 17}.  The total correlator (normalized to the free and massless 
correlator) in the $\phi$ meson channel versus  $x$ (fm), 
for $m_{u}=m_{d}=10$ MeV. The squares are the results of
\cite{SHU1} using 128 instantons and 128 antiinstantons in a periodic box.

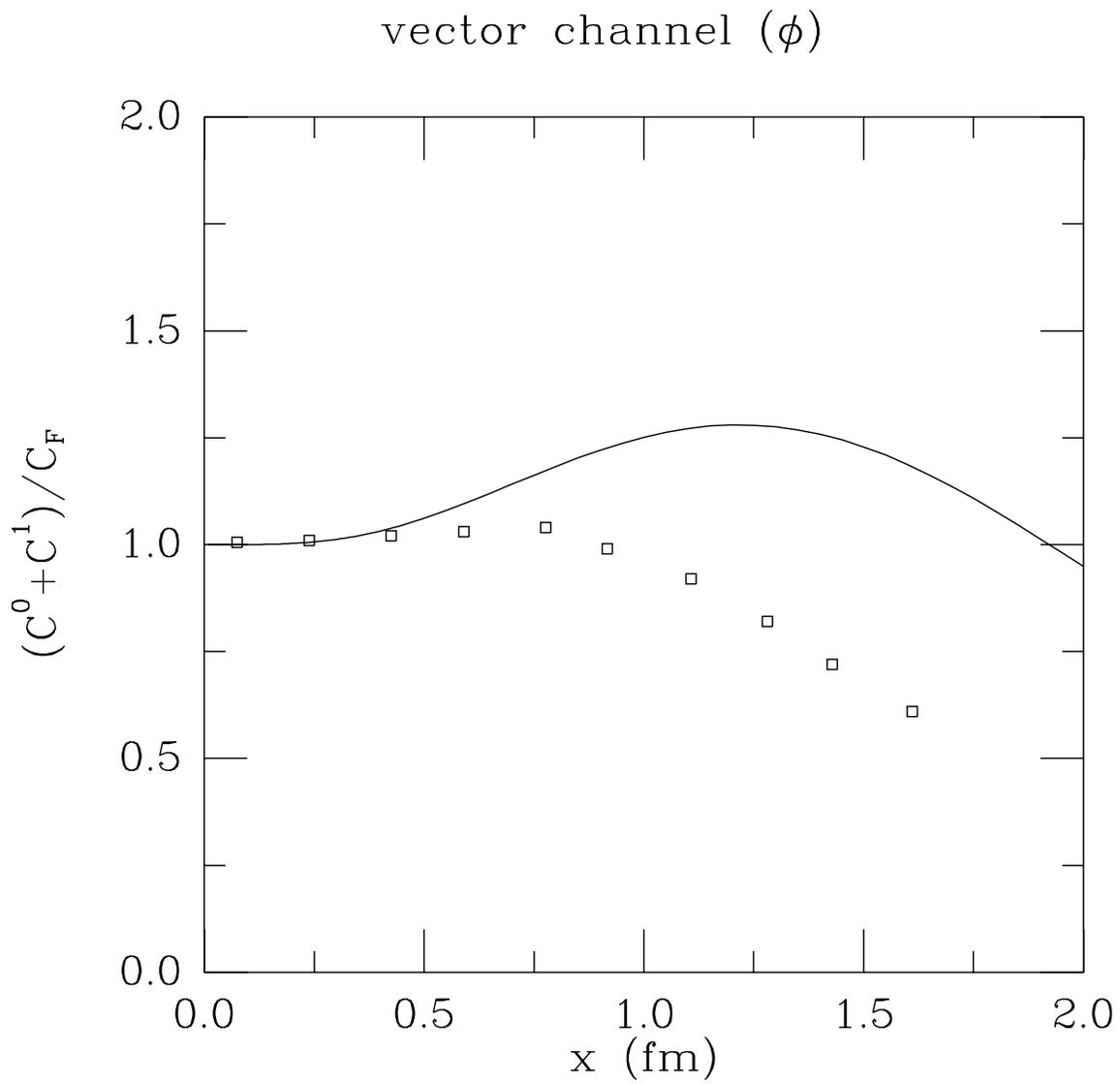
{\bf Fig. 18}.  The total correlator (normalized to the free and massless 
correlator) in the $K^*$ meson channel versus  $x$ (fm), 
for $m_{u}=m_{d}=5$ MeV (upper curve) and $m_{u}=m_{d}=10$ MeV
(lower curve). The squares are the results of
\cite{SHU1} using 128 instantons and 128 antiinstantons in a periodic box.

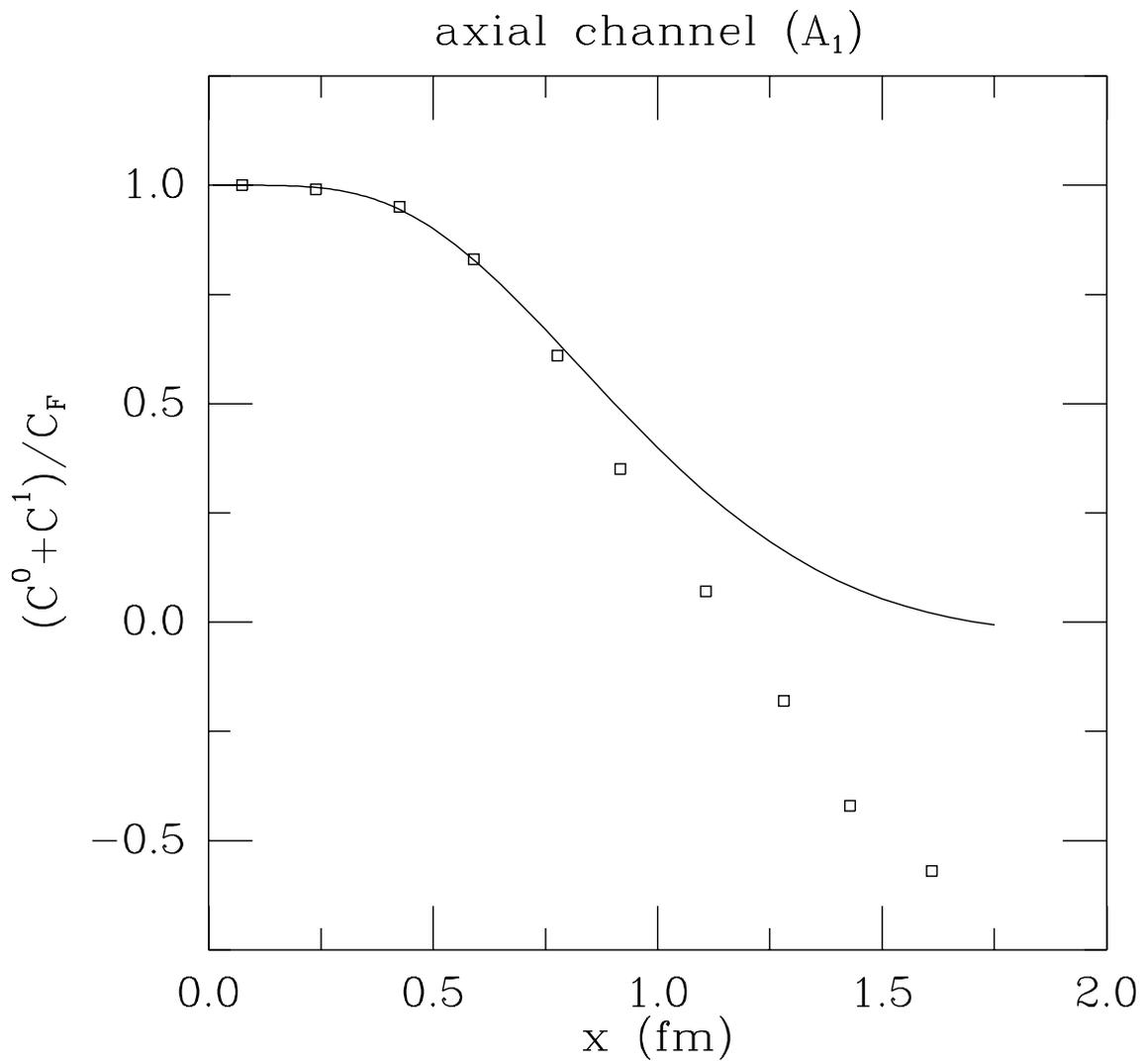
{\bf Fig. 19}.  The total correlator (normalized to the free and massless 
correlator) in the $A_1$ meson channel versus  $x$ (fm), 
for $m_{u}=m_{d}=10$ MeV. The squares are the results of
\cite{SHU1} using 128 instantons and 128 antiinstantons in a periodic box. 

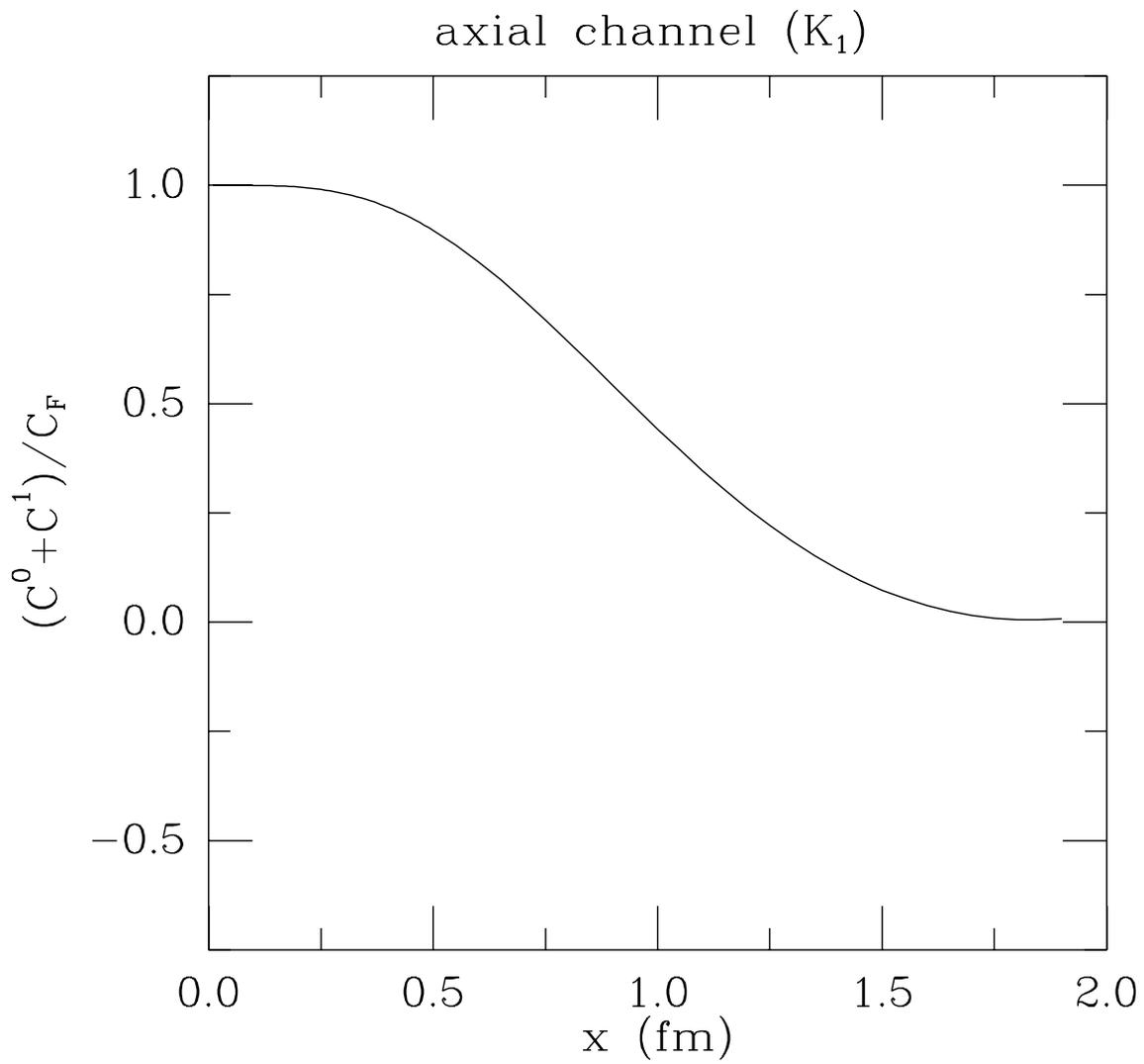
{\bf Fig. 20}.  The total correlator (normalized to the free and massless 
correlator) for the $K_{1}$ meson channel versus  $x $ (fm) 
 for $m_{u}=m_{d}$=\,10 MeV.

{\bf Fig. 21}. The total correlator (normalized to the free and massless
 correlator) in the nucleon channel versus $x$ (fm) is plotted 
 for $m_{u}=m_{d}=5$ MeV (upper curve) and for $m_{u}=m_{d}=10$ 
MeV (lower curve), respectively. 
The squares are the results obtained in \cite{SHU1} using 128 instantons and
128 antiinstantons in a periodic box. The circles are the results obtained
in \cite{NEGELE} from cooled  lattice gauge calculations.

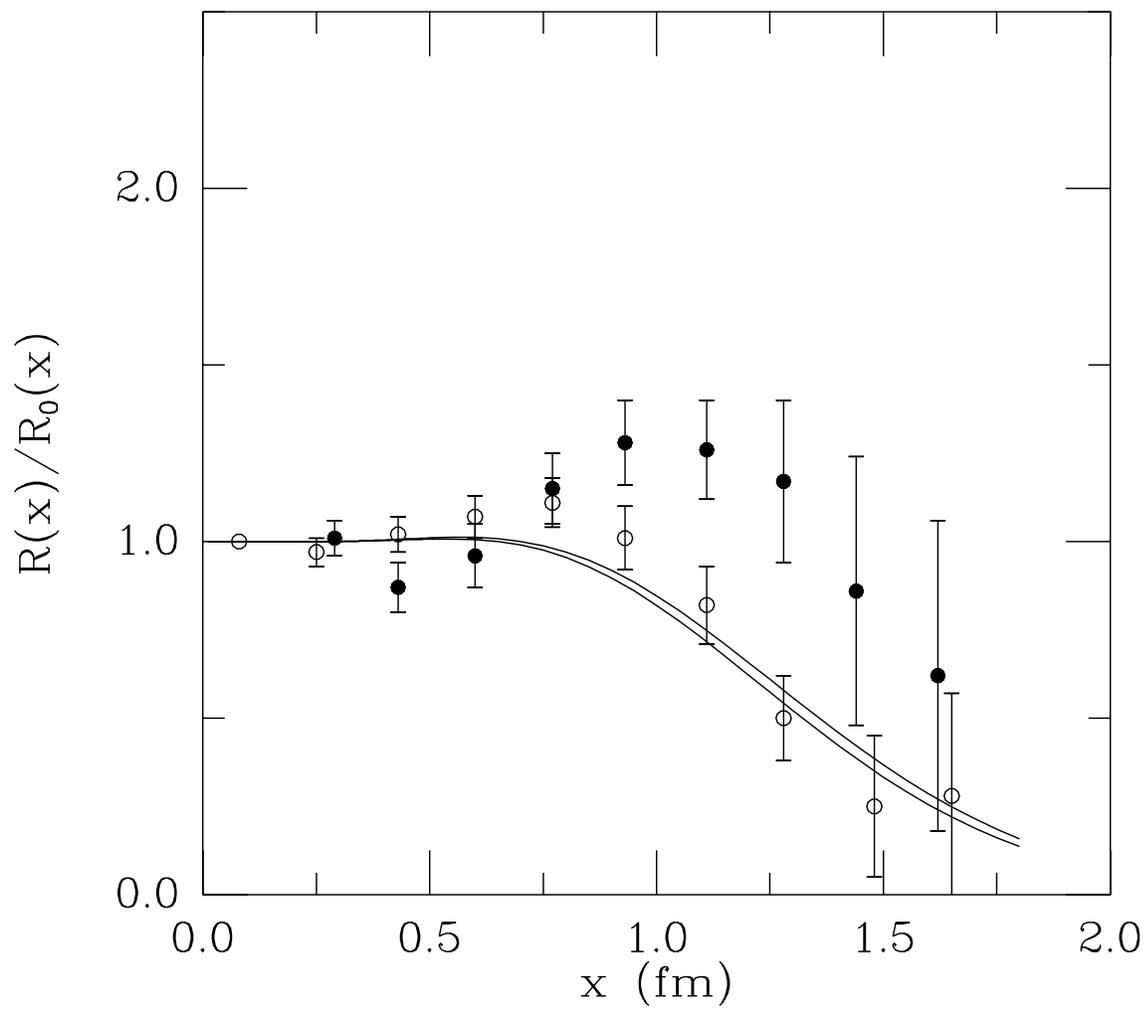
{\bf Fig. 22}.  The total correlator (normalized to the free and massless
 correlator) in the $\Delta$ channel versus  $x$ (fm) is plotted 
 for $m_{u}=m_{d}=5$ MeV (upper curve) and for $m_{u}=m_{d}=10$ 
MeV (lower curve). The squares are the results obtained in \cite{SHU1} 
using 128 instantons and 128 antiinstantons in a periodic box.
  The circles are the results obtained in \cite{NEGELE} from cooled 
 lattice gauge calculations.

{\bf Figs. 23}. The coefficient A(k,m) versus $z=k\rho/2$ 
for up ({\bf 23a}) and for strange ({\bf 23b}) quarks. The solid line is the
unexpanded result and the dotted line is the expanded result.

{\bf Figs. 24}. The coefficient B(k,m) versus $z=k\rho/2$ for
up ({\bf 24a}) and for strange ({\bf 24b}) quarks. The solid line is the
unexpanded result and the dotted line is the expanded result.

{\bf Figs. 25}. The scalar (pseudoscalar) gluonic correlator is plotted
in Fig. 25a (b). The points are  results from simulations in 
(\cite{SHAEFER}).

{\bf Figs. 26}. In (a) and (b),  we display the insertion mechanism
involved in evaluating the gluonic form factor of a constituent quark.
Figs (c) and (d)  show  the mixing that enters in the connected part.

{\bf Figs. 27}.  The insertion mechanism 
for the fermionic form factor of a constituent quark.

{\bf Figs. 28}. The leading contributions to the fermionic form factor. 

{\bf Figs 29}. The form factor in terms of Ioffe's current.

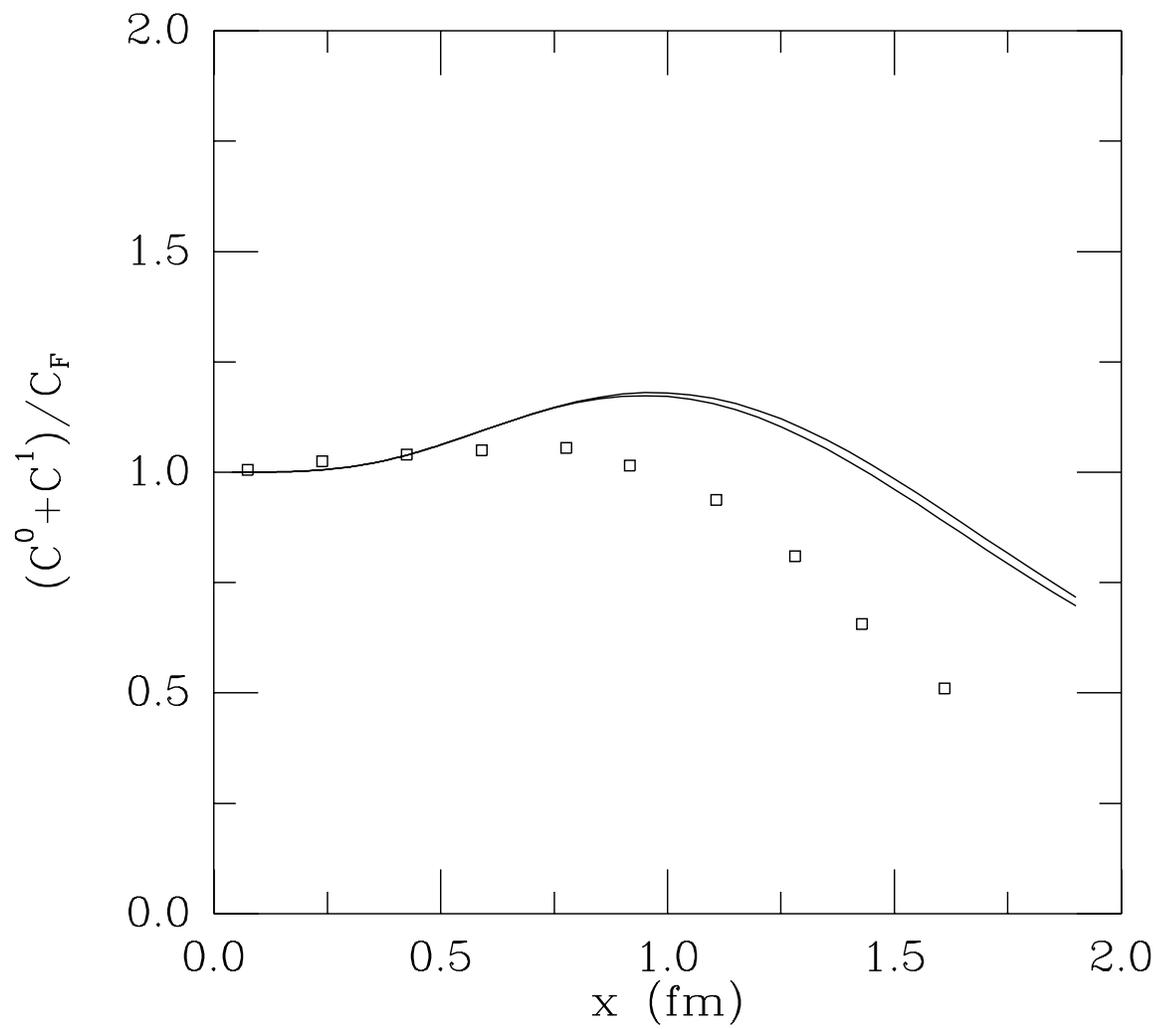
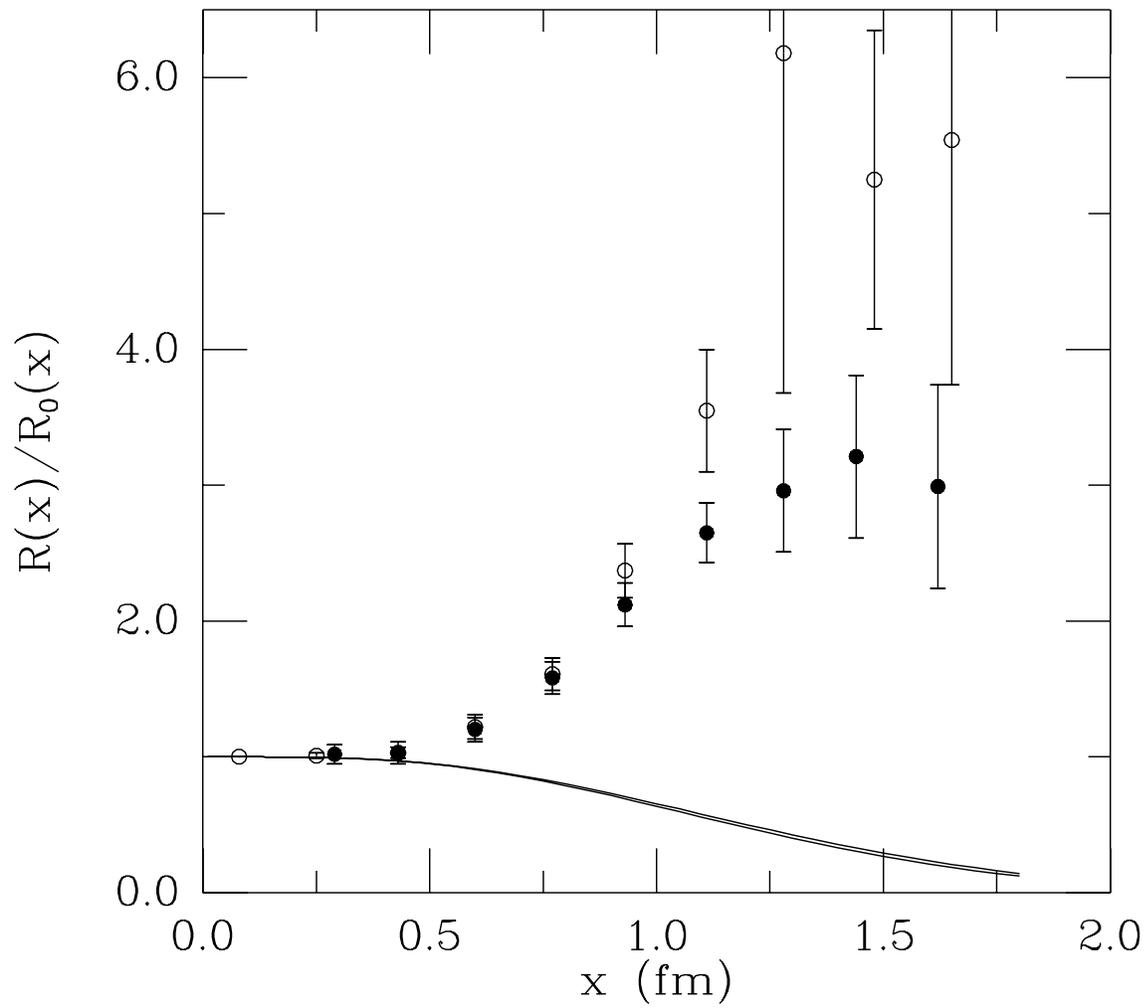
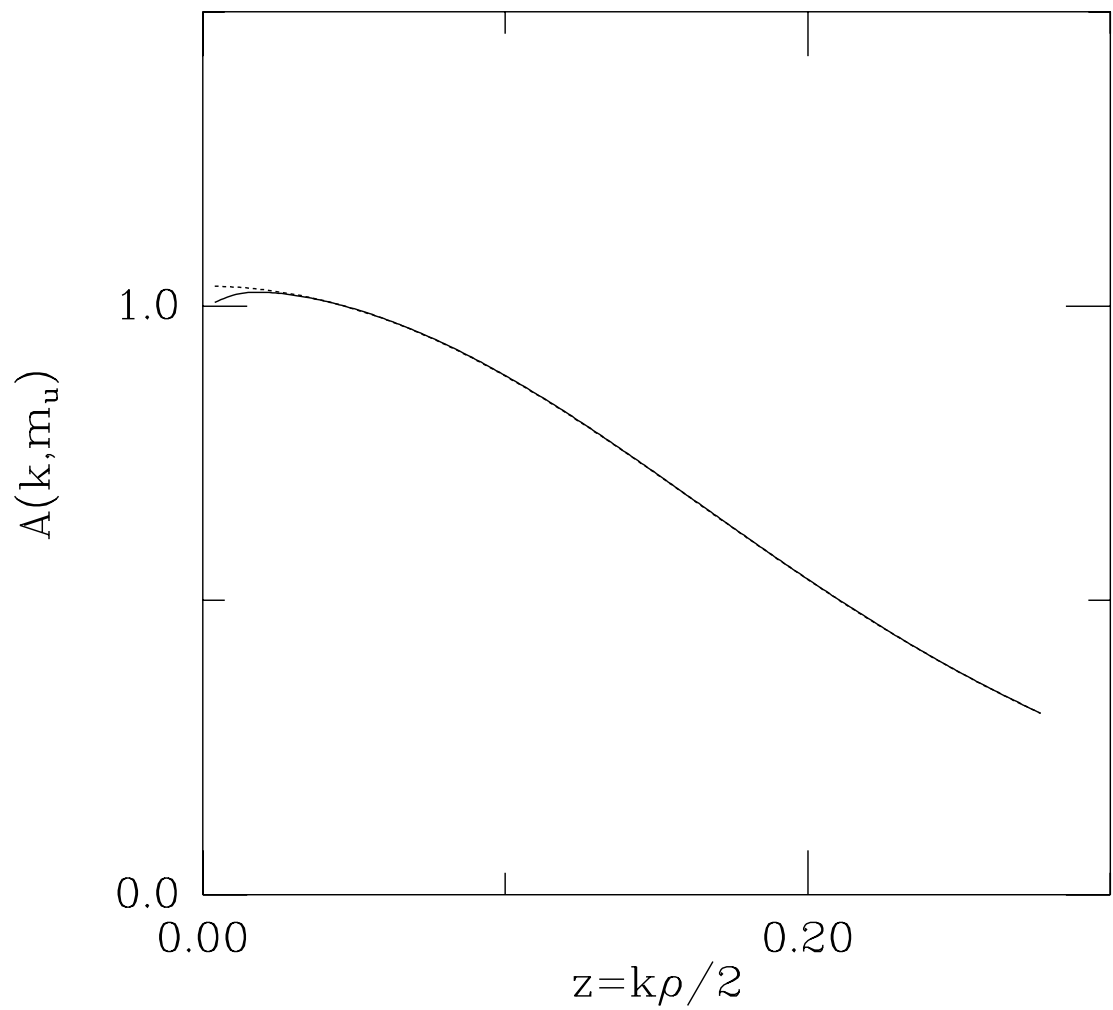
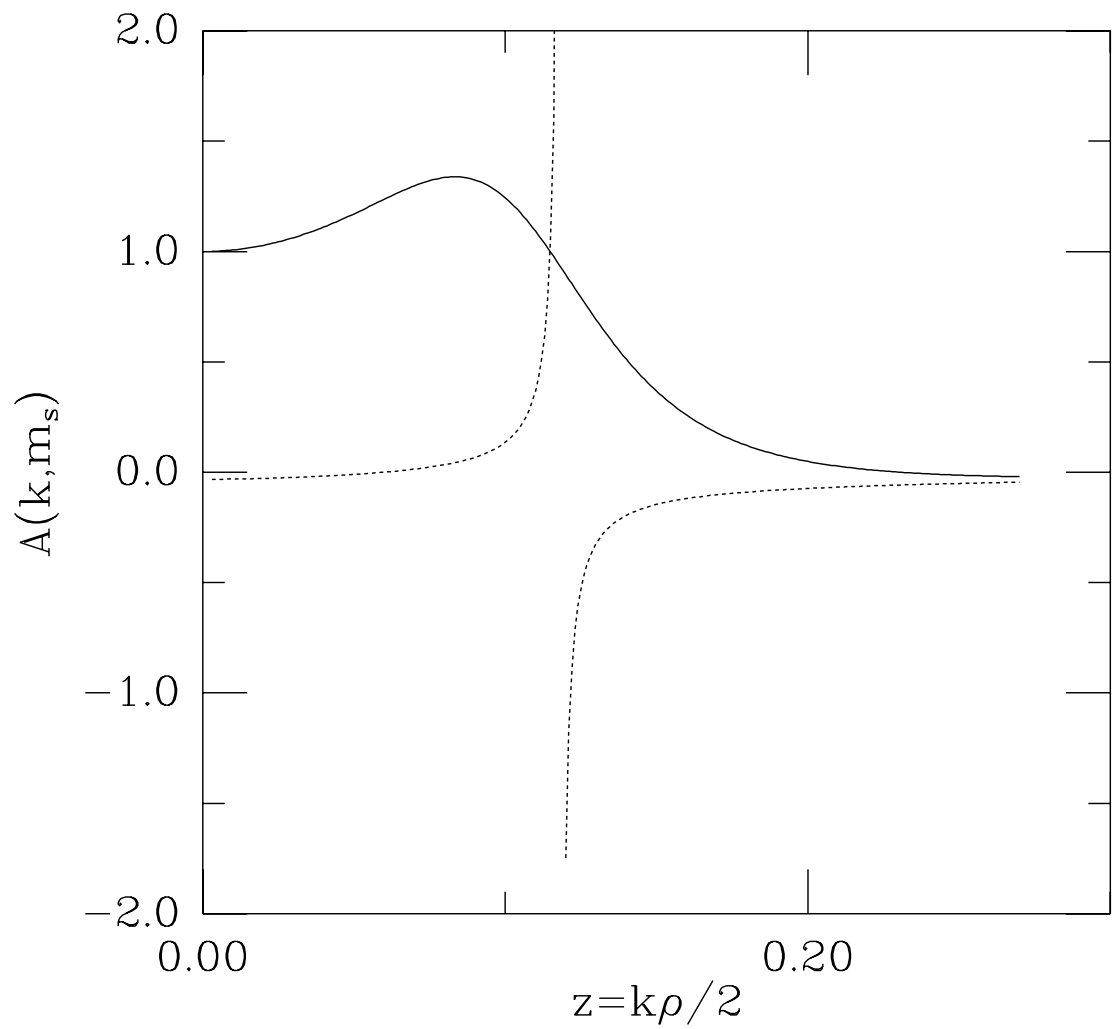
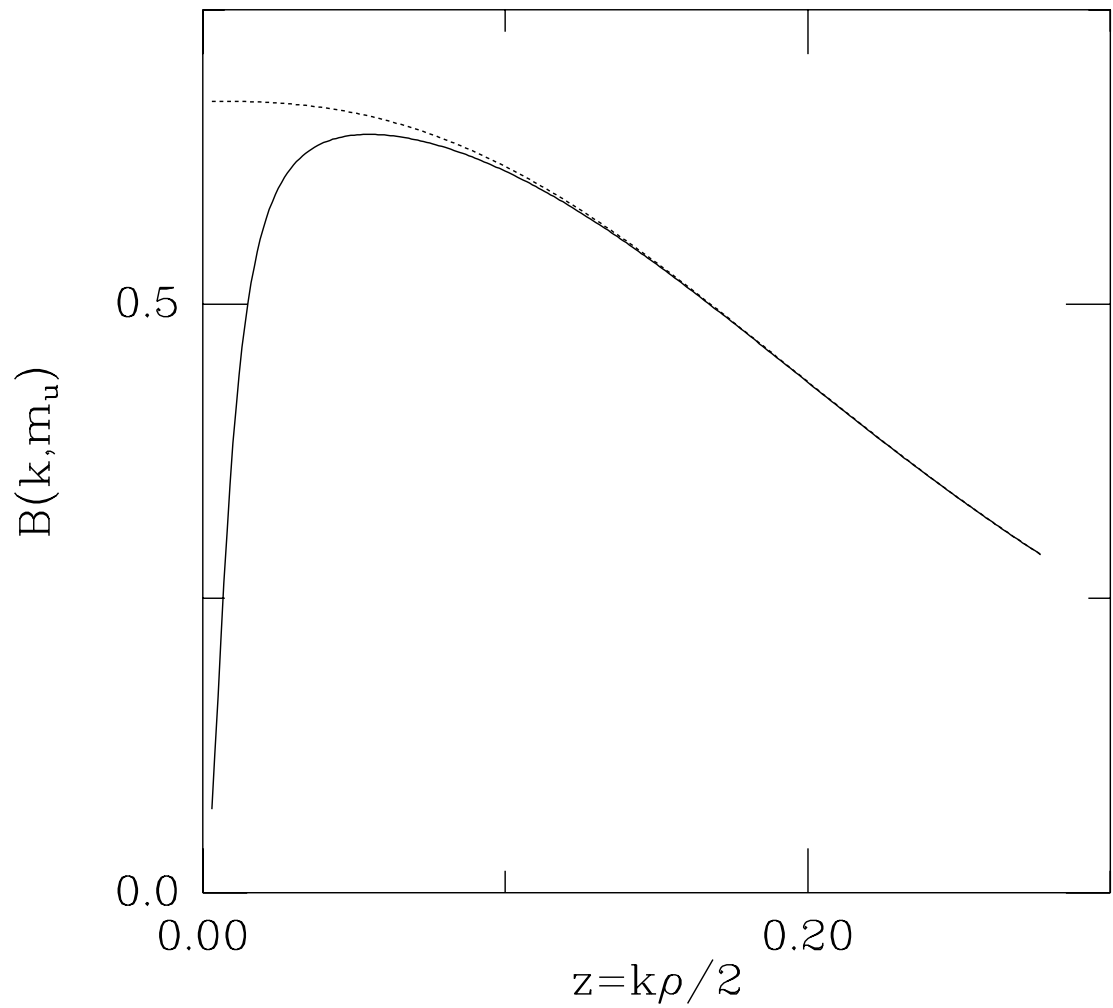
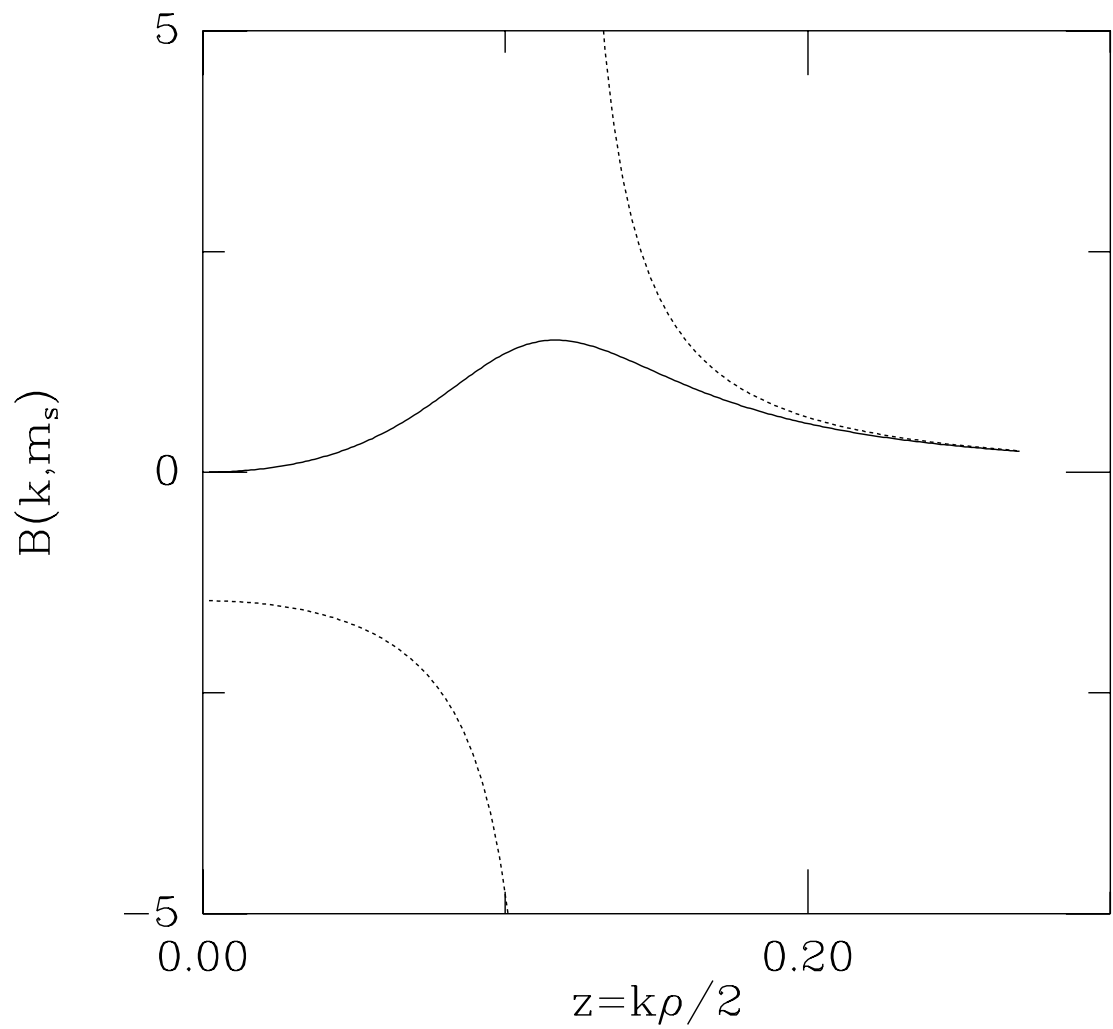
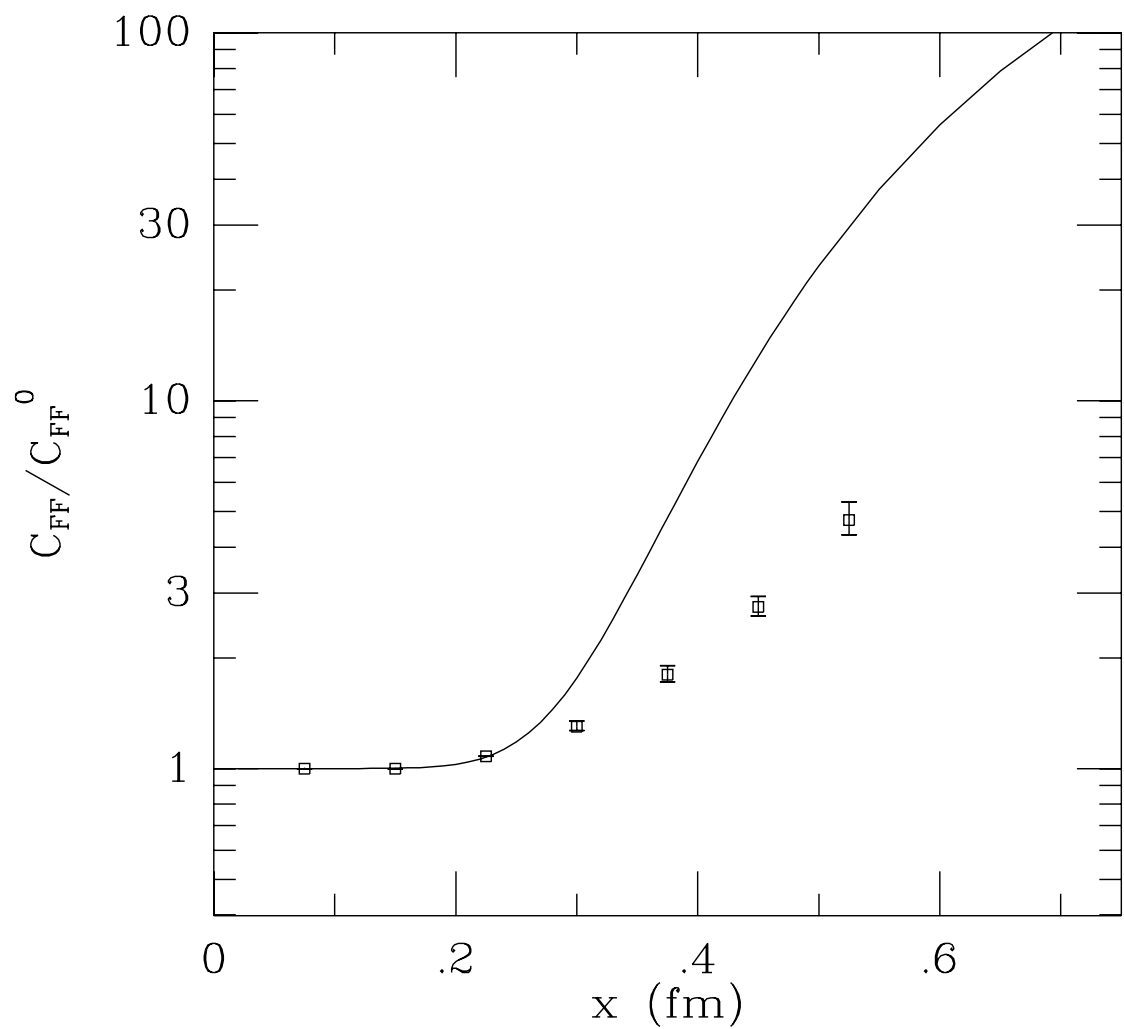
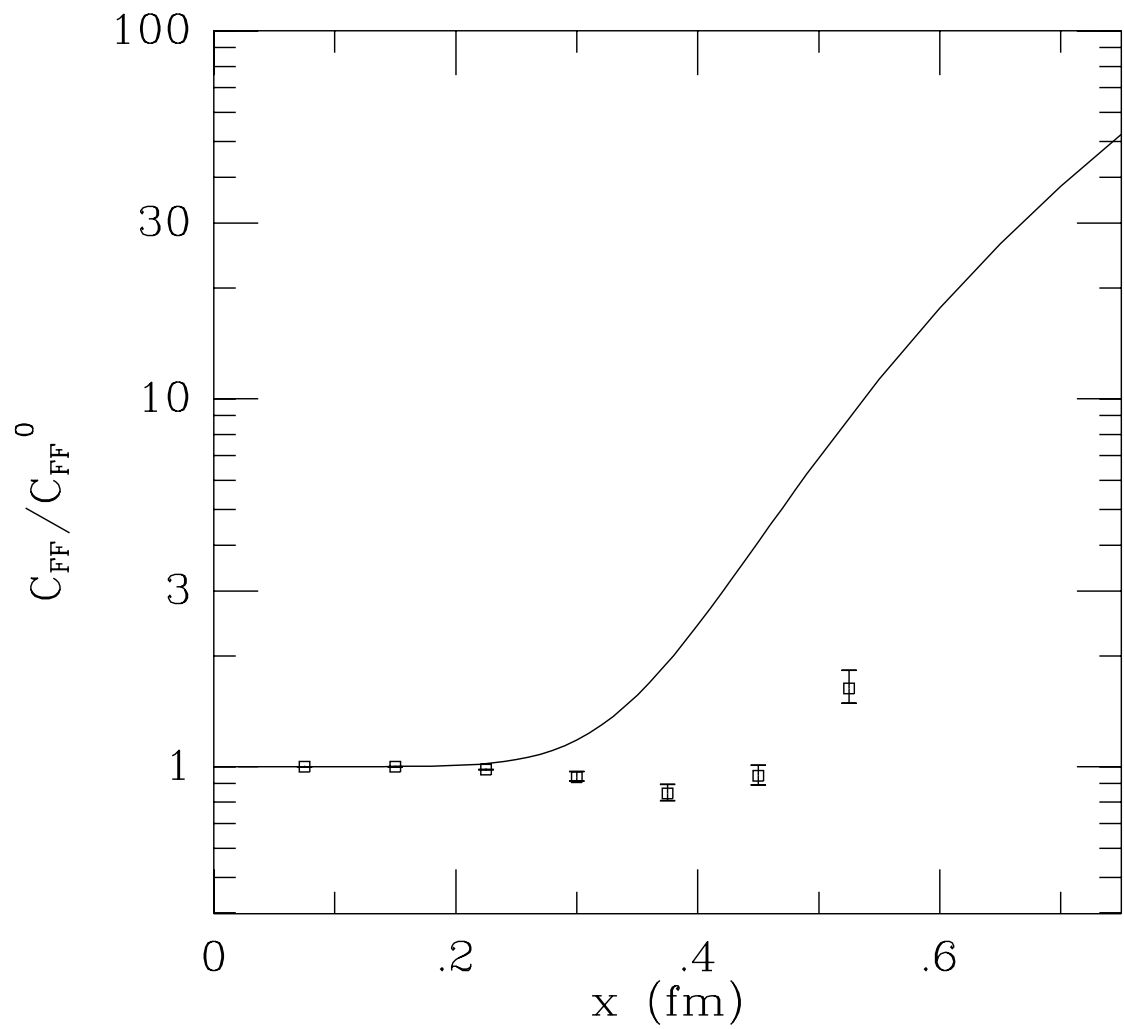
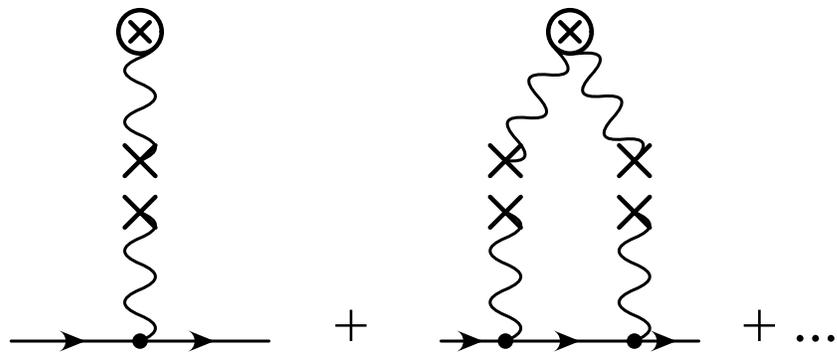
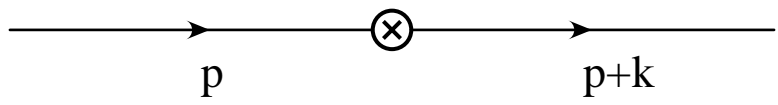
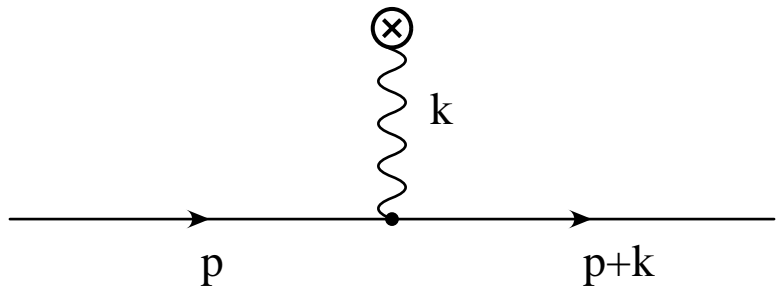
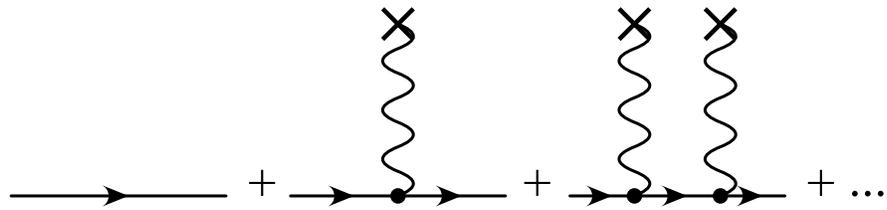
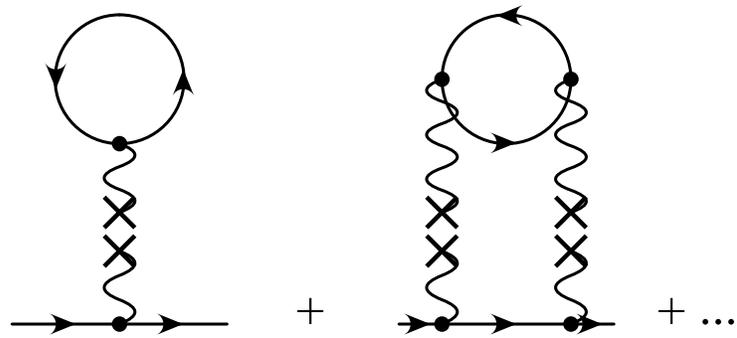
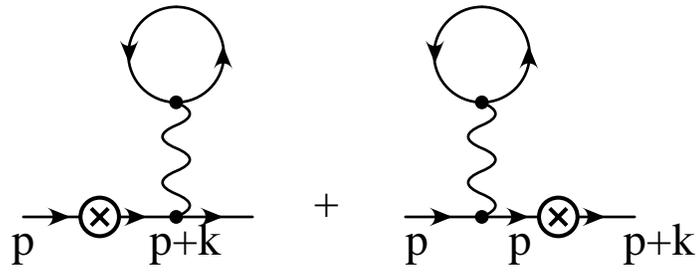
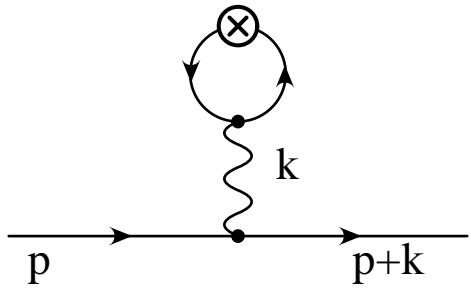
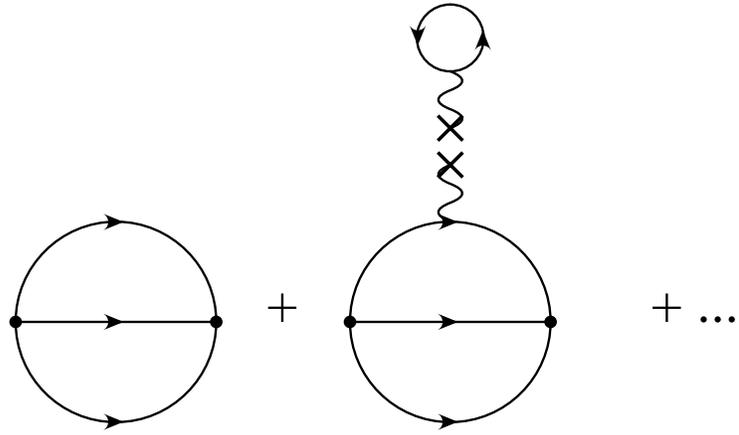
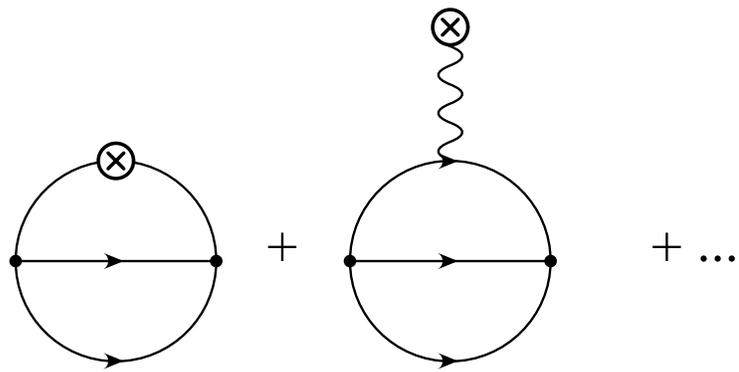
\end{document}